\documentclass[twocolumn]{IEEEtran}
\IEEEoverridecommandlockouts
\normalsize
\usepackage{fixltx2e}
\usepackage{mathrsfs}
\usepackage{amsmath}
\usepackage{amssymb}
\usepackage{mathtools}
\usepackage{graphics}
\usepackage{float}
\usepackage{epstopdf}
\usepackage{graphicx}
\usepackage{bm}
\usepackage{amsthm}
\usepackage{tikz}
\usepackage[lined,boxed,commentsnumbered,linesnumbered,ruled]{algorithm2e}
\usepackage{cite}
\usepackage{comment}
\usepackage{balance}
\usepackage{booktabs} 
\usepackage{enumitem}
\usepackage{mathtools}
\usetikzlibrary{shapes, patterns,decorations.text, decorations.pathreplacing}
\usepackage{ifthen}
\usepackage{multirow}
\usepackage{hyperref} 
\usepackage[capitalize]{cleveref}
\crefname{equation}{\unskip}{\unskip}

\newcommand*{\Scale}[2][4]{\scalebox{#1}{\ensuremath{#2}}}%

\newtheorem{example}{Example}
\newtheorem{theorem}{Theorem}
\newtheorem{definition}{Definition}
\newtheorem{lemma}{Lemma}
\newtheorem{claim}{Claim}
\newtheorem{proposition}{Proposition}
\newtheorem{corollary}{Corollary}
\newtheorem{conjecture}{Conjecture}
\newtheorem{remark}{Remark}

\allowdisplaybreaks

\newcommand{\X}{\bm X}
\newcommand{\BDelta}{\boldsymbol{\Delta}}
\newcommand{\eqdef}{\triangleq}              
\newcommand{\set}[1]{\mathcal{#1}}           
\newcommand{\card}[1]{\left|#1\right|}       
\newcommand{\ecard}[1]{|#1|}
\newcommand{\bigcard}[1]{\bigl|#1\bigr|} 
\newcommand{\mat}[1]{\bm{#1}}                
\newcommand{\vect}[1]{\bm{#1}}               
\newcommand{\trans}[1]{#1^{\top}}            
\newcommand{\GF}{\mathrm{GF}}                
\newcommand{\Nat}[1]{\mathbb{N}_{#1}}        
\newcommand{\code}[1]{\mathcal{#1}}          
\newcommand{\Hwt}[1]{w_\mathsf{H}\left(#1\right)}       
\newcommand{\inv}[1]{#1^{-1}}                
\newcommand{\const}[1]{\textnormal{\usefont{U}{eur}{m}{n}\selectfont #1}} 
\newcommand{\rank}[1]{\mathrm{rank}\left(#1\right)} 

\newcommand{\bigrank}[1]{\operatorname{rank}\bigl(#1\bigr)}
\newcommand{\Bigrank}[1]{\operatorname{rank}\Bigl(#1\Bigr)}
\newcommand{\Dim}[1]{\operatorname{dim}\left(#1\right)} 

\renewcommand{\r}{\color{red}} 
\renewcommand{\b}{\color{blue}} 
\newcommand{\mg}{\color{black}} 
\def\rot#1{\rotatebox{90}{#1}}
\makeatletter
\newcommand*\rel@kern[1]{\kern#1\dimexpr\macc@kerna}
\newcommand*\widebar[1]{%
  \begingroup
  \def\mathaccent##1##2{%
    \rel@kern{0.8}%
    \overline{\rel@kern{-0.8}\macc@nucleus\rel@kern{0.2}}%
    \rel@kern{-0.2}%
  }%
  \macc@depth\@ne
  \let\math@bgroup\@empty \let\math@egroup\macc@set@skewchar
  \mathsurround\z@ \frozen@everymath{\mathgroup\macc@group\relax}%
  \macc@set@skewchar\relax
  \let\mathaccentV\macc@nested@a
  \macc@nested@a\relax111{#1}%
  \endgroup
}
\makeatother

\newcommand{\Ehat}{\hat{\mat{E}}}
\newcommand{\C}{\mathcal{C}}
\newcommand{\Cbar}{\bar{\mathcal{C}}}
\newcommand{\Rub}{\const{R}_{\mathsf{UB}}}
\renewcommand{\trans}[1]{#1^{\textup{\textsf{\tiny T}}}} 

\begin{document}

\title{Achieving Maximum Distance Separable Private Information Retrieval Capacity With Linear Codes}

\author{Siddhartha~Kumar,~\IEEEmembership{Student~Member,~IEEE},
  Hsuan-Yin~Lin,~\IEEEmembership{Senior~Member,~IEEE},\\
  Eirik~Rosnes,~\IEEEmembership{Senior~Member,~IEEE}, and
  Alexandre~Graell~i~Amat,~\IEEEmembership{Senior~Member,~IEEE}
  \thanks{This work was partially funded by the Research Council of Norway (grant 240985/F20) and the Swedish Research
    Council (grant \#2016-04253). This paper was presented in part at the IEEE International Symposium on Information
    Theory (ISIT), Aachen, Germany, June 2017, {\mg at the  IEEE International Symposium on Information
    Theory (ISIT), Vail, CO, USA, June 2018, and at the IEEE Information Theory Workshop (ITW), Guangzhou, China, November 2018.}}%
  \thanks{S.\ Kumar, H.-Y.\ Lin, and E.\ Rosnes are with Simula UiB, N-5008 Bergen, Norway (e-mail: kumarsi@simula.no;
    lin@simula.no; eirikrosnes@simula.no).}  \thanks{A.\ Graell i Amat is with the Department of Electrical Engineering,
    Chalmers University of Technology, SE-41296 Gothenburg, Sweden (e-mail: alexandre.graell@chalmers.se).}  }%

\maketitle

\begin{abstract}
  We propose three private information retrieval (PIR) protocols for distributed storage systems (DSSs) where data is
  stored using an arbitrary linear code. The first two protocols, named Protocol~1 and Protocol~2, achieve privacy for
  the scenario with noncolluding nodes. Protocol~1 requires a file size that is exponential in the number of files in
  the system, while Protocol~2 requires a file size that is independent of the number of files and is hence simpler. We
  prove that, for certain linear codes, Protocol~1 achieves the maximum distance separable (MDS) PIR capacity, i.e., the
  maximum PIR rate (the ratio of the amount of retrieved stored data per unit of downloaded data) for a DSS that uses an
  MDS code to store any given (finite and infinite) number of files, and Protocol~2 achieves the \emph{asymptotic}
  MDS-PIR capacity (with infinitely large number of files in the DSS). In particular, we provide a necessary and a
   sufficient condition for a code to achieve the MDS-PIR capacity with Protocols~1 and 2 and prove that cyclic codes, Reed-Muller (RM) codes,
  and a class of distance-optimal local reconstruction codes achieve both the \emph{finite} MDS-PIR capacity (i.e., with
  any given number of files) and the asymptotic MDS-PIR capacity with Protocols~1 and 2, respectively. Furthermore, we
  present a third protocol, Protocol~3, for the scenario with multiple colluding nodes, which can be seen as an
  improvement of a protocol recently introduced by Freij-Hollanti \emph{et al.}. Similar to the noncolluding case,
    we provide a necessary and a sufficient  condition to achieve the maximum possible PIR rate of Protocol~$3$.
  Moreover, we provide a particular class of codes that is suitable for this protocol and show that RM codes achieve the
  maximum possible PIR rate for the protocol. For all three protocols, we present an algorithm to optimize their PIR
  rates.
\end{abstract}


\section{Introduction}

In data storage applications, besides resilience against disk failures and data protection against illegitimate users,
the privacy may also be of concern. For instance, one may be interested in designing a storage system in which a file
can be downloaded without revealing any information of which file is actually downloaded to the servers storing it. This
form of privacy is usually referred to as \emph{private information retrieval} (PIR). PIR is important to, e.g., protect
users from surveillance and monitoring.

PIR protocols were first studied in the computer science literature by Chor \emph{et al.} in \cite{cho95, cho98}, which
introduced the concept of an $n$-server PIR protocol, where a binary storage node is replicated among $n$ servers
(referred to as nodes) and the aim is to privately retrieve a single bit from the storage nodes while minimizing the
total upload and download communication cost. Additionally, an $n$-server PIR protocol assumes that the $n$ nodes do not
collude in order to reveal the identity of the requested {\mg bit}. The communication cost in \cite{cho95} was further
reduced in \cite{Bei02,Yek08,Efr09}. Since then, coded PIR schemes have been introduced, where data is encoded (as
opposed to simply being replicated) across several nodes \cite{ish04}. With the advent of distributed storage systems
(DSSs), where the user data is encoded and then stored on $n$ nodes, there has been an increasing interest in
implementing coded PIR protocols for these systems.

In recent years PIR has become an active research area in the information theory community with a fundamental difference
in the measurement of efficiency. In the information-theoretic sense, the message sizes are much larger than the size of
all queries sent to the storage nodes. Thus, rather than accounting for both the upload and the download cost,  efficiency is measured in terms of download cost only as the upload cost can be neglected. The ratio of the requested
file size to the amount of downloaded data is referred to as the PIR rate, where a higher PIR rate means a higher
efficiency. The highest achievable PIR rate for any $n$-server PIR protocol is referred to as the PIR capacity.

In the information theory literature, the authors in \cite{sha14} were the first to present PIR protocols for DSSs where
data is stored using codes from two explicit linear code constructions. In \cite{cha15}, the authors presented upper
bounds on the tradeoff between the storage and the PIR rates for a certain class of linear PIR protocols. In
\cite{Faz15}, Fazeli \emph{et al.} introduced PIR codes which, when used in conjunction with traditional $n$-server PIR
protocols, allow to achieve PIR on DSSs. These codes achieve high code rates without sacrificing on the communication
cost of an $n$-server PIR protocol. In \cite{SunJafar17_1}, given an arbitrary number of files, the authors derived the
PIR capacity for noncolluding and replicated databases, where the data can be seen as being encoded by a trivial 
class of maximum distance separable (MDS) codes, i.e., repetition codes. For the case of noncolluding nodes, Banawan and
Ulukus \cite{BanawanUlukus18_1} derived the PIR capacity for DSSs using an $[n,k]$ MDS code to store a given number of
files, referred to as the MDS-PIR capacity. In this paper, we will refer to the MDS-PIR capacity for a given finite
number of files as the \emph{finite MDS-PIR capacity}, and  to 
the MDS-PIR capacity for an infinite number of files as the \emph{asymptotic MDS-PIR capacity}.
In \cite{TajeddineGnilkeElRouayheb18_1}, a PIR protocol for MDS-coded DSSs and noncolluding nodes was proposed and shown
to achieve the asymptotic MDS-PIR capacity. PIR protocols for the case of colluding nodes were proposed in
  \cite{TajeddineGnilkeElRouayheb18_1, Zha17, Zha17b, Oli18}.  {The MDS-PIR capacity for the colluding case
  is still unknown in general, except for some special cases \cite{SunJafar18_1} and for repetition codes
  \cite{SunJafar18_2}}. The problem of \emph{symmetric} PIR for DSSs was recently considered in \cite{WangSkoglund17_2}, where an
expression for the symmetric PIR capacity for linear schemes in the general case of colluding nodes and an MDS linear
storage code was derived. In the symmetric case, the user should not only be able to privately retrieve the requested
file from the system, but also learn nothing about the other files stored from the retrieved data. See also the related
work \cite{WangSkoglund19_1app,sun19}, which deals with replicated databases. The PIR capacity for the case where a given number of storage
nodes fail to respond (so-called robust PIR) was given in \cite{SunJafar18_2} for the scenario of colluding servers with
replication coding.

In the storage community, it is well known that MDS codes are inefficient in the repair of failed
nodes. \textcolor{black}{In particular, they have large repair locality, i.e., the repair of a failed node requires
  contacting a large number of nodes.\footnote{\textcolor{black}{In a parallel line of work, schemes for efficient
      repair (in terms of repair bandwidth) of Reed-Solomon codes have been proposed
      \cite{Gur17,TamoYeBarg17_1}.}}} Repair is essential to maintain the initial state of reliability of the DSS. To
address \textcolor{black}{low repair locality}, Pyramid codes \cite{Hua07}, locally repairable codes \cite{Sat13}, local
reconstruction codes (LRCs) \cite{Hua12,Kam14}, and locally recoverable codes \cite{tam14} \textcolor{black}{are some
  non-MDS codes that} have been proposed. These four classes of codes follow the same design philosophy and for
simplicity, we will refer to them generically as LRCs.
Following the motivation of using non-MDS codes in DSSs, the authors of \cite{Kum17b} presented a PIR protocol for DSSs
that store data using arbitrary linear codes for the scenario of noncolluding nodes. Independently, Freij-Hollanti
\emph{et al.} in \cite{FreijHollantiGnilkeHollantiKarpuk17_1} presented a PIR protocol that ensures privacy even when a
subset of at most $n-k$ nodes collude. {The protocol is based on two codes, the storage code and the \emph{query code},
  which defines the queries. The retrieval process is then characterized by the \emph{retrieval code}, which is the
  Hadamard product of these two codes. The PIR rate of the protocol is upperbounded by $(n-\tilde{k})/n$, where
  $\tilde{k}$ is the dimension of the retrieval code. The authors showed that with generalized Reed-Solomon (GRS) codes for the
  storage and query codes, the upper bound on the PIR rate is achieved. To the best of our knowledge, in the asymptotic
  regime when the number of files tends to infinity, the PIR rate $(n-\tilde{k})/n$ is the highest achievable PIR rate
  known so far.
  Moreover, they showed that their protocol could work with certain non-MDS codes. However, for non-MDS codes (e.g.,
  Reed-Muller (RM) codes where considered in \cite{Hol17}) the PIR rates that can be achieved by the protocol in
  \cite{FreijHollantiGnilkeHollantiKarpuk17_1} are lower than the upper bound $(n-\tilde{k})/n$.}

  
In this paper, as an extension of \cite{Kum17b}, we present three PIR protocols for DSSs using arbitrary linear
codes. These protocols share the fact that all of them are constructed by making use of correctable erasure patterns and
information sets of the underlying storage code. We first focus on the noncolluding scenario and propose two PIR
protocols, referred to as Protocol~1 and Protocol~2. Protocol~1 requires a file size that is exponential in the number
of files in the system, while Protocol~2 requires a file size that is independent of the number of files and is
therefore simpler. Furthermore, Protocol~1 is designed such that its PIR rate depends on the number of files in the
system, while Protocol~2 is such that its PIR rate is independent of the number of files. We prove that, interestingly,
for certain non-MDS code families, Protocol~1 achieves the finite MDS-PIR capacity (and hence the asymptotic MDS-PIR
capacity as well) and Protocol~2 achieves the asymptotic  MDS-PIR
capacity. Thus, we show that the MDS property required to achieve the MDS-PIR capacity in
\cite{SunJafar17_1,BanawanUlukus18_1,TajeddineGnilkeElRouayheb18_1} is not necessary and is overly restrictive. In
particular, we give a sufficient condition based on code automorphisms and a necessary condition connected to the generalized
Hamming weights of the underlying storage code to achieve the MDS-PIR capacity for Protocols~1 and 2. We prove that
cyclic codes, RM codes, and distance-optimal information locality codes achieve the finite MDS-PIR capacity (and thus
the asymptotic MDS-PIR capacity, too) with Protocol~1 and the asymptotic MDS-PIR capacity with Protocol~2. For other
codes, we present an optimization algorithm for Protocols~1 and 2 to optimize their PIR rates. 

We furthermore present a third protocol, Protocol~3, for the scenario of multiple colluding nodes and non-MDS storage
codes. This protocol is based on and improves the protocol in \cite{FreijHollantiGnilkeHollantiKarpuk17_1,Hol17}, 
  in the sense that it achieves higher PIR rates. We extend the necessary and the sufficient condition from the
  noncolluding case to provide joint conditions on the storage and query codes to achieve the upper bound
  $(n-\tilde{k})/n$ on the PIR rate of Protocol~3. Moreover, we show that Protocol~3 achieves the upper bound
  $(n-\tilde{k})/n$ on the PIR rate for RM codes and some non-MDS codes. We also provide an optimization algorithm for
the protocol to optimize the PIR rate. Such an optimization is in itself an extension of the optimization algorithm for
Protocols~1 and 2 for the case of noncolluding nodes. Besides GRS and RM codes as in
\cite{FreijHollantiGnilkeHollantiKarpuk17_1,Hol17}, we also prove that $(\mathcal{U}|\mathcal{U}+\mathcal{V})$ codes
\cite{MacWilliamsSloane77_1} with $\mathcal{U}$ being an arbitrary binary linear code and $\mathcal{V}$ a binary repetition code can
be used in conjunction with Protocol~3. We finally give examples of all-symbol locality LRCs with good PIR rates.

{The main contributions of the paper are summarized in the following:
  \begin{itemize}
  \item For the noncolluding case, we propose two PIR protocols, Protocol~1 and Protocol~2
    (Sections~\ref{sec:file-dep-PIR} and \ref{sec:file-indep-PIR}), and provide a necessary and a sufficient condition
    for a code to achieve the MDS-PIR capacity with these protocols
    (Theorems~\ref{thm:general-d_MDS-PIRcapacity-achieving-codes} and
    \ref{thm:sufficient_MDS-PIRcapacity-achieving-codes}, respectively, in
    Section~\ref{sec:MDS-PIRcapacity-achiving-codes}).
  \item For the noncolluding case, we show that important classes of non-MDS codes, namely cyclic codes, RM codes, and
    distance-optimal information locality codes achieve the finite MDS-PIR capacity and the asymptotic MDS-PIR capacity
    with Protocols~1 and 2, respectively (Corollaries~\ref{cor:MDS-PIRcapacity-achiving_cyclic-codes},
    \ref{cor:MDS-PIRcapacity-achieving_RMcodes}, and Theorem~\ref{th: LRCcap_proof}, respectively, in
    Section~\ref{sec:MDS-PIRcapacity-achiving-codes}).
  \item For the colluding case, we propose Protocol~3 that achieves higher asymptotic PIR rates for non-MDS codes (equal
    to its upper bound) than the best known protocol \cite{FreijHollantiGnilkeHollantiKarpuk17_1,Hol17}. Similar to the
    noncolluding case, a necessary and a sufficient condition for the protocol to achieve PIR rates equal to its upper
    bound is provided (Corollary~\ref{cor:general-d_CstarD-PIRmaxRate-codes} and Theorem~\ref{thm:sufficient_PIRmaxRte},
    respectively, in Section~\ref{sec:MultipleCollNodePIR}). Independently of this work, by using an approach similar to
    ours, in \cite{FreijHollantiGnilkeHollantiHorlemannKarpukKubjas19_1app} the authors modified the protocol in
    \cite{Hol17} and showed that the PIR rate $(n-\tilde{k})/n$ is achievable for \emph{transitive codes}.\footnote{Note
      that the proposed sufficient condition (Theorem~\ref{thm:sufficient_PIRmaxRte}) is not equivalent to the concept
      of transitive codes in \cite{FreijHollantiGnilkeHollantiHorlemannKarpukKubjas19_1app} when there are at least $2$
      colluding nodes (in the noncolluding case the concept of transitive codes in
      \cite{FreijHollantiGnilkeHollantiHorlemannKarpukKubjas19_1app} reduces to our sufficient condition
      (Theorem~\ref{thm:sufficient_MDS-PIRcapacity-achieving-codes})).} However, the protocol in
    \cite{FreijHollantiGnilkeHollantiHorlemannKarpukKubjas19_1app} requires a much larger number of stripes and query
    sizes than our proposed Protocol~3, since it is based on transitive subgroups of the automorphishm groups of the
    storage and query codes, and thus is less practical.
  \item For both the noncolluding and colluding cases, we provide an algorithm that optimizes the PIR rate of the
    underlying code (Sections~\ref{sec:SingleColl_Analysis} and \ref{sec:optC}).
  \end{itemize}
}

The remainder of this paper is organized as follows. We provide some definitions and preliminaries in Section~\ref{sec:
  DefandPrelim}. In Section~\ref{sec:SystemModel}, we provide a general system model for the three PIR protocols proposed in
the paper. In Sections~\ref{sec:file-dep-PIR} and \ref{sec:file-indep-PIR}, we present Protocols~1 and 2 for the scenario with
noncolluding nodes. {\mg In Section~\ref{sec:MDS-PIRcapacity-achiving-codes}, we give a necessary and a  sufficient condition
  for codes to achieve the MDS-PIR capacity with Protocols~1 and 2 and  prove that several families of codes achieve it. In Section~\ref{sec:SingleColl_Analysis}, we give an optimization algorithm to optimize the PIR rate.}
In Section~\ref{sec:MultipleCollNodePIR}, we consider the scenario with colluding nodes and propose Protocol~3. In the same
section, we also present a family of storage codes that can be used with this protocol. Lastly, {\mg we provide a
  necessary and a sufficient condition to achieve an upper bound on the PIR rate for this protocol, and we show that RM
  codes satisfy the sufficient condition and thus achieve the upper bound on the PIR rate of Protocol~3.} In
Section~\ref{sec:results}, we optimize Protocols~1, 2, and 3 to maximize their PIR rates for different storage codes under the
scenarios of noncolluding and colluding nodes. Finally, some conclusions are drawn in Section~\ref{sec:conclusion}.

\subsection{Notation and Conventions}

In this paper, we use the following notation. We use lowercase bold letters to denote vectors, uppercase bold letters
to denote matrices, and calligraphic uppercase letters to denote sets. For example: $\bm x$, $\bm X$, and $\mathcal X$
denote a vector, a matrix, and a set, respectively. An identity matrix of dimensions $m \times m$ is denoted as $\bm I_{m}$. The 
all-zero matrix of dimensions $a\times b$ is denoted as $\bm 0_{a\times b}$, while the all-one matrix of dimensions 
$a\times b$ is referred to as $\bm 1_{a\times b}$. $\trans{(\cdot)}$ represents the transpose of its argument and
$\langle\cdot,\cdot\rangle$ denotes the scalar dot product between two vectors.  The operator $\circ$ represents the
Hadamard product. As such, $\bm x\circ\bm y$ represents the Hadamard product of two length-$n$ vectors $\bm x$ and
$\bm y$.  Consider the column vectors $\bm x_1,\ldots,\bm x_a$, then $(\bm x_1|\ldots|\bm x_a)$ represents the
horizontal concatenation of the column vectors. Similarly, the horizontal concatenation of the matrices
$\bm X_1,\ldots,\bm X_a$, all with the same number of rows, will be denoted by $(\bm X_1|\ldots|\bm X_a)$.  We represent
a submatrix of $\mat{X}$ that is restricted in columns by the set $\set{I}$ and in rows by the set $\set{J}$ by
$\mat{X}|_{\set{I}}^{\set{J}}$, and the matrix rank of $\mat{X}$ by $\rank{\mat{X}}$. The function $\mathsf{LCM}(a,b)$
computes the lowest common multiple of two positive integers $a$ and $b$, and $a\mid b$ denotes that $a$ is a divisor of $b$,
while the function $\mathsf{H}(\cdot)$ represents the entropy of its argument.

In the rest of the paper, $\code{C}$ will denote a linear code over a finite field $\GF(q)$. The operations over
$\GF(q)$, such as addition, multiplication, etc., will be clearly understood from the context. We use the customary code
parameters $[n,k]$ to refer to a code of block length $n$ and dimension $k$, having code rate $R^\code{C}=k/n$. The
dimension of a code $\code{C}$ will sometimes be denoted by $\Dim{\code{C}}$. Furthermore,
$[n,k,d^\code{C}_\mathsf{min}]$ represents an $[n,k]$ code of minimum Hamming distance $d^\code{C}_\mathsf{min}$. Since
a code $\code{C}$ can be seen as a codebook matrix, the shortened and punctured codes are denoted by
$\code{C}|_\set{I}^\set{J}$, with column indices $\set{I}$ and row coordinates $\set{J}$. In addition,
$\bm H^{\mathcal C}$, $\bm G^{\mathcal C}$, and $\mathcal C^\perp$ represent a parity-check matrix, a generator matrix, and the dual code, respectively, of $\mathcal C$. We denote by $\Nat{}$ the set of all positive
integers, $\Nat{a}\eqdef\{1,2,\ldots,a\}$, and $\Nat{n_1:n_2}\eqdef\{n_1,n_1+1,\ldots,n_2\}$ for two positive integers
$n_1\leq n_2$, $n_1,n_2\in\Nat{}$. The Hamming weight of a binary vector $\vect{x}$ is denoted by $\Hwt{\vect{x}}$,
while the support of a vector $\vect{x}$, i.e., the set of nonzero entries of $\vect{x}$,
will be denoted by $\chi(\vect{x})$.  Note that sometimes, for the sake of convenience, we will omit the superscripts
and/or the subscripts if the arguments we refer to are contextually unambiguous. Also, with some abuse of language, the
index of a coordinate of a vector is sometimes referred to simply as the coordinate. An erasure pattern is a binary vector where the ones represent erased positions, while the zeros represent nonerased positions. The weight of an erasure pattern is the number of erased positions, and 
 an erasure pattern  $\bm{x}$ is said to be correctable by a code $\code{C}$ if
   $\bm H^{\code{C}}|_{\chi(\bm{x})}$ has rank $|\chi(\bm{x})|$. Finally, for ease of notation, we will refer to a
  matrix with constant row weight, constant column weight, and constant row and column weight equal to $a$ as an $a$-row
  regular, $a$-column regular, and $a$-regular matrix, respectively.

\section{Definitions and Preliminaries}
\label{sec: DefandPrelim}

In this section, we review some basic notions in coding theory and some classes of codes that will be used throughout
the paper.
\begin{definition}
  \label{def:infoS}
  Let $\code{C}$ be an $[n,k]$ code defined over $\GF(q)$. A set of coordinates of $\mathcal C$,
  $\set{I}\subseteq\Nat{n}$, of size $k$ is said to be an \emph{information set} if and only if
  $\mat{G}^\code{C}|_\set{I}$ is invertible.
\end{definition}

\begin{definition}
  \label{def:suppD}
  Let $\code{D}$ be a subcode of an $[n,k]$ code $\code{C}$. The \emph{support} of $\code{D}$ is defined as
  \begin{IEEEeqnarray*}{c}
    \chi(\code{D})\eqdef\{j\in\Nat{n}\colon\exists\,\vect{x}=(x_1,\ldots,x_n)\in\code{D}, x_j\neq 0\}.
  \end{IEEEeqnarray*}  
\end{definition}
 It is noted that
  \begin{IEEEeqnarray*}{c}
    \chi(\code{D})=\bigcup_{\vect{x}\in\code{D}}\chi(\vect{x}).
  \end{IEEEeqnarray*}

Next, we introduce the concept of generalized Hamming weights \cite{wei91_1}.
\begin{definition}
  \label{def:sth-GHW}
  The $s$-th generalized Hamming weight of an $[n,k]$ code $\code{C}$, denoted by $d_s^{\code{C}}$, $s\in\Nat{k}$, is
  defined as the cardinality of the smallest support of an $s$-dimensional subcode of $\code{C}$, i.e.,
  \begin{IEEEeqnarray*}{rCl}
    d_s^{\code{C}}\eqdef\min\bigl\{\card{\chi(\code{D})}\colon\code{D}\text{ is an } [n,s] \text{ subcode of }
    \code{C}\bigr\}.
  \end{IEEEeqnarray*}
\end{definition}

For the sequel, we introduce the notion of Hadamard product \cite{Hor13} of vector spaces.
{
\begin{definition}
  \label{def: Hadamard Product}
  Let $\mathcal X$ and $\mathcal Y$ be two vector spaces in $\GF(q)^n$. The Hadamard product of $\mathcal X$ and
  $\mathcal Y$, denoted by $\mathcal X\circ\mathcal Y$, is defined as the space in $\GF(q)^n$ generated by the Hadamard products
  $\bm x\circ\bm y$ for all $\bm x\in\mathcal X$ and $\bm y\in\mathcal Y$.
\end{definition}
}

\subsection{Reed-Muller Codes}
\label{sec:reed-muller-codes}

We review the family of binary linear RM codes \cite{Reed54_1} and then quickly summarize a result related to
information sets of an RM code. We adapt the concept and definition from \cite[Ch.~13]{MacWilliamsSloane77_1}, and the
details can be found therein.
\begin{definition}
  \label{def:reed-muller-codes}
  For a given $m\in\Nat{}$, the $v$-th order binary RM code $\code{R}(v,m)$ is an $[n,k]$ code with length
  $n=2^m$ and code dimension $k=\sum_{i=0}^v\binom{m}{i}$ for
  $v\in\{0\}\cup\Nat{m}$, 
  constructed as the linear space spanned by the set of all $m$-variable Boolean monomials of degree at most $v$.
\end{definition}

For example, $\code{R}(2,3)$ can be viewed as the linear space spanned by the set of Boolean monomials
$\{1,z_1,z_2,z_3,z_1z_2,z_1z_3,z_2z_3\}$.

Next, we introduce a way to number the coordinate index of an RM codeword. Without loss of generality, since there are
in total $n=2^m$ codeword coordinates, each coordinate index $i\in\Nat{2^m}$ can be described by a binary column
$m$-tuple $\vect{\mu}=\trans{(\mu_{1},\ldots,\mu_m)}$, $\mu_j\in\GF(2)$, such that
\begin{IEEEeqnarray}{c}
  i\eqdef 1+\sum_{j=1}^{m} \mu_j 2^{j-1}.\label{eq:transformation}
\end{IEEEeqnarray}
For instance, for $m=4$, the $7$-th coordinate of an RM code corresponds to $\trans{(0\;1\;1\;0)}$. Hence, a set of
coordinates $\set{I}\subseteq\Nat{n}$ can alternatively be written as a set of corresponding $m$-tuples for RM codes.

Let $\mat{V}$ be an $m\times m$ invertible matrix over $\GF(2)$ and $\vect{\sigma}\in\GF(2)^{m\times 1}$ be a length-$m$
binary column vector. It is well known that the coordinate transformation mapping $\vect{\mu}$ onto
$g(\vect{\mu})\eqdef\mat{V}\vect{\mu}+\vect{\sigma}$ is an \emph{automorphism} for the RM code
\cite[Ch.~13]{MacWilliamsSloane77_1}.

For the sake of simplicity, throughout the paper we assume $\bm V=\bm I_m$.
\begin{example}
  \label{ex:2-3_RMcode}
  Consider the RM code $\code{R}(1,3)$ with generator matrix
  \begin{IEEEeqnarray*}{rCl}
    \mat{G}^{\code{R}(1,3)}=
    \begin{pmatrix}
      1 &1 &1 &1 &1 &1 &1 &1
      \\
      0 &1 &0 &1 &0 &1 &0 &1
      \\
      0& 0 &1 &1 &0 &0 &1 &1
      \\
      0& 0 &0 &0 &1 &1 &1 &1
    \end{pmatrix}.
  \end{IEEEeqnarray*}
  {The $i$-th row of $\mat{G}^{\code{R}(1,3)}$ corresponds to the $i$-th monomial of the set of Boolean monomials
  $\{1,z_1,z_2,z_3\}$, $i\in\Nat{4}$. It can be seen that any codeword of $\code{R}(1,3)$ corresponds to a linear
  combination of the Boolean monomials as $w_0 1+ w_1 z_1+w_2 z_3 +w_3 z_3$, where $w_i\in\GF(2)$,
  $i\in\Nat{4}$.} Clearly, $\set{I}=\{\trans{(0\;0\;0)},\trans{(1\;0\;0)},\trans{(0\;1\;0)},\trans{(0\;0\;1)}\}$ forms an
  information set for $\code{R}(1,3)$. Pick an automorphism $g$ with $\mat{V}=\bm I_3$  and
  $\vect{\sigma}=\trans{(0\;0\;1)}$. Then,
  \begin{IEEEeqnarray*}{rCl}
    \set{I}'& = &\{g(\vect{\mu})=\vect{\mu}+\vect{\sigma} \colon\vect{\mu}\in\set{I}\}
    \\
    & = &\{\trans{(0\;0\;1)},\trans{(1\;0\;1)},\trans{(0\;1\;1)},\trans{(0\;0\;0)}\}
  \end{IEEEeqnarray*}
  is also an information set of $\code{R}(1,3)$.
\end{example}

The following lemma shows how to determine an information set for an RM code.
\begin{lemma}  
  \label{lem:det-infoSets_RMcode}
  Consider the RM code $\code{R}(v,m)$ with $v\in\{0\}\cup\Nat{m}$, $m\in\Nat{}$. Then, the set of $m$-tuples given by
  \begin{IEEEeqnarray*}{rCl}
    \set{I}\eqdef\{\vect{\mu}\in\GF(2)^{m\times 1}\colon\Hwt{\vect{\mu}}\leq v\}
  \end{IEEEeqnarray*}
  is an information set for $\code{R}(v,m)$.
\end{lemma}
\begin{IEEEproof}
  The proof is based on the definition of RM codes. The details are given in
  Appendix~\ref{sec:proof_det-infoSets_RMcode}.
\end{IEEEproof}

Lemma~\ref{lem:det-infoSets_RMcode} can be extended to nonbinary generalized RM codes (see the comprehensive work in
\cite{KeyMcDonoughMavron06_1} that determines the information sets for generalized RM codes).

\subsection{Local Reconstruction Codes}
\label{sec:local-reconstr-codes}

LRCs are a family of codes that are used in DSSs because of their low repair locality,
i.e., they need to contact a relatively low number of nodes in order to repair a failed node. Systematic codes that
focus on lowering the locality for the systematic nodes (i.e., the nodes that store the systematic code symbols; see the
system model in Section~\ref{sec:SystemModel}) are referred to as \emph{information locality} codes. Examples of such codes are
presented in \cite{Hua07,Hua12,Sat13,Kam14}. On the contrary, LRCs that achieve low locality for all nodes are referred to as
\emph{all-symbol locality} codes. The codes presented in \cite{tam14} are examples of all-symbol locality
codes. Formally, information locality codes are defined as follows.
\begin{definition}[{$(r,\delta)$ information locality code\cite[Def.~2]{Kam14}}]
\label{def: LRC}
An $[n,k]$  code is said to be an $(r,\delta)$ information locality
code if there exist $L_{\mathsf c}$ punctured codes $\code{C}_j\eqdef\set{C}|_{\set{S}_j}$ of $\mathcal C$ with 
column coordinate set $\mathcal S_j\subset\Nat{n}$ for $j\in\Nat{L_{\mathsf{c}}}$. Furthermore,
$\{\code{C}|_{\set{S}_j}\}_{j\in\Nat{L_{\mathsf{c}}}}$ must satisfy the following conditions:
\begin{enumerate}
\item $|\mathcal S_j|\leq r+\delta-1$, $\forall\,j\in\Nat{L_{\mathsf{c}}}$,
\item $d_{\mathsf{min}}^{\mathcal C_j}\geq\delta$, $\forall\,j\in\Nat{L_{\mathsf{c}}}$, and
\item $\bigrank{\bm G|_{\bigcup_j\set{S}_j}}=k$.
\end{enumerate}
\end{definition}

In other words, Definition~\ref{def: LRC} says that there are $L_\mathsf{c}$ local codes in
$\mathcal C$ each having a block length of at most $r+\delta-1$, a minimum Hamming distance at least $\delta$, and
the union of all coordinate sets of the local codes contains an information set. The overall code $\mathcal C$ has minimum Hamming distance
$d_{\mathsf{min}}^{\mathcal C}\leq n-k+1-(\lceil k/r\rceil-1)(\delta-1)$ and can repair up to $\delta-1$ systematic
nodes by contacting $r$ storage nodes. Codes that achieve the upper bound on the minimum Hamming
distance are known as distance-optimal $(r,\delta)$ information locality codes and have the following structure.

\begin{definition}[Distance-optimal $(r,\delta)$ information locality code {\cite[Th.~2.2]{Kam14}}]
\label{def: OptLRCs}
Let $r\mid k$ such that $L_\mathsf{c}=k/r$. An $(r,\delta)$ information locality code $\mathcal C$ as defined in
Definition~\ref{def: LRC} is distance-optimal if:
\begin{enumerate}
\item Each local code $\mathcal C|_{\set{S}_j}$, $j\in\Nat{L_\mathsf{c}}$, is an $[r+\delta-1,r]$ MDS code defined by a
  parity-check matrix $\mat{H}^{\mathcal C|_{\set{S}_j}}=(\bm P_j|\bm I_{\delta-1})$ of dimensions
  $(\delta-1)\times(r+\delta-1)$ and minimum Hamming distance $d_{\mathsf{min}}^{\mathcal C|_{\mathcal S_j}}=\delta$.
\item The sets $\{\mathcal S_j\}_{j\in\Nat{L_\mathsf{c}}}$ are disjoint, i.e., $\set{S}_j\cap\set{S}_{j'}=\emptyset$ for
  all $j,j'\in\Nat{L_\mathsf{c}}$, $j\not=j'$.
\item The code $\mathcal C$ has a parity-check matrix of the form
\end{enumerate}
\begin{align}
  \label{Eq: H_opt_LRC}
  \mat{H}=\left(\begin{array}{ccccccc|c}
    \bm P_1 & \bm I_{\delta-1} &           &                  &       &             &                  & \\
            &                  & \bm P_2   & \bm I_{\delta-1} &       &             &                  & \\
            &                  &           &                  &\ddots &             &                  & \\
            &                  &           &                  &       & \bm P_{L_\mathsf{c}} & \bm I_{\delta-1} & \\
  \hline
  \mat{M}_1 & \bm 0            & \mat{M}_2 & \bm 0            &\cdots & \mat{M}_{L_\mathsf{c}} & \bm 0 & \bm I_a\\ 
  \end{array}\right)
\end{align}
where the matrices $\bm M_1,\ldots, \bm M_{L_\mathsf{c}}$ are arbitrary matrices  in $\GF(q)$ of dimensions $(n-L_\mathsf{c}(r+\delta-1))\times r$,  and $a\eqdef n-L_\mathsf{c}(r+\delta-1)$.
\end{definition}

For ease of exposition, we refer to the local parities as the parity symbols that take part in the local codes, while the
parity symbols that are not part of the $L_{\mathsf c}$ local codes are referred to as global parity symbols. According
to Definition~\ref{def: OptLRCs}, there exist $n-L_{\mathsf{c}}(r+\delta-1)$ global parities and $L_{\mathsf{c}}(\delta-1)$ local
parities. We partition the coordinates of these parities into $L+1$ sets, where
$L\eqdef\bigl\lfloor \frac{n}{r+\delta-1}\bigr\rfloor$. For $i\in\Nat{L+1}$, we have
\begin{align}
  \mathcal P_j
  =\begin{cases}
    \{(j-1)n_\mathsf{c}+r+1,\ldots,j n_\mathsf{c}\} & \text{if }j\in\Nat{L_{\mathsf c}},\\
    \{(j-1)n_\mathsf{c}+1,\ldots,j n_\mathsf{c}\} & \text{if }j\in\Nat{L_{\mathsf c}+1: L},\\
    \{L n_\mathsf{c}+1,\ldots,n\} & \text{if } j=L+1,
  \end{cases}\label{eq:parities_LRC}
\end{align} 
where $n_\mathsf{c}\eqdef r+\delta-1$ is the block length of each local code. The set $\mathcal P_{j}$,
$j\in\Nat{L_{\mathsf c}}$, represents the coordinates of the local parities of the $j$-th local code $\mathcal C_{j}$. The remaining sets $\mathcal P_{j}$, $j\in\Nat{L_{\mathsf c}+1:L+1}$, represent the coordinates of the global
parities of $\mathcal C$. As such, the set $\set{P}=\bigcup_{j=1}^{L+1} \mathcal P_j$ represents the parity
coordinates of $\mathcal C$.

\subsection{UUV Codes}

Consider an $[n_1,k_1]$ code $\mathcal U$ and an $[n_1,k_2]$ code $\mathcal V$ both over $\GF(q)$. An
$[n=2n_1,k=k_1+k_2]$ $(\mathcal U\mid\mathcal U+\mathcal V)$ code \cite{MacWilliamsSloane77_1} (herein referred to as a
UUV code) has codewords of the form $(\bm u\mid \bm u+\bm v)$, where $\bm u\in\mathcal U$ and $\bm v\in\mathcal V$. A
UUV code has generator matrix
\begin{align*}
  \mat{G}^{\mathsf{UUV}}=
  \begin{pmatrix}
    \mat{G}^{\mathcal U} & \mat{G}^{\mathcal U}\\
    \bm 0_{k_2\times n_1}  & \mat{G}^{\mathcal V}
  \end{pmatrix},
\end{align*}
where $\mat{G}^{\mathcal U}$ and $\mat{G}^{\mathcal V}$ are the generator matrices of $\mathcal U$ and $\mathcal V$,
respectively. One can construct RM codes using UUV codes in an iterative manner \cite[p.~374]{MacWilliamsSloane77_1}.

\section{System Model}
\label{sec:SystemModel}

\begin{figure}[t]
\centering
\begin{tikzpicture}[scale=0.7, every node/.style={transform shape}, thick]

\node[minimum width=3cm, minimum height=2cm] (user) at (5.5,10.5) {};
\foreach \i in {1,4,7,10}{
\ifthenelse{\equal{\i}{1}\OR\equal{\i}{4}}
{\node[draw, black, fill=gray!10, cylinder, thick, aspect=0.9,minimum height=5 cm, minimum width=1.75 cm, rotate=90] (n\i) at (\i,4.5) {};}
{\node[draw, black, fill=white, cylinder, thick, aspect=0.9,minimum height=5 cm, minimum width=1.75 cm, rotate=90] (n\i) at (\i,4.5) {};}
}
\node at (2.55,4.5) {\LARGE $\cdots$};
\node at (8.55,4.5) {\LARGE $\cdots$};
\node at (10,3.1) {$\vdots$};
\node at (7,3.1) {$\vdots$};
\foreach \i in {2,3,3.5,4,4.5,5,5.5,6}{
	\pgfmathsetmacro\Gval{\i*40}
	\definecolor{mycolor\i}{RGB}{255,\Gval,100}
	\node[fill=mycolor\i, minimum height=0.15cm, minimum width=1.5cm, rounded corners=1pt] at (10,\i+0.5) {};
	\node[fill=mycolor\i, minimum height=0.15cm, minimum width=1.5cm, rounded corners=1pt] at (7,\i+0.5) {};
}
\definecolor{file}{RGB}{255,220,100}
\node[fill=file, rounded corners=2pt, minimum width=1.5cm, minimum height=3.5cm] (f) at (12.25,9.25) {};
\foreach \i/\j in {7.5/3,9/2,10/1}{
	\ifthenelse{\equal{\j}{3}}
	{\node[fill=white, minimum height=0.7cm, minimum width=1.25cm, rounded corners=1pt] at (12.25,\i+0.5) {$c^{(2)}_{\beta, n}$};}
	{\node[fill=white, minimum height=0.7cm, minimum width=1.25cm, rounded corners=1pt] at (12.25,\i+0.5) {$c^{(2)}_{\j, n}$};}
}
\node at (12.25,8.85) {\color{white}{$\vdots$}};
\node[draw, thin, ellipse, minimum width=2cm, minimum height=0.5cm] (ell) at (10,6) {};
\draw[>=stealth,->,thin] (ell.east) to[in=270,out=0] (f.south);
\begin{scope}[yshift=0cm]
	
\clip[shift={(-1,0.5)},rotate around={-30:(7,4.5)}] (6,2)rectangle(8,7.5);
\node[draw, black,fill=gray!10, cylinder, thick, aspect=0.9,minimum height=5 cm, minimum width=1.75 cm, rotate=90] at (7,4.5) {};	
\end{scope}
\begin{scope}[shift={(4,9.5)}]
	\draw[rounded corners=1.2pt] (0,0.5)rectangle(1.5,1.5);
	\path[fill=black] (0,0.5)rectangle(1.5,0.75);
	\draw[rounded corners=2pt] (2,0.25)rectangle(2.65,1.5);
	\path[fill=black] (2,0.35)rectangle(2.65,1.4);
	\path[fill=white] (2.325,0.65) circle (1pt);
	\path[fill=black] (0.65,0.25)rectangle(0.85,0.5);
	\draw[thick] (0.5,0.25)--(1,0.25);
\end{scope}

\draw[>=stealth,<->] (n1.east) -- (user.south west);
\draw[>=stealth,<->] (n4.east) -- (user.south);
\draw[>=stealth,<->] (n7.east) -- (user.south);
\draw[>=stealth,<->] (n10.east) -- (user.south east);
\node[shift={(0,0.5)}] at (n1.east) {$\bm Q^{(1)}$};
\node[shift={(-0.1,0.5)}] at (n4.east) {$\bm Q^{(k)}$};
\node[shift={(0.4,0.5)}] at (n7.east) {$\bm Q^{(k+1)}$};
\node[shift={(0,0.5)}] at (n10.east) {$\bm Q^{(n)}$};

\node[shift={(-0.75,-0.25)}] at (user.south west) {$\bm r_1$};
\node[shift={(-0.45,-0.25)}] at (user.south) {$\bm r_k$};
\node[shift={(0.7,-0.25)}] at (user.south) {$\bm r_{k+1}$};
\node[shift={(0.7,-0.25)}] at (user.south east) {$\bm r_n$};

\draw[decorate,decoration={brace,mirror}, thick] (0,1.9)--node[below]{$n$ storage  nodes}(11,1.9);

\node at (5.5,1) {(a)};
\node at (12.25,11.5) {(b)};
\node at (5.5,11.5) {(c)};

\end{tikzpicture}
\vspace{-2ex}
\caption{System Model. (a) The colored boxes in each storage node represent the $f$ coded chunks pertaining to the $f$
  files. (b) Coded chunk corresponding to the $2$nd file in the $n$-th node. It consists of $\beta$ code symbols,
  $c^{(2)}_{i,n}$, $i\in\Nat{\beta}$. (c) The user sends the queries $\bm Q^{(l)}$, $l\in\Nat{n}$, to the storage nodes and
  receives responses $\bm r_l$.}
\label{Fig: System Model}
\vskip -3ex
\end{figure}

We consider a DSS that stores $f$ files $\bm X^{(1)},\ldots, \bm X^{(f)}$, where each file
$\bm{X}^{(m)}=(x_{i,j}^{(m)})$, $m\in\Nat{f}$, can be seen as a $\beta \times k$ matrix over $\GF(p^{\alpha \ell})$,
with $\beta$, $k$, $\alpha$, $\ell\in\Nat{}$, and $p$ being some prime number. {Each file is encoded} using a linear code
as follows. Let $\bm x^{(m)}_i=(x^{(m)}_{i,1},\ldots,x^{(m)}_{i,k})$, $i\in\Nat{\beta}$, be a message vector
corresponding to the $i$-th row of $\X^{(m)}$. Each $\bm x_i^{(m)}$ is encoded by an $[n,k]$ code $\mathcal{C}$ over
$\GF(q)$ with $q\eqdef p^\alpha$, having \emph{subpacketization} $\alpha$, into a length-$n$ codeword
$\bm c^{(m)}_i=\bigl(c^{(m)}_{i,1},\ldots,c^{(m)}_{i,n}\bigr)$, where $c_{i,j}^{(m)}\in\GF(q^\ell)$, $j\in\Nat{n}$. For
$\alpha=1$, the code $\mathcal{C}$ is referred to as a \emph{scalar} code. Otherwise, the code is called a \emph{vector}
code \cite{Bla95}. The $\beta f$ generated codewords $\bm c_i^{(m)}$ are then arranged in the array
$\bm C= \bigl((\mat{C}^{(1)})^{\textup{\textsf{\tiny T}}}|\ldots|(\mat{C}^{(f)})^{\textup{\textsf{\tiny T}}}\bigr)^{\textup{\textsf{\tiny T}}}$ of dimensions $\beta f \times n$, where
$ \mat{C}^{(m)} = \bigl((\bm c^{(m)}_1)^{\textup{\textsf{\tiny T}}}|\ldots|(\bm c^{(m)}_{\beta})^{\textup{\textsf{\tiny T}}} \bigl)^{\textup{\textsf{\tiny T}}}$ for $m \in \Nat{f}$. For a
given column $j$ of $\bm C$, we denote the column vector $\bigl(c_{1,j}^{(m)},\ldots,c_{\beta,j}^{(m)}\bigr)^{\textup{\textsf{\tiny T}}}$ as a coded
chunk pertaining to file $\bm{X}^{(m)}$. The $f$ coded chunks in column $j$ are stored in the $j$-th storage node,
$j\in\Nat{n}$, as shown in Fig.~\ref{Fig: System Model}(a). In case the $[n,k]$ code $\code{C}$ is systematic, the nodes
that store the systematic code symbols are referred to as systematic nodes.

\subsection{Privacy Model}
\label{sec:privacy}

We consider a DSS where a set of $T$ nodes may act as spies. Further, they may collude and hence they are referred to
as colluding nodes. In addition, it is assumed that the remaining nonspy nodes do not collaborate with the spy
nodes. The scenario of a single spy node ($T=1$) in the DSS is analogous to having a system with no colluding nodes. Let
$\set{T}\subset\Nat{n}$, $|\set{T}|=T$, denote the set of spy nodes in the DSS. The role of the spy nodes is to determine
which file $\bm{X}^{(m)}$ is accessed by the user. We assume that the user does not know $\set{T}$, since otherwise it can
trivially achieve PIR by not contacting the spy nodes. To retrieve  file $\bm X^{(m)}$ from the DSS, the user
sends a $d\times\beta f$ matrix query $\bm Q^{(l)}$ over $\GF(q)\subseteq\GF(q^\ell)$ to the $l$-th node for all
$l\in \Nat{n}$. The query matrices are represented in the form of $d$ subquery vectors $\bm q^{(l)}_i$ of length
$\beta f$ as
\begin{align*}
	\bm Q^{(l)}=\left(\begin{matrix}
		\bm q^{(l)}_1\\
		\vdots\\
		\bm q^{(l)}_d\\
	\end{matrix}\right)=\left(\begin{matrix}
		q^{(l)}_{1,1}  & \cdots & q^{(l)}_{1,\beta f}\\
		\vdots &  \cdots & \vdots\\
		q^{(l)}_{d,1}  & \cdots & q^{(l)}_{d,\beta f}\\
	\end{matrix}\right).
\end{align*}
The $i$-th subqueries $\bm q^{(l)}_i$, $l\in\Nat{n}$, of the $n$ queries aim at recovering $\Gamma$ unique code
symbols\footnote{In general, the $i$-th subqueries recover $\Gamma_i$ unique code symbols such that among the
  $\sum_i \Gamma_i$ recovered code symbols there are $\beta k$ distinct information symbols. However, for the sake of
  simplicity, we assume $\Gamma_i=\Gamma$ for all $i$  for Protocols~2 and 3.} of the file $\bm X^{(m)}$. In response to the received query
matrix, node $l$ sends the column vector
\begin{equation}
  \label{eq:response}
  \bm r_l=(r_{l,1},\ldots,r_{l,d})^{\textup{\textsf{\tiny T}}} = \bm
  Q^{(l)}\bigl(c^{(1)}_{1,l},\ldots,c^{(1)}_{\beta,l},\ldots,c^{(f)}_{\beta,l}\bigr)^{\textup{\textsf{\tiny T}}},
\end{equation}
referred to as the response vector, back to the user as illustrated in Fig.~\ref{Fig: System Model}(c). We refer to
$r_{l,i}$ as the $i$-th subresponse of the $l$-th node. Perfect information-theoretic PIR for such a scheme is defined
in the following.

\begin{definition}
\label{def:cond}
Consider a DSS with $n$ nodes storing $f$ files in which a set of $T$ nodes
$\set{T}=\{t_1,\ldots,t_T\}\subset\Nat{n}$, $1\leq|\set{T}|=T\leq n-k$, act as colluding spies. A user who wishes to
retrieve the $m$-th file sends the queries $\bm Q^{(l)}$, $l\in\Nat{n}$, to the storage nodes, which return the
responses $\bm r_l$. This scheme achieves perfect information-theoretic PIR if and only if
\begin{subequations} 
  \begin{align}
    \label{eq:cond1}
    \text{Privacy:}\qquad  &\mathsf H\bigl(m|\bm Q^{(t_1)},\ldots,\bm Q^{(t_T)}\bigr)=\mathsf H(m);
    \\
    \label{eq:cond2}
    \text{Recovery:}\qquad &\mathsf H\bigl(\bm X^{(m)}|\bm r_1,\ldots, \bm r_n\bigr)=0.
  \end{align}
\end{subequations}
\end{definition}

Queries satisfying \eqref{eq:cond1} ensure that the file requested by the user is independent of the queries. Thus, the
colluding nodes in $\set{T}$ do not gain any additional information regarding which file is requested by the user by
observing the queries. The recovery constraint in \eqref{eq:cond2} ensures that the user is able to recover the
requested file from the responses sent by the DSS.

The efficiency of a PIR protocol is defined as the amount of retrieved data per unit of total amount of downloaded data,
since it is assumed that the content of the retrieved file dominates the total communication cost
\cite{cha15, TajeddineGnilkeElRouayheb18_1}.
\begin{definition}
  The PIR rate of a PIR protocol, denoted by $\const{R}$, is the amount of information retrieved per downloaded symbol,
  i.e.,
  \begin{align*}
    \const{R}\eqdef\frac{\beta k}{nd}.
  \end{align*}
\end{definition}

Since the size of each file is $\beta k$, the parameters $d$ and $\Gamma$ should be chosen such that $\beta k=\Gamma
d$.\footnote{For Protocol~$1$, $d$ and $\Gamma$ should be chosen such that $\beta k= \sum_{i=1}^d \Gamma_i$.}  For  Protocols~2 and 3 in Sections~\ref{sec:file-indep-PIR} and \ref{sec:MultipleCollNodePIR} {to be
  practical, we may select} 
\begin{align}
\label{Eq: Beta_and_d_DEF}
  \beta=\frac{\mathsf{LCM}(k,\Gamma)}{k}\quad\text{and}\quad d=\frac{\mathsf{LCM}(k,\Gamma)}{\Gamma},
\end{align}
{as it ensures the smallest values of $\beta$ and $d$.}
This is not the case for Protocol~1 in Section~\ref{sec:file-dep-PIR}, where $\beta$ is exponential in the number of files in
order achieve optimal PIR rates. By choosing the values above for $\beta$ and $d$, the PIR rate for Protocols~2 and 3
becomes 
\begin{align*}
  \const{R}=\frac{\Gamma}{n}.
\end{align*}
We will write $\const{R}(\mathcal C)$ to highlight that the PIR rate depends on the underlying storage code
$\mathcal C$. The maximum achievable PIR rate is the PIR capacity. It was shown in \cite{BanawanUlukus18_1}
that for the noncolluding case and for a given number of files $f$ stored using an $[n,k]$ MDS code, the MDS-PIR
capacity, denoted by $\const{C}_f$, is 
\begin{IEEEeqnarray}{rCl}
    \const{C}_f\eqdef\frac{n-k}{n}\inv{\left[1-\Bigl(\frac{k}{n}\Bigr)^f\right]}.
  \label{eq:MDS-PIRcapacity}
\end{IEEEeqnarray}
Throughout the paper we refer to the capacity in \eqref{eq:MDS-PIRcapacity} as the \emph{finite MDS-PIR capacity} as it
depends on the number of files.
On the contrary, when the number of files $f\rightarrow\infty$, the \emph{asymptotic MDS-PIR capacity} is
\begin{IEEEeqnarray}{rCl}
  \const{C}_\infty\eqdef\frac{n-k}{n}.\label{eq:PIRasympt-capacity}
\end{IEEEeqnarray}
It was shown in \cite[Th.~3]{cha15} that the PIR rate for a DSS with noncolluding nodes is upperbounded by
$\const{C}_\infty$ for a special class of linear information retrieval schemes. In the case of colluding nodes, an explicit upper
bound is currently unknown, as well as an expression for the MDS-PIR capacity. Some initial work for the case of two
colluding nodes has recently been presented in \cite{SunJafar18_1}.

\section{Finite MDS-PIR Capacity-Achieving Protocol for the Noncolluding Case}
\label{sec:file-dep-PIR}

In this section, we propose a capacity-achieving protocol, named Protocol~1, that achieves the finite MDS-PIR capacity
in \eqref{eq:MDS-PIRcapacity} for the scenario of noncolluding nodes. The protocol is inspired by the protocol introduced
in \cite{BanawanUlukus18_1}.

\subsection{PIR Achievable Rate Matrix}
\label{sec:PIRachievable-rate-matrix}

In \cite{SunJafar17_1}, the concept of exploiting \emph{side information} for PIR problems was introduced. By side
information we mean additional redundant symbols not related to the requested file but downloaded by the user in order
to maintain privacy. These symbols can be exploited by the user to retrieve the requested file from the responses of the
storage nodes. In \cite[Sec.~V.A]{BanawanUlukus18_1}, it was shown that a $[5,3,3]$ MDS storage code can be used to
achieve the finite MDS-PIR capacity, where the side information is decoded by utilizing other code coordinates forming
an information set in the code array. For instance, the authors chose the $\nu=5$ information sets $\set{I}_1=\{1,2,3\}$,
$\set{I}_2=\{1,4,5\}$, $\set{I}_3=\{2,3,4\}$, $\set{I}_4=\{1,2,5\}$, and $\set{I}_5=\{3,4,5\}$ of the $[5,3,3]$ MDS code
in their PIR achievable scheme. Observe that in $\{\set{I}_i\}_{i\in\Nat{5}}$ each coordinate of the $[5,3,3]$ code
appears exactly $\kappa	=3$ times. This motivates the following definition.

\begin{definition}
  \label{def:PIRachievable-rate-matrix}
  Let $\code{C}$ be an arbitrary $[n,k]$ code. A $\nu\times n$ binary matrix $\mat{\Lambda}_{\kappa,\nu}(\code{C})$ is
  said to be a \emph{PIR achievable rate matrix} for $\code{C}$ if the following conditions are satisfied.
  \begin{enumerate}
  \item \label{item:1} The Hamming weight of each column of $\mat{\Lambda}_{\kappa,\nu}$ is $\kappa$, and
  \item \label{item:2} for each matrix row $\vect{\lambda}_i$, $i\in\Nat{\nu}$, $\chi(\vect{\lambda}_i)$ always contains an
    information set.
  \end{enumerate}
  In other words, each coordinate $j$ of $\code{C}$, $j\in\Nat{n}$, appears exactly $\kappa$ times in
  $\{\chi(\vect{\lambda}_i)\}_{i\in\Nat{\nu}}$, and every set $\chi(\vect{\lambda}_i)$ contains an information set.
\end{definition}

\begin{lemma}
  \label{lem:PIRrate_upper-bound}
  If a matrix $\mat{\Lambda}_{\nu,\kappa}(\code{C})$ exists for an $[n,k]$ code $\code{C}$, then we have
  \begin{IEEEeqnarray*}{rCl}
    \frac{\kappa}{\nu}\geq \frac{k}{n},
  \end{IEEEeqnarray*}
  where equality holds if $\chi(\vect{\lambda}_i)$, $i\in\Nat{\nu}$, are all information sets.
\end{lemma}
\begin{IEEEproof}
  Since by definition each row $\vect{\lambda}_i$ of $\mat{\Lambda}_{\nu,\kappa}$ always contains an information set, we
  have $\Hwt{\vect{\lambda}_i}\geq k$, $i\in\Nat{\nu}$. Let $\vect{v}_j$, $j\in\Nat{n}$, be the $j$-th column of
  $\mat{\Lambda}_{\kappa,\nu}$. If we look at $\mat{\Lambda}_{\kappa,\nu}$ from both a row-wise and a column-wise point
  of view, we obtain
  \begin{IEEEeqnarray*}{c}
    \nu k \leq\sum_{i=1}^{\nu}\Hwt{\vect{\lambda}_i}=\sum_{j=1}^n\Hwt{\vect{v}_j}=\kappa n,
  \end{IEEEeqnarray*}  
  from which the result follows. Clearly, equality holds if $\chi(\vect{\lambda}_i)$, $i\in\Nat{\nu}$,
    are all information sets.
\end{IEEEproof}

\begin{example}
  \label{ex:n5k3_bad}
  Consider the $[5,3,2]$ systematic code with generator matrix 
    \begin{IEEEeqnarray*}{rCl}
    \mat{G}=
    \begin{pmatrix}
      1 & 0 & 0 & 1 & 0
      \\
      0 & 1 & 0 & 1 & 0
      \\
      0 & 0 & 1 & 0 & 1
    \end{pmatrix}.
  \end{IEEEeqnarray*}
  One can easily verify that
  \begin{IEEEeqnarray*}{rCl}
    \mat{\Lambda}_{2,3}=
    \begin{pmatrix}
      0 & 1 & 1 & 1 & 1
      \\
      1 & 0 & 0 & 1 & 1
      \\
      1 & 1 & 1 & 0 & 0     
    \end{pmatrix}
  \end{IEEEeqnarray*}
  is a PIR achievable rate matrix for this code.
\end{example}

Before we state our main results, in order to clearly illustrate our example and the following achievability proof, we
first introduce the following definition. 
\begin{definition}
  \label{def:PIRinterference-matrices}
  For a given $\nu\times n$ PIR achievable rate matrix $\mat{\Lambda}_{\kappa,\nu}(\code{C})=(\lambda_{u,j})$, we define
  the PIR interference matrices $\mat{A}_{\kappa{\times}n}=(a_{i,j})$ and $\mat{B}_{(\nu-\kappa){\times}n}=(b_{i,j})$
  for the code $\code{C}$ with
  \begin{IEEEeqnarray*}{rCl}
    a_{i,j}& \eqdef &u \text{ if } \lambda_{u,j}=1,\,\forall\,j\in\Nat{n},\,i\in\Nat{\kappa},u\in\Nat{\nu},
    \\
    b_{i,j}& \eqdef &u \text{ if } \lambda_{u,j}=0,\,\forall\,j\in\Nat{n},\,i\in\Nat{\nu-\kappa},u\in\Nat{\nu}.
  \end{IEEEeqnarray*}
\end{definition}
{\mg
Note that in Definition~\ref{def:PIRinterference-matrices}, for each $j\in\Nat{n}$, distinct values of $u\in\Nat{\nu}$ should be
assigned for all $i$. Thus, the assignment is not unique in the sense that the order of the entries of each column of
$\mat{A}$ and $\mat{B}$ can be permuted. For $j\in\Nat{n}$, let $\set{A}_j\eqdef\{a_{i,j}\colon i\in\Nat{\kappa}\}$ and
$\set{B}_j\eqdef\{b_{i,j}\colon i\in\Nat{\nu-\kappa}\}$. Note that the $j$-th column of $\mat{A}$ contains the row
indices of $\mat{\Lambda}$ whose entries in the $j$-th column are equal to $1$, while $\mat{B}$ contains the remaining
row indices of $\mat{\Lambda}$. Hence, it can be observed that $\set{B}_j=\Nat{\nu}\setminus\set{A}_j$,
$\forall\,j\in\Nat{n}$.}
\begin{definition}
  \label{def:aSet_A}
  By $\set{S}(a|\mat{A}_{\kappa \times n})$ we denote the set of column coordinates of matrix
  $\mat{A}_{\kappa{\times}n}=(a_{i,j})$ in which at least one of its entries is equal to $a$, i.e.,
  \begin{IEEEeqnarray*}{rCl}
    \set{S}(a|\mat{A}_{\kappa{\times}n})\eqdef\{j\in\Nat{n}\colon\exists\,a_{i,j}=a,i\in\Nat{\kappa}\}.
  \end{IEEEeqnarray*}
\end{definition}

The following claim can be directly verified.
\begin{claim}
  \label{clm:property_PIRinterference-matrices}
  $\set{S}(a|\mat{A}_{\kappa\times n})$ contains an information set of code $\code{C}$,
  $\forall\,a\in\Nat{\nu}$. {\textcolor{black}{Moreover, for an arbitrary entry $b_{i,j}$ of $\mat{B}_{(\nu-\kappa)\times n}$,
  $\set{S}(b_{i,j}|\mat{A}_{\kappa\times n})=\set{S}(a|\mat{A}_{\kappa\times n})\subseteq\Nat{n}\setminus\{j\}$ if
  $b_{i,j}=a$.}}
\end{claim}

We illustrate the previous points in the following example.
\begin{example}
  \label{ex:A-B_n5k3_bad}
  Continuing with Example~\ref{ex:n5k3_bad} and following Definition~\ref{def:PIRinterference-matrices}, we obtain
  \begin{IEEEeqnarray*}{rCl}
    \mat{A}_{2\times 5}=
    \begin{pmatrix}
      2 &1 &1 &1 & 1
      \\
      3 &3 &3 &2 & 2
    \end{pmatrix}
    \text{ and } \mat{B}_{1\times 5}=
    \begin{pmatrix}
      1 &2 &2 &3 & 3
    \end{pmatrix}
  \end{IEEEeqnarray*}
  for $\mat{\Lambda}_{2,3}$. One can see that $\set{A}_j\cup\set{B}_j=\Nat{3}$, $\forall\,j\in\Nat{5}$. Moreover, for
  instance, take $a=1$, then $\set{S}(1|\mat{A}_{2\times 5})=\{2,3,4,5\}$ contains an information set of the $[5,3,2]$
  systematic code of Example~\ref{ex:n5k3_bad}.
\end{example}

Now consider the two matrices
\begin{IEEEeqnarray*}{rCl}
  &\begin{pmatrix}
    c^{(m)}_{\mu + {\r a_{1,1}},1} & c^{(m)}_{\mu + {\r a_{1,2}},2} &\cdots & c^{(m)}_{\mu + {\r a_{1,n}},n}
    \\
    \vdots & \cdots & & \vdots
    \\
    c^{(m)}_{\mu + {\r a_{\kappa,1}},1} & c^{(m)}_{\mu + {\r a_{\kappa,2}},2} &\cdots & c^{(m)}_{\mu + {\r a_{\kappa,n}},n}
  \end{pmatrix}&
  \text{and}\\[2mm]
  &\begin{pmatrix}
    c^{(m)}_{\mu + {\b b_{1,1}},1} & c^{(m)}_{\mu + {\b b_{1,2}},2} &\cdots & c^{(m)}_{\mu + {\b b_{1,n}},n}
    \\
    \vdots & \cdots & & \vdots
    \\
    c^{(m)}_{\mu + {\b b_{\nu-\kappa,1}},1} & c^{(m)}_{\mu + {\b b_{\nu-\kappa,2}},2} &\cdots
    & c^{(m)}_{\mu + {\b b_{\nu-\kappa,n}},n}
  \end{pmatrix}&
\end{IEEEeqnarray*}
of code symbols of the $m$-th file, where $\mu\in\Nat{\beta-\nu}\cup\{0\}$. Observe that if the user knows the first
matrix of code symbols, from Claim~\ref{clm:property_PIRinterference-matrices}, since the coordinate set
$\set{S}(b_{i,j}|\mat{A}_{\kappa\times n})\subseteq\Nat{n}\setminus\{j\}$ contains an information set and the user knows
the structure of the storage code $\code{C}$, the code symbols $c^{(m)}_{\mu+b_{i,j},j}$ of the second matrix can be
obtained. {The intuition behind the definition of the interference matrices $\mat{A}$ and $\mat{B}$ is as
  follows. Assume that $\bm X^{(1)}$ is requested. Protocol~1 requires the user to download the side information
  $\sum_{m\neq 1} c^{(m)}_{\mu+a_{i,j},j}$ based on $\mat{A}$ and also to download code symbols as sums of code symbols
  from the requested file and the side information $\sum_{m\neq 1} c^{(m)}_{\mu+b_{i,j},j}$ based on
  $\mat{B}$. Claim~\ref{clm:property_PIRinterference-matrices} then indicates that the side information
  $\sum_{m\neq 1} c^{(m)}_{\mu+b_{i,j},j}$ based on $\mat{B}$ can be reliably decoded and hence we can obtain the
  requested file by cancelling the side information.} Here, the entries of $\mat{A}$ and $\mat{B}$ are respectively
marked in red and blue. We are now ready to state Protocol~1.

\subsection{Protocol~1}
\label{sec:file-dep-PIRachievable-scheme}

The proposed Protocol~1 generalizes the MDS-coded PIR protocol in \cite{BanawanUlukus18_1} to DSSs where  files
are stored using an arbitrary linear code. Inspired by \cite{SunJafar17_1} and \cite{BanawanUlukus18_1}, a PIR
capacity-achievable scheme should follow three important principles: 1) enforcing symmetry across storage nodes, 2)
enforcing file symmetry within each storage node, and 3) exploiting side information of undesired symbols to retrieve
new desired symbols. Note that principle 1) is in general not a necessary requirement for a feasible PIR protocol.
However, as pointed out in \cite{BanawanUlukus18_1} and \cite{SunJafar18_1}, any PIR scheme can be made symmetric, hence we
keep this principle for the purpose of simplifying the implementation.

The PIR achievable rate matrix $\mat{\Lambda}_{\kappa,\nu}$ for the given storage code $\code{C}$ plays a central role
in the proposed PIR protocol. Moreover, the protocol requires $\beta=\nu^{f}$ stripes and exploits the corresponding PIR
interference matrices $\mat{A}_{\kappa\times n}$ and $\mat{B}_{(\nu-\kappa)\times n}$. Note that the number of stripes
depends on the number of files $f$, hence Protocol~1 depends on $f$ as well. We first outline the steps of the protocol,
and then we will prove that the proposed protocol satisfies the perfect privacy condition of \eqref{eq:cond1} and
results in the PIR rate of Theorem~\ref{thm:PIRachievable-rate_code} below. Without loss of generality, we assume that
the user wants to download the first file, i.e., $m=1$. The algorithm is composed of four steps as described below. {\mg
  In Appendix~\ref{sec:achievablity-proof}, we show that the algorithm generates $d\times\beta f$ query matrices
  $\mat{Q}^{(l)}$, $l\in\Nat{n}$, with
\begin{IEEEeqnarray*}{c}
  d=\frac{\kappa}{\nu-\kappa}\Bigl[\nu^f-\kappa^f\Bigr].
\end{IEEEeqnarray*}
}

\subsubsection*{$\mathsf{Step~1}$. Index Preparation}

For all files, the user interleaves the query indices for requesting the rows of $\mat{C}^{(m)}$ randomly and
independently of each other. This is equivalent to generating the interleaved code array
$\mat{Y}^{(m)}=\trans{\bigl(\trans{(\vect{y}^{(m)}_1)}|\ldots|\trans{(\vect{y}^{(m)}_\beta)}\bigr)}$,
$\forall\,m\in\Nat{f}$, with rows
\begin{IEEEeqnarray*}{rCl}
  \vect{y}^{(m)}_i=\vect{c}^{(m)}_{\pi(i)}, \quad i\in\Nat{\beta},
\end{IEEEeqnarray*}
where $\pi(\cdot):\Nat{\beta}\to\Nat{\beta}$ is a random permutation, which is privately known to the user
only. Therefore, when the user requests code symbols from each storage node, this procedure is designed to make the
requested row indices to be random and independent of the requested file index.

\smallskip

\subsubsection*{$\mathsf{Step~2}$. Download Symbols in the $i$-th Repetition}
\label{sec:download-symbols_ith}

The user downloads the needed symbols in $\kappa$ repetitions. In the $i$-th repetition, $i\in\Nat{\kappa}$, the user
downloads the required symbols in a total of $f$ rounds. Each repetition comprises $f$ rounds. In the $m$-th round,
the user  downloads symbols that are linear sums of code symbols from any $m$ files, $m\in\Nat{f}$. Using the terminology
in \cite{BanawanUlukus18_1}, the user downloads two types of symbols in each round, \emph{desired symbols}, which are
directly related to the requested file index $m=1$, and \emph{undesired symbols}, which are not related to the requested
file index $m=1$, but are exploited to decode the requested file from the desired symbols. For the desired symbols, we
will distinguish between round $\ell=1$ and round $\ell\in\Nat{2:f}$.

\begin{description}[leftmargin=0cm]
\setlength\itemsep{1ex}
\item[Undesired symbols.] The undesired symbols refer to sums of code symbols which do not contain symbols from the
  requested file. For every round $\ell$, $\ell\in\Nat{f-1}$, the user downloads the code symbols
  \begin{IEEEeqnarray}{rCl}
    \Biggl\{
    &&\quad\sum_{m'\in\set{M}}
    y^{(m')}_{((i-1)\const{U}(f-1)+\const{U}(\ell-1))\cdot\nu+{\r a_{1,j}},j},
    \nonumber\\
    &&\ldots,\sum_{m'\in\set{M}}
    y^{(m')}_{((i-1)\const{U}(f-1)+\const{U}(\ell-1))\cdot\nu+{\r a_{\kappa,j}},j},
    \nonumber\\
    &&\ldots,\sum_{m'\in\set{M}}
    y^{(m')}_{((i-1)\const{U}(f-1)+\const{U}(\ell)-1)\cdot\nu+{\r a_{1,j}},j},
    \nonumber\\
    &&\ldots,\sum_{m'\in\set{M}}
    y^{(m')}_{((i-1)\const{U}(f-1)+\const{U}(\ell)-1)\cdot\nu+{\r a_{\kappa,j}},j}\Biggr\}
    \label{eq:undesired-files_l}
  \end{IEEEeqnarray}
  for all $j \in \Nat{n}$ and for all possible subsets $\set{M}\subseteq\Nat{2:f}$, where $\card{\set{M}}=\ell$ and
  \begin{IEEEeqnarray*}{rCl}
    \const{U}(\ell)\eqdef\sum_{h=1}^\ell\kappa^{f-(h+1)}(\nu-\kappa)^{h-1}.
  \end{IEEEeqnarray*}
  In contrast to undesired symbols, desired symbols are sums of code symbols which contain symbols of the requested
  file. The main idea of the protocol is that the user downloads desired symbols that are linear sums of requested
  symbols and undesired symbols from the previous round.

\item[Desired symbols in the first round.] In the first round, the user downloads
  $\kappa\cdot\const{U}(1)=\kappa\kappa^{f-(1+1)}(\nu-\kappa)^{1-1}=\kappa^{f-1}$ undesired symbols from each storage
  node. However, these symbols cannot be exploited directly. Hence, due to symmetry, in round $\ell=1$, the user
  downloads the $\kappa^{f-1}$ desired symbols
  \begin{IEEEeqnarray}{rCl}
    \Bigl\{y^{(1)}_{\kappa^{f-1}({\r a_{i,j}}-1)+1,j},\ldots,y^{(1)}_{\kappa^{f-1}({\r
        a_{i,j}}-1)+\kappa^{f-1},j}\Bigr\}
    \label{eq:desired-files_1}
  \end{IEEEeqnarray}
  from the $j$-th storage node, $j\in\Nat{n}$, i.e., the user also downloads $\kappa^{f-1}$ symbols for $m=1$ from each
  storage node.

\item[Desired symbols in higher rounds.] In the $(\ell+1)$-th round, $\ell\in\Nat{f-1}$, in order to exploit the side
  information, i.e., the undesired symbols from the previous round, the user downloads the symbols
  \begin{IEEEeqnarray}{rCl}
    \Biggl\{
    &&\quad y^{(1)}_{\const{D}(\ell-1)\cdot\nu+{\r a_{i,j}},j}\nonumber\\
    &&\quad \>+
    \sum_{m'\in\set{M}_1}
    y^{(m')}_{((i-1)\const{U}(f-1)+\const{U}(\ell-1))\cdot\nu+{\b b_{1,j}},j},
    \nonumber\\[1mm]
    &&\quad  y^{(1)}_{(\const{D}(\ell-1)+1)\cdot\nu+{\r a_{i,j}},j}\nonumber\\
    &&\quad \>+
    \sum_{m'\in\set{M}_1}
    y^{(m')}_{((i-1)\const{U}(f-1)+\const{U}(\ell-1))\cdot\nu+{\b b_{2,j}},j},
    \nonumber\\[1mm]
    &&\ldots,y^{(1)}_{(\const{D}(\ell-1)+(\nu-\kappa)-1)\cdot\nu+{\r a_{i,j}},j}\nonumber\\
    &&\quad \>+
    \sum_{m'\in\set{M}_1}
    y^{(m')}_{((i-1)\const{U}(f-1)+\const{U}(\ell-1))\cdot\nu+{\b b_{\nu-\kappa,j}},j},
    \nonumber\\[1mm]
    &&\quad y^{(1)}_{(\const{D}(\ell-1)+(\nu-\kappa))\cdot\nu+{\r a_{i,j}},j}\nonumber\\
    &&\quad \>+
    \sum_{m'\in\set{M}_1}
    y^{(m')}_{((i-1)\const{U}(f-1)+\const{U}(\ell-1)+1)\cdot\nu+{\b b_{1,j}},j},
    \nonumber\\[1mm]
    &&\ldots,y^{(1)}_{\bigl[\const{D}(\ell-1)+(\const{U}(\ell)-\const{U}(\ell-1))(\nu-\kappa)-1\bigr]\cdot\nu+{\r a_{i,j}},j}
    \nonumber\\[1mm]
    &&\quad \>+
    \sum_{m'\in\set{M}_1}
    y^{(m')}_{((i-1)\const{U}(f-1)+\const{U}(\ell)-1)\cdot\nu+{\b b_{\nu-\kappa,j}},j},
    \nonumber\\[1mm]
    &&\ldots,y^{(1)}_{(\const{D}(\ell)-(\nu-\kappa))\cdot\nu+{\r a_{i,j}},j}\nonumber\\
    &&\quad \>+
    \sum_{m'\in\set{M}_{\const{N}(\ell)}}
    y^{(m')}_{((i-1)\const{U}(f-1)+\const{U}(\ell)-1)\cdot\nu+{\b b_{1,j}},j},
    \nonumber\\[1mm]
    &&\ldots,y^{(1)}_{(\const{D}(\ell)-1)\cdot\nu+{\r a_{i,j}},j}\nonumber\\
    &&\quad \>+
    \sum_{m'\in\set{M}_{\const{N}(\ell)}}
    y^{(m')}_{((i-1)\const{U}(f-1)+\const{U}(\ell)-1)\cdot\nu+{\b b_{\nu-\kappa,j}},j}\Biggr\}
    \IEEEeqnarraynumspace\label{eq:desired-files_l}
  \end{IEEEeqnarray}
  for all distinct $\ell$-sized subsets $\set{M}_1,\ldots,\set{M}_{\const{N}(\ell)}\subseteq\Nat{2:f}$, where
  $j\in\Nat{n}$, $\const{N}(\ell)\eqdef\binom{f-1}{\ell}$, and 
  \begin{IEEEeqnarray*}{rCl}
    \const{D}(\ell)& \eqdef &\kappa^{f-1}+\sum_{h=1}^\ell\binom{f-1}{h}\kappa^{f-(h+1)}(\nu-\kappa)^h.
  \end{IEEEeqnarray*}
  This indicates that for each combination of files indexed by $\set{M}_l$, $l\in\Nat{\const{N}(\ell)}$, the user downloads
  $\bigl[\const{U}(\ell)-1-\const{U}(\ell-1)+1\bigr](\nu-\kappa)$ new desired symbols from each storage node, and since
  there are in total $\const{N}(\ell)$ combinations of files, in each round $\const{D}(\ell)-1-\const{D}(\ell-1)+1$
  extra desired symbols are downloaded from each storage node.

\item[Exploiting the side information.] Using the fact that for a linear code $\code{C}$ any linear combination of
  codewords is also a codeword, and together with Claim~\ref{clm:property_PIRinterference-matrices}, it is not too hard to
  see that by fixing an arbitrary coordinate $j\in\Nat{n}$, there always exist some coordinates
  $\set{S}\subset\Nat{n}\setminus\{j\}$ (see Claim~\ref{clm:property_PIRinterference-matrices}) such that for a subset
  $\set{M}\subseteq\Nat{2:f}$ with $\card{\set{M}}=\ell$, the so-called \emph{aligned sum}
  \begin{IEEEeqnarray*}{rCl}
    \Biggl\{
    &&\sum_{m'\in\set{M}}
    y^{(m')}_{((i-1)\const{U}(f-1)+\const{U}(\ell-1))\cdot\nu+{\b
        b_{1,j}},j},\nonumber\\
    &&\quad \>\ldots,
    \sum_{m'\in\set{M}}
    y^{(m')}_{((i-1)\const{U}(f-1)+\const{U}(\ell)-1)\cdot\nu+{\b b_{\nu-\kappa,j}},j}
    \Biggr\}
  \end{IEEEeqnarray*}
  for $\ell\in\Nat{f-1}$ and $i\in\Nat{\kappa}$, can be decoded. Consequently, in the $(\ell+1)$-th round, from each
  storage node $j$ we can collect code symbols related to $m=1$ from the desired symbols, i.e.,
  \begin{IEEEeqnarray}{rCl}
    \Bigl\{y^{(1)}_{\const{D}(\ell-1)\cdot\nu+{\r a_{i,j}},j},\ldots, y^{(1)}_{(\const{D}(\ell)-1)\cdot\nu+{\r
        a_{i,j}},j}\Bigr\}
    \label{eq:desired-interference}
  \end{IEEEeqnarray}
  is obtained.

\item[Symmetry across storage nodes.] In the previous steps, since the user downloads the same amount of required
  symbols for each $j\in\Nat{n}$ and for every round, symmetry across storage nodes is ensured.
  
\item[File symmetry within each storage node.] To ensure that the privacy condition \eqref{eq:cond1} is fulfilled, we
  have to make sure that in each round $\ell\in\Nat{f}$ of each repetition, for each storage node and for every
  combination of files indexed by $\set{M}\subseteq\Nat{f}$ with $\card{\set{M}}=\ell$, the user requests the same number of linear
  sums $\eta(\set{M})\eqdef\sum_{m\in\set{M}}y^{(m)}_{\eta_m,j}$, where $\eta_m$ depends on $m$. This will be shown to
  be inherent from the protocol (see proof of Theorem~\ref{thm:PIRachievable-rate_code} in Appendix
  \ref{sec:achievablity-proof}). In addition, since the user always requests the same number of linear sums for each
  combination of files, the scheme also implies that the frequencies of requested code symbols pertaining to each
  individual file index $m\in\Nat{f}$ among all the linear sums are the same for each storage node.
\end{description}

\smallskip

\subsubsection*{$\mathsf{Step~3}$. Complete $\kappa$ Repetitions} The user repeats Step~2 until $i=\kappa$. We will show
that by our designed parameters $\const{U}(\ell)$ and $\const{D}(\ell)$, the user indeed downloads in total
$\beta=\nu^f$ stripes for the requested file (see again Appendix \ref{sec:achievablity-proof}).

\smallskip

\subsubsection*{$\mathsf{Step~4}$. Shuffling the Order of Queries to Each Node} The order of the queries to each storage
node is uniformly shuffled to prevent the storage node to be able to identify which file is requested from the index of
the first downloaded symbol.

\subsection{Achievable PIR Rate}
\label{sec:PIRrate_codes}

The PIR rate, $\const{R}(\code{C})$, of Protocol~1 in Section~\ref{sec:file-dep-PIRachievable-scheme} for a DSS where
$f$ files are stored using an arbitrary $[n,k]$ code $\code{C}$ is given in the following theorem.
\begin{theorem}
  \label{thm:PIRachievable-rate_code}
  Consider a DSS that uses an $[n,k]$ code $\code{C}$ to store $f$ files. If a PIR achievable rate matrix
  $\mat{\Lambda}_{\kappa,\nu}(\code{C})$ exists, then the PIR rate
  \begin{IEEEeqnarray}{rCl}
    \const{R}(\code{C})& = &\frac{(\nu-\kappa)k}{\kappa n} \inv{\left[1-\Bigl(\frac{\kappa}{\nu}\Bigr)^f\right]}
    \label{eq:PIRachievable-rate_code}
  \end{IEEEeqnarray}
  is achievable.
\end{theorem}
\begin{IEEEproof}
  See Appendix~\ref{sec:achievablity-proof}.
\end{IEEEproof}

We remark that from Lemma~\ref{lem:PIRrate_upper-bound}, \eqref{eq:PIRachievable-rate_code} is smaller than or equal to the
finite MDS-PIR capacity in \eqref{eq:MDS-PIRcapacity} since
\begin{IEEEeqnarray}{rCl}
  \const{R}(\code{C})& = & \frac{\frac{\nu k}{\kappa n}\Bigl[1-\frac{\kappa}{\nu}\Bigr]}
  {\Bigl[1-\bigl(\frac{\kappa}{\nu}\bigr)^f\Bigr]} =\frac{\nu k}{\kappa n}\inv{\left[1+\frac{\kappa}{\nu}
      +\cdots+\Bigl(\frac{\kappa}{\nu}\Bigr)^{f-1}\right]}
  \nonumber\\
  & \leq &\inv{\left[1+\frac{k}{n}
      +\cdots+\Bigl(\frac{k}{n}\Bigr)^{f-1}\right]},\label{eq:upperbound_PIRrate}
\end{IEEEeqnarray}
and it becomes the finite MDS-PIR capacity  in \eqref{eq:MDS-PIRcapacity} if there exists a matrix 
$\mat{\Lambda}_{\kappa,\nu}$ for $\code{C}$ with $\frac{\kappa}{\nu}=\frac{k}{n}$. The inequality in
\eqref{eq:upperbound_PIRrate} follows from Lemma~\ref{lem:PIRrate_upper-bound}.
{
\begin{corollary}
  \label{cor:MDS-PIRcapacity-achieving-codes}
  If a PIR achievable rate matrix $\mat{\Lambda}_{\kappa,\nu}(\code{C})$ with $\frac{\kappa}{\nu}=\frac{k}{n}$ exists for an
  $[n,k]$ code $\code{C}$, then the finite MDS-PIR capacity in \eqref{eq:MDS-PIRcapacity} is achievable.
\end{corollary}
}

This gives rise to the following definition.
\begin{definition}
  \label{def:MDS-PIRcapacity-achieving-codes}
  A PIR achievable rate matrix $\mat{\Lambda}_{\kappa,\nu}(\code{C})$ with $\frac{\kappa}{\nu}=\frac{k}{n}$ for an
  $[n,k]$ code $\code{C}$ is called an \emph{MDS-PIR capacity-achieving} matrix, and $\code{C}$ is referred to as an
  \emph{MDS-PIR capacity-achieving} code.
\end{definition}

{We remark that there might exist codes that are MDS-PIR capacity-achieving for which an MDS-PIR capacity-achieving matrix does not exist.}

Note that the largest achievable PIR rate in the noncolluding case where data is stored using an arbitrary linear code
is still unknown. Interestingly, it is observed from Lemma~\ref{lem:PIRrate_upper-bound} and \eqref{eq:upperbound_PIRrate}
that the largest possible achievable PIR rate for an arbitrary linear code with Protocol~1 strongly depends on the
smallest possible value of $\frac{\kappa}{\nu}$ for which a PIR achievable rate matrix $\mat{\Lambda}_{\kappa,\nu}$
exists. We stress that the existence of an MDS-PIR capacity-achieving matrix $\mat{\Lambda}_{\kappa,\nu}$ does not
necessarily require $(\nu,\kappa)=(n,k)$, but $\frac{\kappa}{\nu}=\frac{k}{n}$.

Since the existence of a PIR achievable rate matrix is connected to the information sets of a code, we review a widely
known result in coding theory.
\begin{proposition}[\hspace{-1sp}{\cite[Th.~1.4.15]{HuffmanPless10_1}}]
  \label{prop:infoS_nkd}
  Let $\code{C}$ be an $[n,k,d^{\code{C}}_\mathsf{min}]$ code. Then, every set of $n-d^{\code{C}}_\mathsf{min}+1$ coordinates of $\code{C}$
  contains an information set. Furthermore, $n-d^{\code{C}}_\mathsf{min}+1$ is the smallest number of coordinates with this
  property.
\end{proposition}

\begin{lemma}
  \label{lem:lower-nu_k}
  For a given $[n,k,d^{\code{C}}_\mathsf{min}]$ code $\code{C}$, there always exists a PIR achievable rate matrix
  $\mat{\Lambda}_{k,\nu}$ with 
  \begin{IEEEeqnarray*}{rCl}
    \nu=k+\min\left(k,d^{\code{C}}_\mathsf{min}-1\right).
  \end{IEEEeqnarray*}
\end{lemma}
\begin{IEEEproof}
  See Appendix~\ref{sec:proof_lower-nu_k}.
\end{IEEEproof}

A lower bound on the largest possible achievable PIR rate obtained from Theorem~\ref{thm:PIRachievable-rate_code} and
Lemma~\ref{lem:lower-nu_k} is given as follows.
{
\begin{corollary}
  Consider a DSS that uses an $[n,k,d^\code{C}_\mathsf{min}]$ code $\code{C}$ to store $f$ files. Then, the PIR rate
  \begin{IEEEeqnarray*}{c}
    \const{R}(\code{C})=\frac{\min\left(k,d^{\code{C}}_\mathsf{min}-1\right)}{n}
    \inv{\left[1-\Bigl(\frac{k}{k+\min\left(k,d^{\code{C}}_\mathsf{min}-1\right)}\Bigr)^f\right]}
  \end{IEEEeqnarray*}
  is achievable. 
\end{corollary}
}

We remark that because every set of $k$ coordinates of an $[n,k]$ MDS code is an information set, we can construct $n$
information sets by cyclically shifting an arbitrary information set $n$ times, hence an MDS-PIR capacity-achieving
matrix $\mat{\Lambda}_{k,n}$ of an MDS code can be easily constructed. In other words, Protocol~1 with MDS codes is
MDS-PIR capacity-achieving (see Corollary~\ref{cor:MDS-PIRcapacity-achieving-codes}) and MDS codes are a class of
MDS-PIR capacity-achieving codes.

{\mg
\begin{remark}
  Since minimum storage regenerating (MSR) codes are MDS codes \cite{Dim10} and can be viewed as scalar linear codes over a larger extension field, it follows that MSR codes are also MDS-PIR
  capacity-achieving codes.
\end{remark}}

In Section~\ref{sec:MDS-PIRcapacity-achiving-codes}, we provide a necessary and a sufficient condition for an
arbitrary linear code to achieve the MDS-PIR capacity with Protocol~1 and give certain families of MDS-PIR
capacity-achieving codes. For illustration purposes, in the next subsection, we give an example of an MDS-PIR
capacity-achieving code.

\subsection{A $[5,3,2]$ MDS-PIR Capacity-Achieving Code for $f=2$}
\label{sec:example_n5k3}

In this subsection, we compute the PIR achievable rate of a $[5,3,2]$ non-MDS code for a DSS that stores two files, $f=2$,
and show that it is MDS-PIR capacity-achieving.

Let $\code{C}$ be a non-MDS $[5,3,2]$ binary code with generator matrix
\begin{IEEEeqnarray}{rCl}
  \mat{G}=
  \begin{pmatrix}
    1 & 0 & 0 & 1 & 0
    \\
    0 & 1 & 0 & 1 & 1
    \\
    0 & 0 & 1 & 0 & 1
  \end{pmatrix}.\label{eq:goodG_n5k3}
\end{IEEEeqnarray}
One can see that the $\nu\times n=5\times 5$ matrix
\begin{IEEEeqnarray*}{rCl}
  \mat{\Lambda}_{3,5}=
  \begin{pmatrix}
    1 & 1 & 1 & 0 & 0
    \\
    1 & 0 & 0 & 1 & 1
    \\
    0 & 1 & 0 & 1 & 1
    \\
    0 & 1 & 1 & 1 & 0
    \\
    1 & 0 & 1 & 0 & 1
  \end{pmatrix}
\end{IEEEeqnarray*}
is a PIR achievable rate matrix. From $\mat{\Lambda}_{3,5}$, we obtain the following sets:
\begin{IEEEeqnarray*}{rCl}
  \chi(\vect{\lambda}_1)& = &\{1,2,3\},
  \chi(\vect{\lambda}_2)=\{1,4,5\},\chi(\vect{\lambda}_3)=\{2,4,5\},\nonumber\\
  \chi(\vect{\lambda}_4)& = &\{2,3,4\},\chi(\vect{\lambda}_5)=\{1,3,5\}.
\end{IEEEeqnarray*}
All of these sets contain an information set of $\code{C}$ (see
Definition~\ref{def:PIRachievable-rate-matrix}). Furthermore, we get the following PIR interference matrices
\begin{IEEEeqnarray*}{rCl}
  &&{\r\mat{A}_{3{\times}5}=
    \begin{pmatrix}
      1& 1 &1 &2 &2
      \\
      2& 3 &4 &3 &3
      \\
      5& 4 &5 &4 &5
    \end{pmatrix}},
  \\[2mm]
  &&{\b\mat{B}_{2{\times}5}=
    \begin{pmatrix}
      3& 2 &2 &1 &1
      \\
      4& 5 &3 &5 &4
    \end{pmatrix}}.
\end{IEEEeqnarray*}
One can see that Claim~\ref{clm:property_PIRinterference-matrices} holds. For example,
$\set{S}(3|\mat{A}_{3\times 5})=\{2,4,5\}$ contains an information set for $\code{C}$.

In the next step, for each $m\in\Nat{2}$ and for $\beta=\nu^f=5^2$, we first generate the interleaved code array
$\mat{Y}^{(m)}$ with row vectors $\vect{y}^{(m)}_i=\vect{c}^{(m)}_{\pi(i)}$, $i\in\Nat{5^2}$, by a randomly selected
permutation function $\pi(\cdot)$. Suppose that the user wishes to obtain $\mat{X}^{(1)}$. We list all downloaded sums
of code symbols in Table~\ref{tab:MDS-PIRcapacity-achieving-code_n5k3}, which is similar to
\cite[Table~II]{BanawanUlukus18_1}. Similar to the PIR protocol in \cite{BanawanUlukus18_1}, Protocol~1 requires $f=2$
rounds in each repetition, and the scheme needs to be repeated $\kappa=3$ times. Note that since the protocol requests
an equal amount of code symbols associated with $\mat{X}^{(1)}$ and $\mat{X}^{(2)}$, it is straightforward to see that
the privacy constraint is satisfied.

\begin{table*}[htbp]
  \centering
  \caption{{Protocol~1 with a $[5,3,2]$ non-MDS code for $f=2$.}}
  \label{tab:MDS-PIRcapacity-achieving-code_n5k3}
   \vspace{-7.5mm}
  \begin{IEEEeqnarray*}{rCl}
    \begin{IEEEeqnarraybox}[
      \IEEEeqnarraystrutmode
      \IEEEeqnarraystrutsizeadd{3pt}{1pt}]{v/c/V/c/v/c/v/c/v/c/v/c/v/c/v}
      \IEEEeqnarrayrulerow\\
      & && && \text{Server } 1  && \text{Server } 2 &&
      \text{Server } 3 && \text{Server } 4 &&
      \text{Server } 5 &\\
      \hline\hline
      & && && y^{(1)}_{3({\r 1}-1)+1,1} &&  y^{(1)}_{3({\r 1}-1)+1,2} && y^{(1)}_{3({\r 1}-1)+1,3} &&
      y^{(1)}_{3({\r 2}-1)+1,4} && y^{(1)}_{3({\r 2}-1)+1,5} &
      \\
      & && && y^{(1)}_{3({\r 1}-1)+2,1} &&  y^{(1)}_{3({\r 1}-1)+2,2} && y^{(1)}_{3({\r 1}-1)+2,3} &&
      y^{(1)}_{3({\r 2}-1)+2,4} && y^{(1)}_{3({\r 2}-1)+2,5} &
      \\
      &  && && y^{(1)}_{3({\r 1}-1)+3,1}
      &&  y^{(1)}_{3({\r 1}-1)+3,2}       && y^{(1)}_{3({\r 1}-1)+3,3}
      &&  y^{(1)}_{3({\r 2}-1)+3,4}       && y^{(1)}_{3({\r 2}-1)+3,5} &
      \\*\cline{5-15}
      & && \rot{\rlap{\text{round} 1}}&&y^{(2)}_{5\cdot 0+{\r 1},1} &&  y^{(2)}_{5\cdot 0+{\r 1},2} && y^{(2)}_{5\cdot 0+{\r 1},3} &&
      y^{(2)}_{5\cdot 0+{\r 2},4} && y^{(2)}_{5\cdot 0+{\r 2},5} &
      \\
      & && &&y^{(2)}_{5\cdot 0+{\r 2},1} && y^{(2)}_{5\cdot 0+{\r 3},2} && y^{(2)}_{5\cdot 0+{\r 4},3} &&
      y^{(2)}_{5\cdot 0+{\r 3},4} && y^{(2)}_{5\cdot 0+{\r 3},5}
      \\
      & \rot{\rlap{\text{repetition} 1}}&& &&
      y^{(2)}_{5\cdot 0+{\r 5},1} && y^{(2)}_{5\cdot 0+{\r 4},2} && y^{(2)}_{5\cdot 0+{\r 5},3} &&
      y^{(2)}_{5\cdot 0+{\r 4},4} && y^{(2)}_{5\cdot 0+{\r 5},5}      
      \\*\cline{3-15}
      & && &&y^{(1)}_{3\cdot 5+{\r 1},1}+y^{(2)}_{5\cdot 0+{\b 3},1} &&
      y^{(1)}_{3\cdot 5+{\r 1},2}+y^{(2)}_{5\cdot 0+{\b 2},2}
      && y^{(1)}_{3\cdot 5+{\r 1},3}+y^{(2)}_{5\cdot 0+{\b 2},3} &&
      y^{(1)}_{3\cdot 5+{\r 2},4}+y^{(2)}_{5\cdot 0+{\b 1},4}&& y^{(1)}_{3\cdot 5+{\r 2},5}+y^{(2)}_{5\cdot 0+{\b 1},5} &
      \\
      & && \rot{\rlap{\text{rnd}. 2}}&&y^{(1)}_{4\cdot 5+{\r 1},1}+y^{(2)}_{5\cdot 0+{\b 4},1} &&
      y^{(1)}_{4\cdot 5+{\r 1},2}+y^{(2)}_{5\cdot 0+{\b 5},2}
      && y^{(1)}_{4\cdot 5+{\r 1},3}+y^{(2)}_{5\cdot 0+{\b 3},3}  && y^{(1)}_{4\cdot 5+{\r 2},4}+y^{(2)}_{5\cdot 0+{\b 5},4}
      && y^{(1)}_{4\cdot 5+{\r 2},5}+y^{(2)}_{5\cdot 0+{\b 4},5} &
      \\*\hline\hline
      & && && y^{(1)}_{3({\r 2}-1)+1,1} &&  y^{(1)}_{3({\r 3}-1)+1,2} && y^{(1)}_{3({\r 4}-1)+1,3} &&
      y^{(1)}_{3({\r 3}-1)+1,4} && y^{(1)}_{3({\r 3}-1)+1,5} &
      \\
      & && && y^{(1)}_{3({\r 2}-1)+2,1} &&  y^{(1)}_{3({\r 3}-1)+2,2} && y^{(1)}_{3({\r 4}-1)+2,3} &&
      y^{(1)}_{3({\r 3}-1)+2,4} && y^{(1)}_{3({\r 3}-1)+2,5} &
      \\
      & && && y^{(1)}_{3({\r 2}-1)+3,1} &&
      y^{(1)}_{3({\r 3}-1)+3,2} && y^{(1)}_{3({\r 4}-1)+3,3} &&
      y^{(1)}_{3({\r 3}-1)+3,4} && y^{(1)}_{3({\r 3}-1)+3,5} &
      \\*\cline{5-15}
      & &&  \rot{\rlap{\text{round} 1}}&&y^{(2)}_{5\cdot 1+{\r 1},1} &&  y^{(2)}_{5\cdot 1+{\r 1},2} && y^{(2)}_{5\cdot 1+{\r 1},3} &&
      y^{(2)}_{5\cdot 1+{\r 2},4} && y^{(2)}_{5\cdot 1+{\r 2},5} &
      \\
      & && &&y^{(2)}_{5\cdot 1+{\r 2},1} && y^{(2)}_{5\cdot 1+{\r 3},2} && y^{(2)}_{5\cdot 1+{\r 4},3} &&
      y^{(2)}_{5\cdot 1+{\r 3},4} && y^{(2)}_{5\cdot 1+{\r 3},5}
      \\
      & \rot{\rlap{\text{repetition} 2}} && &&y^{(2)}_{5\cdot 1+{\r 5},1} &&
      y^{(2)}_{5\cdot 1+{\r 4},2} && y^{(2)}_{5\cdot 1+{\r 5},3} &&
      y^{(2)}_{5\cdot 1+{\r 4},4} && y^{(2)}_{5\cdot 1+{\r 5},5}      
      \\*\cline{3-15}
      & && &&y^{(1)}_{3\cdot 5+{\r 2},1}+y^{(2)}_{5+{\b 3},1} &&
      y^{(1)}_{3\cdot 5+{\r 3},2}+y^{(2)}_{5+{\b 2},2}
      && y^{(1)}_{3\cdot 5+{\r 4},3}+y^{(2)}_{5+{\b 2},3} && y^{(1)}_{3\cdot 5+{\r 3},4}+y^{(2)}_{5+{\b 1},4}
      && y^{(1)}_{3\cdot 5+{\r 3},5}+y^{(2)}_{5+{\b 1},5} &
      \\
      & && \rot{\rlap{\text{rnd}. 2}}&&y^{(1)}_{4\cdot 5+{\r 2},1}+y^{(2)}_{5+{\b 4},1} && y^{(1)}_{4\cdot 5+{\r 3},2}+y^{(2)}_{5+{\b 5},2}
      && y^{(1)}_{4\cdot 5+{\r 4},3}+y^{(2)}_{5+{\b 3},3}  && y^{(1)}_{4\cdot 5+{\r 3},4}+y^{(2)}_{5+{\b 5},4}
      && y^{(1)}_{4\cdot 5+{\r 3},5}+y^{(2)}_{5+{\b 4},5} &
      \\*\hline\hline
      & && && y^{(1)}_{3({\r 5}-1)+1,1} &&  y^{(1)}_{3({\r 4}-1)+1,2} && y^{(1)}_{3({\r 5}-1)+1,3} &&
      y^{(1)}_{3({\r 4}-1)+1,4} && y^{(1)}_{3({\r 5}-1)+1,5} &
      \\
      & && && y^{(1)}_{3({\r 5}-1)+2,1} &&  y^{(1)}_{3({\r 4}-1)+2,2} && y^{(1)}_{3({\r 5}-1)+2,3} &&
      y^{(1)}_{3({\r 4}-1)+2,4} && y^{(1)}_{3({\r 5}-1)+2,5} &
      \\
      & && && y^{(1)}_{3({\r 5}-1)+3,1} &&
      y^{(1)}_{3({\r 4}-1)+3,2} && y^{(1)}_{3({\r 5}-1)+3,3} &&
      y^{(1)}_{3({\r 4}-1)+3,4} && y^{(1)}_{3({\r 5}-1)+3,5} &
      \\*\cline{5-15}
      & && \rot{\rlap{\text{round} 1}}&&y^{(2)}_{5\cdot 2+{\r 1},1} &&  y^{(2)}_{5\cdot 2+{\r 1},2} && y^{(2)}_{5\cdot 2+{\r 1},3} &&
      y^{(2)}_{5\cdot 2+{\r 2},4} && y^{(2)}_{5\cdot 2+{\r 2},5} &
      \\
      & && &&y^{(2)}_{5\cdot 2+{\r 2},1} && y^{(2)}_{5\cdot 2+{\r 3},2} && y^{(2)}_{5\cdot 2+{\r 4},3} &&
      y^{(2)}_{5\cdot 2+{\r 3},4} && y^{(2)}_{5\cdot 2+{\r 3},5}
      \\
      & \rot{\rlap{~\text{repetition} 3}}&& &&y^{(2)}_{5\cdot 2+{\r 5},1} &&
      y^{(2)}_{5\cdot 2+{\r 4},2} && y^{(2)}_{5\cdot 2+{\r 5},3} &&
      y^{(2)}_{5\cdot 2+{\r 4},4} && y^{(2)}_{5\cdot 2+{\r 5},5}      
      \\*\cline{3-15}
      & && &&y^{(1)}_{3\cdot 5+{\r 5},1}+y^{(2)}_{5\cdot 2+{\b 3},1} &&
      y^{(1)}_{3\cdot 5+{\r 4},2}+y^{(2)}_{5\cdot 2+{\b 2},2}
      && y^{(1)}_{3\cdot 5+{\r 5},3}+y^{(2)}_{5\cdot 2+{\b 2},3} && y^{(1)}_{3\cdot 5+{\r 4},4}+y^{(2)}_{5\cdot 2+{\b 1},4}
      && y^{(1)}_{3\cdot 5+{\r 5},5}+y^{(2)}_{5\cdot 2+{\b 1},5} &
      \\
      & && \rot{\rlap{\text{rnd}. 2}}&&y^{(1)}_{4\cdot 5+{\r 5},1}+y^{(2)}_{5\cdot 2+{\b 4},1} &&
      y^{(1)}_{4\cdot 5+{\r 4},2}+y^{(2)}_{5\cdot 2+{\b 5},2}
      && y^{(1)}_{4\cdot 5+{\r 5},3}+y^{(2)}_{5\cdot 2+{\b 3},3} && y^{(1)}_{4\cdot 5+{\r 4},4}+y^{(2)}_{5\cdot 2+{\b 5},4}
      && y^{(1)}_{4\cdot 5+{\r 5},5}+y^{(2)}_{5\cdot 2+{\b 4},5} &
      \\*\IEEEeqnarrayrulerow
    \end{IEEEeqnarraybox}
  \end{IEEEeqnarray*}
\end{table*}

It should be mentioned that here we strongly make use of the PIR interference matrices. For example, in round $2$ of
repetition $1$ (see Table~\ref{tab:MDS-PIRcapacity-achieving-code_n5k3}), since the user knows $\code{C}$, the code symbols
$y^{(2)}_{5\cdot 0+{\b 3},1}$, $y^{(2)}_{5\cdot 0+{\b 2},2}$, and $y^{(2)}_{5\cdot 0+{\b 2},3}$ can be obtained by knowing
$\{y^{(2)}_{5\cdot 0+{\r 3},2},y^{(2)}_{5\cdot 0+{\r 3},4}, y^{(2)}_{5\cdot 0+{\r 3},5}\}$ and
$\{y^{(2)}_{5\cdot 0+{\r 2},1},y^{(2)}_{5\cdot 0+{\r 2},4}, y^{(2)}_{5\cdot 0+{\r 2},5}\}$, from which the corresponding
coded symbols $\{y^{(1)}_{3\cdot 5+{\r 1},1},y^{(1)}_{3\cdot 5+{\r 1},2}, y^{(1)}_{3\cdot 5+{\r 1},3}\}$ can be obtained
by cancelling the side information. Since $\{1,2,3\}$ is an information set, the corresponding requested file vector of
length $k=3$ can also be decoded. Hence, in summary, it is sufficient to reliably decode $5^2=25$ different length-$k$
requested file vectors for $m=1$. In summary, for $f=2$, the user downloads $3\times 5$ undesired symbols based
on~\eqref{eq:undesired-files_l} and $(3+2)\times 5=25$ desired symbols according to~\eqref{eq:desired-files_1} and
\eqref{eq:desired-files_l} in each repetition. Hence, the PIR achievable rate is equal to
\begin{IEEEeqnarray*}{rCl}
  \const{R}(\code{C})=\frac{3\cdot 25}{3\cdot (25+15)}=\frac{5}{8}=\frac{1-\frac{3}{5}}{1-\bigl(\frac{3}{5}\bigr)^2},
\end{IEEEeqnarray*}
which corresponds to the finite MDS-PIR capacity in \eqref{eq:MDS-PIRcapacity} with $f=2$, i.e., the $[5,3,2]$ non-MDS
code given by \eqref{eq:goodG_n5k3} is MDS-PIR capacity-achieving.

\section{Asymptotic MDS-PIR Capacity-Achieving Protocol for the Noncolluding Case}
\label{sec:file-indep-PIR}

In this section, we present Protocol~2, a PIR protocol with PIR rate independent of the number of files that achieves the asymptotic MDS-PIR capacity in
\eqref{eq:PIRasympt-capacity} for the case of noncolluding nodes. We assume that the DSS uses an $[n,k]$ code $\code{C}$
over $\GF(q)$ of rate $R^{\mathcal C}$ and subpacketization $\alpha$. For such a code $\mathcal C$, the user designs the
$l$-th, $l\in\Nat{n}$, query as
\begin{align}
  \label{Eq: GenQuery_design}
  \bm Q^{(l)}=\bm U+\bm V^{(l)},
\end{align}
where $\bm U=(u_{i,j})$ is a $d \times \beta f$ matrix whose elements $u_{i,j}$ are chosen independently and uniformly
at random from $\GF(q)$ {\mg and whose purpose is to make $\bm Q^{(l)}$ appear random and thus ensure privacy.}
$\bm V^{(l)}=\bigl(v_{i,j}^{(l)}\bigr)$ is a $d\times\beta f$ deterministic binary matrix over $\GF(q)$, where
$v^{(l)}_{i,j}=1$ means that the $j$-th symbol in node $l$ is accessed by the $i$-th subquery of $\bm Q^{(l)}$, {\mg
  that allows recovery of the requested data by the user.}  {Matrix $\bm V^{(l)}$ is constructed from a $d\times n$
  matrix $\hat{\bm E}$, as explained below.

Let $\set{I}_1,\ldots,\set{I}_\beta$ be $\beta$ information sets for $\code{C}$ (which are implicitly linked to the
  $\beta$ stripes of each file) and define $\set{F}_l\eqdef\{i\in\Nat{\beta}\colon l\in\set{I}_i\}$ to be the set of
indices of the information sets $\set I_1,\ldots,\set I_{\beta}$ containing the $l$-th coordinate of $\C$. Then, 
$\hat{\bm E}=(\hat e_{i,l})$ is a binary matrix of size $d\times n$ that has the following structure.
\begin{enumerate}
\item[$\mathsf{C1.}$] Each row, denoted by $\hat{\bm e}_i$, $i\in\Nat{d}$, has Hamming weight
  $\Hwt{\hat{\bm e}_i}=\Gamma$.
\item[$\mathsf{C2.}$] Each row $\hat{\bm e}_i$ is an erasure pattern that is correctable by $\mathcal C$.
\item[$\mathsf{C3.}$] Each column, denoted by $\bm t_l$, $l\in\Nat{n}$, has weight $\Hwt{\vect{t}_l}=\card{\set{F}_l}$,
  i.e., the weight of the $l$-th column of $\hat{\bm E}$ is the number of times the $l$-th coordinate of the storage
  code $\code{C}$ appears in the $\beta$ information sets $\set{I}_1,\ldots,\set{I}_\beta$.
\end{enumerate}
For later use, we call the vector $(\Hwt{\vect{t}_1},\ldots,\Hwt{\vect{t}_n})$ the \emph{column weight profile of
  $\hat{\mat{E}}$}.
    
Matrix $\bm V^{(l)}$ is constructed from $\Ehat$ such that if $\hat e_{i,l}=1$, then the $i$-th subquery of the $l$-th
query, $\bm q_i^{(l)}$, accesses a code symbol stored in the $l$-th node. Additionally, $\hat{\mat{E}}$ is a matrix
having strictly $\Gamma d$ nonzero entries, ensuring that $\Gamma d$ code symbols are downloaded by the protocol. We
defer the intuition behind the three conditions above until later in this section.
More precisely, matrix $\bm V^{(l)}$ is constructed from $\Ehat$ as follows.} For $l\in\Nat{n}$, $\bm V^{(l)}$ has the form
\begin{align*}
  \bm V^{(l)}=
  \left(\begin{matrix}
      \bm 0_{d\times(m-1)\beta} \mid \Scale[1]{\BDelta}_l \mid \bm 0_{d\times(f-m)\beta} 
    \end{matrix}\right),
\end{align*}
where $\bm \Delta_l$ is the $d\times\beta$ binary matrix
\begin{align} 
  \label{Eq: delta_design}
  \Scale[1]{\BDelta}_l=
  \left(\begin{matrix}
      \bm \omega^{\textup{\textsf{\tiny T}}}_{j_1^{(l)}} & | &  \bm \omega^{\textup{\textsf{\tiny T}}}_{j_2^{(l)}} & | & \ldots & | & \bm \omega^{\textup{\textsf{\tiny T}}}_{j_d^{(l)}} 
    \end{matrix}\right)^{\textup{\textsf{\tiny T}}},
\end{align}%
with $\vect{\omega}_j$, $j\in\Nat{\beta}$, being the $j$-th $\beta$-dimensional unit vector, i.e., a length-$\beta$
weight-$1$ binary vector with a single $1$ at the $j$-th position and $\vect{\omega}_0=\vect{0}_{1\times\beta}$. Also,
given a chosen $d\times n$ matrix $\hat{\mat{E}}$,
\begin{align}
  \label{Eq: IndexDownload}
  j_i^{(l)}=\begin{cases}
    s_{i}^{(l)} & \text{if $\hat e_{i,l}=1$},
    \\
    0 & \text{otherwise},
  \end{cases}
\end{align}
where $s_i^{(l)}\in \mathcal{F}_l$ and $s_i^{(l)}\neq s_{i'}^{(l)}$ for $i\neq i'$, $i,i'\in\Nat{d}$. This completes the
construction of the protocol.

{Now, we provide the intuition behind conditions $\mathsf{C1}$, $\mathsf{C2}$, and $\mathsf{C3}$ above.
\begin{itemize}
\item Condition $\mathsf{C1}$ stems from the fact that the user should be able to recover $\Gamma$ unique code symbols
  of the requested file $\bm{X}^{(m)}$ from the $i$-th subqueries $\bm q^{(l)}_i$ that are sent to the $n$ nodes. Thus,
  each row of $\hat{\bm E}$ should have exactly $\Gamma$ ones.
\item For $\mathsf{C2}$, consider an arbitrary row $\hat{\bm e}_i$ of $\Ehat$. The corresponding set of $n$ subqueries
  $\{\bm q_{i}^{(1)},\ldots,\bm q_{i}^{(n)}\}$ trigger a response from the $n$ nodes of the form
  \begin{align*}
    r_{l,i}=\begin{cases}
      Y_l+\phi_l & \text{if } \hat{e}_{i,l}=1,\\
      Y_l     & \text{otherwise}, 
    \end{cases}
  \end{align*}  
  where $\phi_l$ represents a code symbol present in the $l$-th node, and $Y_l$ is some interference symbol
  generated due to the product between $\bm q_{i}^{(l)}$ and the content of the $l$-th node. The vector
  $(Y_1,\ldots,Y_n)$ represents a codeword of $\mathcal C$ (see also Theorem~\ref{th:GenPIR} below and its proof in
  Appendix~\ref{Appendix: Proof1} for further details). In order to recover $\phi_l$, $l\in\chi(\hat{\bm e}_i)$, we need
  to know $Y_l$. This can be seen as a decoding problem over the binary erasure channel. In other words, the $i$-th row
  of $\hat{\bm E}$ should be an erasure pattern 
  that is correctable by $\mathcal C$.
\item Condition $\mathsf{C3}$ comes from the fact that the protocol should be able to recover $\Hwt{\vect{t}_l}$ unique
  code symbols from the $l$-th node.
\end{itemize}
}

The idea behind the construction of $\mat{V}^{(l)}$ from $\hat{\mat{E}}$ is that the retrieval process can be cast as the correction of an erasure pattern. Thus, we design $\mat{V}^{(l)}$ (and subsequently the responses) so that erasure correction is possible.

We remark that for a code $\code{C}$, $\hat{\mat{E}}$ and $\{\set{I}_i\}_{i\in\Nat{\beta}}$ need not be
unique. Furthermore, each set $\set{I}_i$, $i\in\Nat{\beta}$, can alternatively be represented as a correctable erasure
pattern $\bar{\vect{e}}_i=(\bar{e}_{i,1},\ldots,\bar{e}_{i,n})$, where $\bar{e}_{i,l}= 0$,
$\forall\,l\in\set{I}_i$. Also, the information sets $\{\set{I}_i\}_{i\in\Nat{\beta}}$ can alternatively be
defined by a matrix $\bar{\bm E}$ of size $\beta\times n$ as
\begin{align*}
  \bar{\bm E}=\left(\begin{matrix}
      \bar{\vect{e}}_{1}
      \\
      \vdots
      \\
      \bar{\vect{e}}_{\beta}
    \end{matrix}\right).
\end{align*}
The two matrices $\hat{\mat{E}}$ and $\bar{\bm E}$ can be stacked into the matrix $\bm E=(e_{i,l})$ as
\begin{align}
  \bm E=\left(\begin{matrix}
      \hat{\bm E}
      \\
      \bar{\bm E}
    \end{matrix}\right).\label{eq:PIRerasure-matrix}
\end{align}
To meet  condition $\mathsf{C3}$, for each $l\in\Nat{n}$,
$\Hwt{\vect{t}_l}=\beta-\Hwt{\vect{w}_l}$, where $\vect{t}_l$ and $\vect{w}_l$ are columns of $\hat{\mat{E}}$ and
$\bar{\mat{E}}$, respectively. It follows that meeting all three conditions $\mathsf{C1}$, $\mathsf{C2}$, and $\mathsf{C3}$ is equivalent to finding a $(\beta+d)\times n$ $\beta$-column regular matrix
$\mat{E}$ in which each row is a correctable erasure pattern. Hence, we conclude that the requirements for $\mat{E}$ are
equivalent to finding a PIR achievable rate matrix
\begin{IEEEeqnarray}{rCl}
  \mat{\Lambda}_{d,\beta+d}(\code{C})=\vect{1}_{(\beta+d)\times n}-\mat{E}_{(\beta+d)\times n}, \label{eq:equivalent_R-E}
\end{IEEEeqnarray}
where $\beta$ and $d$ are chosen according to $\beta k = \Gamma d$.

In the following lemma, we prove that our construction of the queries ensures that the privacy condition
\eqref{eq:cond1} is satisfied.
\begin{lemma}
  \label{Lem: Privacy}
  Consider a DSS that uses an $[n,k]$ code with subpacketization $\alpha$ to store $f$ files, each divided into $\beta$
  stripes. Then, the queries $\bm Q^{(l)}$, $l\in\Nat{n}$, designed as in \eqref{Eq: GenQuery_design} satisfy
  $\mathsf H\bigl(m|\bm Q^{(l)}\bigr)=\mathsf H(m)$, where $l\in\Nat{n}$ represents the spy node.
\end{lemma}
\begin{IEEEproof}
  The queries $\bm Q^{(l)}$, $l\in\Nat{n}$, are a sum of a random matrix $\bm U$ and a deterministic matrix
  $\bm V^{(l)}$. The resulting queries have elements that are independently and uniformly distributed at random from
  $\GF(q)$. 
  Hence, any $\bm Q^{(l)}$ obtained by
  the spy node is statistically independent of $m$. This ensures that $\mathsf H\bigl(m|\bm Q^{(l)}\bigr)=\mathsf H(m)$.
\end{IEEEproof}

The following theorem shows that Protocol~2 achieves perfect information-theoretic PIR, and it gives its achievable PIR
rate, $\const{R}(\mathcal C)$. Note that to prove perfect information-theoretic PIR it remains to be shown that from the
responses $\bm r_l$ in \eqref{eq:response} sent by the nodes back to the user, one can recover the requested file, i.e.,
that the constructed PIR protocol satisfies the recovery condition in \eqref{eq:cond2}.

{\mg
\begin{theorem}
  \label{th:GenPIR}
  Consider a DSS that uses an $[n,k]$ code with subpacketization $\alpha$ to store $f$ files, each divided into $\beta$
  stripes. If there exists a $\Gamma$-row regular matrix $\hat{\bm E}$ satisfying conditions $\mathsf{C1}$, $\mathsf{C2}$, and
  $\mathsf{C3}$, then $\mathsf H\bigl(\bm X^{(m)}|\bm r_1,\ldots, \bm r_n\bigr)=0$ and the PIR rate
  \begin{align*}
    \const{R}(\mathcal C)=\frac{\Gamma}{n}\le \frac{n-k}{n}
  \end{align*}
  is achievable.
\end{theorem}
}
\begin{IEEEproof}
  See Appendix~\ref{Appendix: Proof1}.
\end{IEEEproof}

Theorem~\ref{th:GenPIR} generalizes \cite[Th.~1]{TajeddineGnilkeElRouayheb18_1} to any linear code.

{\mg
\begin{corollary}
  If for an $[n,k]$ code $\code{C}$ there exists an $(n-k)$-regular matrix $\bm E$ satisfying conditions $\mathsf{C1}$,
  $\mathsf{C2}$, and $\mathsf{C3}$, then Protocol~2 achieves the asymptotic MDS-PIR capacity $\const{C}_\infty$ in
  \eqref{eq:PIRasympt-capacity}.
\end{corollary}}

{\mg
\begin{remark}
  \label{remark:1}
  From \eqref{eq:equivalent_R-E}, if there exists an $(n-k)$-regular matrix $\bm E$ satisfying conditions $\mathsf{C1}$,
  $\mathsf{C2}$, and $\mathsf{C3}$, a $\mat{\Lambda}_{\kappa,\nu}$ MDS-PIR capacity-achieving matrix with
  $\frac{\kappa}{\nu}=\frac{k}{n}$ exists. Thus, if a code achieves the asymptotic MDS-PIR capacity $\const{C}_\infty$
  with Protocol~2, it also achieves the finite MDS-PIR capacity $\const{C}_f$ with Protocol~1.
\end{remark}}

Note that the parameters $\Gamma$, $\beta$ mentioned in Theorem~\ref{th:GenPIR}, and $d$ (which is not explicitly mentioned) have to be carefully
selected such that $\beta k = \Gamma d$ and such that a $\Gamma$-row regular matrix $\hat{\bm E}$ (satisfying condition $\mathsf{C}3$) actually exists with
a valid collection of information sets $\{\set{I}_i\}_{i\in\Nat{\beta}}$. In the following corollary, we provide a valid
set of values.

\begin{corollary}
  \label{cor:PIR_d}
  Let $\mathcal C$ be an $[n,k,d^{\code{C}}_\mathsf{min}]$ code. For
  $\Gamma=\min(k,d^{\code{C}}_{\mathsf{min}}-1)$, it holds that
  \begin{align}
    \label{eq:Hth_d}
    \mathsf H\bigl(\bm X^{(m)}|\bm r_1,\ldots, \bm r_n\bigr)=0,
  \end{align}
  and the PIR rate {\mg $\const{R}(\code{C})=\frac{\min\left(k,d^{\code{C}}_{\mathsf{min}}-1\right)}{n}$} is achievable.
\end{corollary}
\begin{IEEEproof}
{Let  $d=k$ and $\beta = \Gamma$.}
  Then, \eqref{eq:Hth_d} follows directly from Theorem~\ref{th:GenPIR}, since we have shown in Lemma~\ref{lem:lower-nu_k} that the
  required matrix $\mat{\Lambda}_{k,\Gamma+k}(\code{C})$ exists for $\code{C}$, and the existence of
  $\mat{E}_{(\Gamma+k)\times n}$ follows from \eqref{eq:equivalent_R-E}. 
\end{IEEEproof}
{\mg  
  The above corollary provides a lower bound on the value of $\Gamma$ for any code. In other words,
  it allows us to design a PIR protocol with PIR rate greater than or equal to
  $\min(k,d^{\code{C}}_{\mathsf{min}}-1)/n$. We remark that with a better designed $\hat{\bm E}$, it
  may be possible to achieve a higher PIR rate.} For systematic codes with rate $R^{\mathcal C}>1/2$, a better lower
bound on the maximum achievable PIR rate compared to that of Corollary~\ref{cor:PIR_d} is given below.
\begin{corollary}
  \label{cor:PIR}
  Let $\mathcal C$ be an $[n,k]$ systematic code with $R^{\mathcal C}>1/2$ and
  $\bm H^{\mathcal C}=(\bm P\mid\bm I_{n-k})$. Consider the $[n=k,k']$ code $\code{C}'$  with parity-check matrix 
  $\bm H^{\code{C}'}=\bm P$. For $\Gamma=d_{\mathsf{min}}^{\code{C}'}-1$, it holds that
  \begin{align}
    \label{eq:Hth}
    \mathsf H\bigl(\bm X^{(m)}|\bm r_1,\ldots, \bm r_n\bigr)=0,
  \end{align}
  and the PIR rate {\mg $\const{R}(\code{C})=\frac{d_\mathsf{min}^{\code{C}'}-1}{n}$} is achievable.
\end{corollary}
\begin{IEEEproof}
  As for the proof of Corollary~\ref{cor:PIR_d}, let $d=k$ and $\beta=\Gamma$. Then, \eqref{eq:Hth} follows directly from
  Theorem~\ref{th:GenPIR}. Select $k$ erasure patterns $\hat{\bm e}_i'$, $i\in\Nat{k}$, of length $k$, such that 
  $\Hwt{\hat{\vect{e}}_i'}=d_{\mathsf{min}}^{\code{C}'}-1$ and $\hat{\bm e}_{i+1}'$ is a right cyclic shift of $\hat{\bm e}_{i}'$, $i\in\Nat{k-1}$. The patterns are all correctable by the code $\mathcal C'$. Thus,
  the erasure patterns
  \begin{align*}
    \hat{\bm e}_i=(\hat{\bm e}_i',\underbrace{0,\cdots,0}_{n-k})
  \end{align*}
  are also correctable by $\mathcal C$. Choosing the information sets $\mathcal I_i=\Nat{k}$, $i\in\Nat{\Gamma}$, the
  required $\Gamma$-row regular matrix $\hat{\mat{E}}$ satisfying conditions $\mathsf{C1}$, $\mathsf{C2}$, and
  $\mathsf{C3}$ can then be constructed from
  $\{\hat{\vect{e}}_i\}_{i\in\Nat{k}}$ and $\{\set{I}_i\}_{i\in\Nat{\Gamma}}$.
\end{IEEEproof}

Observe that $R^\code{C}>\frac{1}{2}$ implies $k> d^\code{C}_\mathsf{min}-1$. In
\cite{FreijHollantiGnilkeHollantiKarpuk17_1}, under the assumption that $k>d^\code{C}_\mathsf{min}-1$, a PIR protocol
achieving a PIR rate of $(d^{\code{C}}_\mathsf{min}-1)/n$ was given. Note that
$d^{\code{C}}_\mathsf{min}\leq d^{\code{C}'}_\mathsf{min}$, and thus
$\const{R}(\mathcal C)\geq (d^{\code{C}}_\mathsf{min}-1)/n$ for our construction.

Below we give two examples to elucidate Protocol~2. Example~\ref{ex:5_3_code} illustrates the PIR protocol when the underlying
code has rate $R^{\mathcal C}>1/2$, with parameters $d=k$ and $\beta=\Gamma$. On the other hand, Example~\ref{ex: 7_3_code}
uses an underlying code that has rate $R^{\mathcal C} < 1/2$, again with parameters $d=k$ and $\beta=\Gamma$.

\begin{example}
  \label{ex:5_3_code}
  Consider a DSS that uses the $[5,3,2]$ scalar ($\alpha=1$) binary code $\code{C}$ in Section~\ref{sec:example_n5k3} (with
  generator matrix given in \eqref{eq:goodG_n5k3}) to store a single file by dividing it into $\beta$ stripes. Its
  parity-check matrix is given by
  \begin{align*}
    \bm H^{\mathcal C}=(\bm P|\bm I_{n-k})=\left(
    \begin{array}{ccc|cc}
        1 & 1 & 0 & 1 &0\\
        0 & 1 & 1 & 0 &1
      \end{array}\right).
  \end{align*}
  To determine the value of {the} parameter $\beta$, we compute the minimum Hamming distance
  $d^{\code{C}'}_\mathsf{min}$ of the $[n'=3,k'=1]$ code $\code{C}'$ with parity-check matrix $\bm H^{\code{C}'}=\bm
  P$. From $\bm H^{\code{C}'}$ it follows that $d_\mathsf{min}^{\code{C}'}=3$. Hence, from Corollary~\ref{cor:PIR},
  $\beta=2$. Let the file to be stored be denoted by the $2\times 3$ matrix $\bm X=(x_{i,j})$, where the message symbols
  $x_{i,j}\in\GF(2^\ell)$ for $\ell\in\Nat{}$. Then,
  \begin{align*}
    \bm C=\left(\begin{matrix}
        x_{1,1} & x_{1,2} & x_{1,3} & x_{1,1}+x_{1,2} & x_{1,2}+x_{1,3}\\
        x_{2,1} & x_{2,2} & x_{2,3} & x_{2,1}+x_{2,2} & x_{2,2}+x_{2,3}
      \end{matrix}\right).
  \end{align*}
  The user wants to download the file $\bm X$ from the DSS and sends a query $\bm Q^{(l)}$, $l\in\Nat{5}$, to the $l$-th
  storage node. The queries take the form shown in \eqref{Eq: GenQuery_design}. For $l\in\Nat{5}$, we construct the
  matrix $\bm V^{(l)}=\BDelta_l$ by choosing an appropriate matrix $\hat{\bm E}$. To do this, we carefully choose the
  information sets $\set{I}_1=\{1,2,3\}$ and $\set{I}_2=\{1,2,3\}$ (and hence 
    $\mat{V}^{(4)}=\mat{V}^{(5)}=\mat{0}_{d\times\beta}$). This allows us to generate a column weight
  profile in $\hat{\bm E}$. More specifically, let $\vect{t}_l$ be the $l$-th column of $\hat{\mat{E}}$,
  $l\in\Nat{5}$. We have $\Hwt{\vect{t}_1}=\Hwt{\vect{t}_2}=\Hwt{\vect{t}_3}=2$ and
  $\Hwt{\vect{t}_4}=\Hwt{\vect{t}_5}=0$. A valid matrix
  $\hat{\bm E}$ is
  \begin{align*}
    \hat{\bm E}
    =\left(\begin{matrix}
        1 & 0 & 1 & 0 & 0\\
        1 & 1 & 0 & 0 & 0\\
        0 & 1 & 1 & 0 & 0
      \end{matrix}\right)
  \end{align*}
  and we construct $\BDelta_1$ according to \eqref{Eq: delta_design}. Focusing on the first column of
  $\hat{\bm E}$, we can see that the first two rows have a one in the first position. Thus, we choose
  $j_1^{(1)}=s_1^{(1)}$, $j_2^{(1)}=s_2^{(1)}$, and $j_3^{(1)}=0$, since $\hat e_{1,1}=1$, $\hat e_{2,1}=1$, and $\hat e_{3,1}=0$. We
  take $s_1^{(1)},s_2^{(1)}\in\Nat{2}$. We arbitrarily choose $s_1^{(1)}=1$ and $s_2^{(1)}=2$ to get
  \begin{align*}
    \BDelta_1 =\left(\begin{matrix}
        \bm \omega_{1}\\
        \bm \omega_{2}\\
        \bm \omega_{0}
      \end{matrix}\right)=\left(\begin{matrix}
        1 & 0\\
        0 & 1\\
        0 & 0
      \end{matrix}\right).
  \end{align*}
  Similarly, we construct
  \begin{align*}
    &\BDelta_2=\left(\begin{matrix}
        0&0\\
        1&0\\
        0&1
      \end{matrix}\right) \text{ and }\;
           \BDelta_3=\left(\begin{matrix}
               0&1\\
               0&0\\
               1&0
             \end{matrix}\right).
  \end{align*}
  The queries $\bm Q^{(l)}$ are sent to the respective nodes and the responses
  \begin{align*}
    \begin{split}
      \bm r_1
      &=\left(\begin{smallmatrix}
          u_{1,1}x_{1,1}+u_{1,2}x_{2,1}+x_{1,1}\\
          u_{2,1}x_{1,1}+u_{2,2}x_{2,1}+x_{2,1}\\
          u_{3,1}x_{1,1}+u_{3,2}x_{2,1}
        \end{smallmatrix}\right)=\left(\begin{smallmatrix}
          I_1+x_{1,1}\\
          I_4+x_{2,1}\\
          I_7
        \end{smallmatrix}\right),
    \end{split}
  \end{align*}
  \vskip -12pt
  \begin{align*}
    \begin{split}
      \bm r_2
      &=\left(\begin{smallmatrix}
          u_{1,1}x_{1,2}+u_{1,2}x_{2,2}\\
          u_{2,1}x_{1,2}+u_{2,2}x_{2,2}+x_{1,2}\\
          u_{3,1}x_{1,2}+u_{3,2}x_{2,2}+x_{2,2}
        \end{smallmatrix}\right)=\left(\begin{smallmatrix}
          I_2\\
          I_5+x_{1,2}\\
          I_8+x_{2,2}
        \end{smallmatrix}\right),
    \end{split}
  \end{align*}
  \vskip -12pt
  \begin{align*}
    \begin{split}
      \bm r_3
      &=\left(\begin{smallmatrix}
          u_{1,1}x_{1,3}+u_{1,2}x_{2,3}+x_{2,3}\\
          u_{2,1}x_{1,3}+u_{2,2}x_{2,3}\\
          u_{3,1}x_{1,3}+u_{3,2}x_{2,3}+x_{1,3}
        \end{smallmatrix}\right)=\left(\begin{smallmatrix}
          I_3+x_{2,3}\\
          I_6\\
          I_9+x_{1,3}
        \end{smallmatrix}\right),
    \end{split}
  \end{align*}
  \vskip -12pt
  \begin{align*}
    \bm r_4=\left(\begin{smallmatrix}
        u_{1,1}&u_{1,2}\\
        u_{2,1}&u_{2,2}\\
        u_{3,1}&u_{3,2}
      \end{smallmatrix}\right)\left(\begin{smallmatrix}
        x_{1,1}+x_{1,2}\\
        x_{2,1}+x_{2,2}
      \end{smallmatrix}\right)=\left(\begin{smallmatrix}
        I_1+I_2\\
        I_4+I_5\\
        I_7+I_8
      \end{smallmatrix}\right),
  \end{align*}
  \vskip -12pt
  \begin{align*}
    \bm r_5=\left(\begin{smallmatrix}
        u_{1,1}&u_{1,2}\\
        u_{2,1}&u_{2,2}\\
        u_{3,1}&u_{3,2}
      \end{smallmatrix}\right)\left(\begin{smallmatrix}
        x_{1,2}+x_{1,3}\\
        x_{2,2}+x_{2,3}
      \end{smallmatrix}\right)=\left(\begin{smallmatrix}
        I_2+I_3\\
        I_5+I_6\\
        I_8+I_9
      \end{smallmatrix}\right),
  \end{align*}
  where $I_{i}=\sum_{j=1}^2 u_{h,j}x_{j,h'}$ and $i=3(h-1)+h'$, with $h,h'\in\Nat{3}$, are collected by the user. Notice
  that each storage node sends back $d=k=3$ symbols. The user obtains the requested file as follows. Knowing $I_2$, the
  user obtains $I_1$ and $I_3$ from the first components of $\bm r_4$ and $\bm r_5$. This allows the user to obtain
  $x_{1,1}$ and $x_{2,3}$. In a similar fashion, knowing $I_6$ the user gets $I_5$ from the second component of
  $\bm r_5$, then uses this to obtain $I_4$ from the second component of $\bm r_4$. This allows the user to obtain
  $x_{2,1}$ and $x_{1,2}$. Similarly, knowing $I_7$ allows the user to get $I_8$ from the third component of $\bm
  r_4$. Knowing $I_8$ allows the user to obtain $I_9$ from the third component of $\bm r_5$, which then allows to
  recover the symbols $x_{2,2}$ and $x_{1,3}$. In this way, the user recovers all symbols of the file and hence recovers
  $\bm X$. Note that $\const{R}(\code{C})=\frac{2\cdot3}{5\cdot3}=\frac{2}{5}$, which is equal to the asymptotic MDS-PIR
  capacity $\const{C}_\infty$ in \eqref{eq:PIRasympt-capacity}.
\end{example}

\begin{example}
  \label{ex: 7_3_code}
  Consider a DSS consisting of $n=7$ storage nodes that {store} a single file $\bm X$. The DSS uses a $[7,3,4]$ scalar
  binary code $\mathcal C$. The parity-check matrix of the code is
  \begin{align*}
    \setcounter{MaxMatrixCols}{20}
    \bm H^{\mathcal C}=\left(
    \begin{matrix}
      0 & 1 & 1 & 1 & 0 & 0 & 0
      \\
      1 & 0 & 1 & 0 & 1 & 0 & 0
      \\
      1 & 1 & 0 & 0 & 0 & 1 & 0
      \\
      1 & 1 & 1 & 0 & 0 & 0 & 1
    \end{matrix}\right).
  \end{align*}
  We take $\beta=\Gamma=n-k=4$. File $\bm X$ is of size $\beta\times k$ and hence consists of $\beta k$ symbols in
  $\GF(2^\ell)$. Accordingly, the code array is
  \begin{IEEEeqnarray*}{rCl}
    \Scale[0.7]{\mat{C} =
      \begin{pmatrix}
        x_{1,1} & x_{1,2} & x_{1,3} & x_{1,2}+x_{1,3} & x_{1,1}+x_{1,3} & x_{1,1}+x_{1,2} & x_{1,1}+x_{1,2}+x_{1,3}\\
        x_{2,1} & x_{2,2} & x_{2,3} & x_{2,2}+x_{2,3} & x_{2,1}+x_{2,3} & x_{2,1}+x_{2,2} & x_{2,1}+x_{2,2}+x_{2,3}\\
        x_{3,1} & x_{3,2} & x_{3,3} & x_{3,2}+x_{3,3} & x_{3,1}+x_{3,3} & x_{3,1}+x_{3,2} & x_{3,1}+x_{3,2}+x_{3,3}\\
        x_{4,1} & x_{4,2} & x_{4,3} & x_{4,2}+x_{4,3} & x_{4,1}+x_{4,3} & x_{4,1}+x_{4,2} & x_{4,1}+x_{4,2}+x_{4,3}          
    \end{pmatrix}.}
  \end{IEEEeqnarray*}
  The queries sent to each node, each consisting of $d=k=3$ subqueries, take the form in \eqref{Eq: GenQuery_design}. The
  aim of each subquery is to recover $\Gamma$ code symbols using the PIR protocol. In order to do so, we construct the
  information sets $\{\set{I}_i\}_{i\in\Nat{4}}$. With careful consideration, we choose $\set{I}_1=\{3,4,6\}$,
  $\set{I}_2=\{2,6,7\}$, $\set{I}_3=\{1,3,4\}$, and $\set{I}_4=\{1,5,6\}$. The column weight profile of
  $\hat{\bm E}$ is $(2,1,2,2,1,3,1)$. A valid matrix
  $\hat{\bm E}$ is
  \begin{align*}
    \hat{\bm E}=\left(\begin{matrix}
        0 & 0 & 1 & 1 & 1 & 1 & 0\\
        1 & 1 & 0 & 1 & 0 & 1 & 0\\
        1 & 0 & 1 & 0 & 0 & 1 & 1
      \end{matrix}\right).
  \end{align*}
  Note that each erasure pattern in $\hat{\bm E}$ (each row) is correctable by the code $\mathcal C$. As in
  Example~\ref{ex:5_3_code}, we map the columns of $\hat{\bm E}$ and $\{\set{I}_i\}_{i\in\Nat{4}}$ to the matrix
  $\bm V^{(l)}$ and obtain
  \begin{align*}
    \begin{split}
      &\bm \Delta_1=\left(\begin{matrix}
          0&0&0&0\\
          0&0&1&0\\
          0&0&0&1
        \end{matrix}\right),\,
      \bm \Delta_2=\left(\begin{matrix}
          0&0&0&0\\
          0&1&0&0\\
          0&0&0&0
        \end{matrix}\right),\\
      &\bm \Delta_3=\left(\begin{matrix}
          1&0&0&0\\
          0&0&0&0\\
          0&0&1&0
        \end{matrix}\right),\,
      \bm \Delta_4=\left(\begin{matrix}
          1&0&0&0\\
          0&0&1&0\\
          0&0&0&0
        \end{matrix}\right),\\
      &\bm \Delta_5=\left(\begin{matrix}
          0&0&0&1\\
          0&0&0&0\\
          0&0&0&0
        \end{matrix}\right),\,
      \bm \Delta_6=\left(\begin{matrix}
          1&0&0&0\\
          0&1&0&0\\
          0&0&0&1
        \end{matrix}\right),\\
      &\bm \Delta_7=\left(\begin{matrix}
          0&0&0&0\\
          0&0&0&0\\
          0&1&0&0
        \end{matrix}\right).
    \end{split}
  \end{align*}
  As an example, we next show the reconstruction of symbols from the first pair of subqueries and subresponses. We have
\begin{align*}
  \begin{split}
    &r_{1,1}=I_1,\, r_{2,1}=I_2,\, r_{3,1}=I_3+x_{1,3},
    \\
    &r_{4,1}=I_2+I_3+x_{1,2}+x_{1,3},\,
    r_{5,1}=I_1+I_3+x_{4,1}+x_{4,3},
    \\
    &r_{6,1}=I_1+I_2+x_{1,1}+x_{1,2},\,
    r_{7,1}=I_1+I_2+I_3,
  \end{split}
\end{align*}
where $r_{l,1}$ denotes the first subresponse from the $l$-th node, $I_{i}=\sum_{j=1}^4 u_{h,j}x_{j,h'}$ and
$i=3(h-1)+h'$, $h,h'\in\Nat{3}$. Clearly, the subresponses $r_{1,1}$, $r_{2,1}$, and $r_{7,1}$ allow the user to obtain
the three interference symbols $I_1$, $I_2$, and $I_3$. This is solely because the first row of $\hat{\bm E}$
(pertaining to the first subqueries)  is an erasure pattern correctable by $\mathcal C$. Having this knowledge, the user
obtains the symbols $x_{1,3}$, $x_{1,2}+x_{1,3}$, $x_{4,1}+x_{4,3}$, and $x_{1,1}+x_{1,2}$ from the remaining
subresponses. From the obtained code symbols the user can decode $x_{1,3}$, $x_{1,2}$, and $x_{1,1}$, hence obtaining
the message symbols in the first row of $\bm C$. The code symbol $x_{4,1}+x_{4,3}$ is used to decode $x_{4,1}$,
$x_{4,2}$, and $x_{4,3}$ from the code symbols that are further obtained from the third subresponse. In the same way,
the remaining two subresponses allow the recovery of $\beta k =12$ message symbols.

The PIR rate is $\const{R}(\code{C})=\frac{4\cdot3}{7\cdot3}=\frac{4}{7}$, which is equal to the asymptotic MDS-PIR
capacity $\const{C}_\infty$ in \eqref{eq:PIRasympt-capacity}.
\end{example}

\section{MDS-PIR Capacity-Achieving Codes}
\label{sec:MDS-PIRcapacity-achiving-codes}

For given values of $n$ and $k$, whether an $[n,k]$ code is MDS-PIR capacity-achieving or not is of great interest. In
this section, we provide a necessary condition for an arbitrary linear code to achieve the MDS-PIR capacities
$\const{C}_f$ and $\const{C}_\infty$ with Protocols~1 and 2, respectively. Furthermore, we prove that certain important families of codes, namely cyclic
codes, RM codes, and a class of distance-optimal LRCs are MDS-PIR capacity-achieving. For
Protocol~2, the MDS-PIR capacity-achieving proofs for these classes of codes assume $\beta=n-k$ and $d=k$, which are not
necessary the minimum values given in \eqref{Eq: Beta_and_d_DEF}. However, in the numerical results section (see
Tables~\ref{table_of_codes_t1_above12} and~\ref{table_of_codes_t1_below12}) we show examples for which Protocol~2 also
achieves the MDS-PIR capacity for $\beta$ and $d$ in \eqref{Eq: Beta_and_d_DEF}.

As shown in the previous sections, the only requirement for a code $\code{C}$ to achieve the MDS-PIR capacity with Protocol~1 is that there exists an
MDS-PIR capacity-achieving matrix $\mat{\Lambda}_{\kappa,\nu}(\code{C})$ (or a $(\Gamma=n-k)$-regular matrix $\bm E$ of size
$(\beta+d)\times n$). In other words, the code $\mathcal C$ should be able to correct $\beta+d$ erasure patterns of
$n-k$ erasures that satisfy the regularity condition of $\bm E$.

Let us first consider a fact for any information set of an $[n,k]$ code.
\begin{proposition}[\hspace{-1sp}{\cite[Th.~1.6.2]{HuffmanPless10_1}}]
  \label{prop:infoS_dual-code}
  If $\set{I}$ is an information set of an $[n,k]$ code $\code{C}$, then $\Nat{n}\setminus\set{I}$ is an information set
  of its $[n,n-k]$ dual code $\code{C}^\perp$.
\end{proposition}

Based on Proposition~\ref{prop:infoS_dual-code}, the subsequent result follows.
\begin{corollary}  
  \label{cor:dual_MDS-PIRcapacity-achieving-code}
  The dual of an $[n,k]$ MDS-PIR capacity-achieving code is an $[n,n-k]$ MDS-PIR capacity-achieving code.
\end{corollary}

To check if a linear code achieves the MDS-PIR capacity with Protocol~1, sometimes it might be easier to verify the MDS-PIR
capacity-achieving condition for its dual code.

Next, we derive a useful result that gives the relation between an information set and a subcode of dimension $s$.
\begin{lemma}
  \label{lem:infoS-suppD}
  Given an $[n,k]$ code $\code{C}$, for any information set $\set{I}$ and an $s$-dimensional subcode
  $\code{D}\subseteq\code{C}$, we have
  \begin{IEEEeqnarray*}{rCl}
    \bigcard{\set{I}\cap\chi(\code{D})}\geq s.
  \end{IEEEeqnarray*}
\end{lemma}
\begin{IEEEproof}
  See Appendix~\ref{sec:proof_infoS-suppD}.
\end{IEEEproof}
Now, we are able to provide a necessary condition for a code to achieve the MDS-PIR capacity with Protocol~1.
\begin{theorem}
  \label{thm:general-d_MDS-PIRcapacity-achieving-codes}
  If an MDS-PIR capacity-achieving matrix exists for an $[n,k]$ code $\code{C}$, then
  \begin{IEEEeqnarray}{rCl}
    d_s^{\code{C}}\geq\frac{n}{k}s,\quad\forall\,s\in\Nat{k}.
    \label{eq:general-d_MDS-PIRcapacity-achieving-codes}
  \end{IEEEeqnarray}
\end{theorem}
\begin{IEEEproof}
  By definition there exists a PIR achievable rate matrix $\mat{\Lambda}_{\kappa,\nu}(\code{C})$ with
  $\frac{\kappa}{\nu}=\frac{k}{n}$. This means that there exist information sets $\set{I}_i$, $i\in\Nat{\nu}$, such that
  in $\{\set{I}_i\}_{i\in\Nat{\nu}}$ each coordinate $j$ of $\code{C}$, $j\in\Nat{n}$, appears exactly $\kappa$
  times. Let $\code{D}$ be any subcode of dimension $s$ of the $[n,k]$ code $\code{C}$. This implies that
  \begin{IEEEeqnarray*}{rClCl}
    \kappa\card{\chi(\code{D})}& = &\sum_{i=1}^\nu\bigcard{\set{I}_i\cap\chi(\code{D})} & \overset{(a)}{\geq} & \nu s,
  \end{IEEEeqnarray*}
  where $(a)$ follows from Lemma~\ref{lem:infoS-suppD}. Based on the definition of $d_s^{\code{C}}$, $s\in\Nat{k}$, there
  exists a rank-$s$ subcode $\code{D}^\ast$ that achieves $d_s^{\code{C}}$. We then have
  \begin{IEEEeqnarray*}{rCl}
    d_s^{\code{C}}\geq\frac{\nu}{\kappa}s=\frac{n}{k}s,\quad\,\forall\,s\in\Nat{k}.
  \end{IEEEeqnarray*}
\end{IEEEproof}

Based on the necessary condition, it can be shown that the code $\code{C}$ in Example~\ref{ex:n5k3_bad} is not MDS-PIR
capacity-achieving with Protocol~1, since $d_2^\code{C}=3 < \frac{5}{3}\cdot 2$, i.e., it is impossible to find an MDS-PIR capacity-achieving
matrix $\mat{\Lambda}_{\kappa,\nu}$ for this code.

We would like to emphasize that it seems that the necessary condition for MDS-PIR capacity-achieving matrices in
Theorem~\ref{thm:general-d_MDS-PIRcapacity-achieving-codes} is also a sufficient condition. We have performed an exhaustive
search for codes with parameters $k\in\Nat{n}$ and $n\in\Nat{11}$ (except for $[n,k]=[10,5]$ and
$[n,k]=[11,4\leq k\leq 7]$) and seen that for codes that satisfy the necessary condition, there always exists an MDS-PIR
capacity-achieving matrix. Therefore, we conjecture that \eqref{eq:general-d_MDS-PIRcapacity-achieving-codes} in Theorem
\ref{thm:general-d_MDS-PIRcapacity-achieving-codes} is an if and only if condition for the existence of an MDS-PIR
capacity-achieving matrix.
\begin{conjecture}
  \label{conj:general-d_MDS-PIRcapacity-achiving-codes}
  An MDS-PIR capacity-achieving matrix 
  exists
  for an $[n,k]$ code $\code{C}$ if and only if
  \begin{IEEEeqnarray*}{rCl}
    d_s^{\code{C}}\geq\frac{n}{k}s,\quad\,\forall\,s\in\Nat{k}.
  \end{IEEEeqnarray*}
\end{conjecture}
In the following, we provide a sufficient condition for an $[n,k]$ code $\code{C}$ to achieve the MDS-PIR capacity with Protocol~1 by using code automorphisms
 \cite[Ch.~8]{MacWilliamsSloane77_1}.
{\mg
\begin{theorem}
  \label{thm:sufficient_MDS-PIRcapacity-achieving-codes}
  Given an $[n,k]$ code $\code{C}$, if there exist $n$ distinct automorphisms $\pi_1,\ldots,\pi_n$ of $\code{C}$ such
  that for every code coordinate $j\in\Nat{n}$, $\{\pi_1(j),\ldots,\pi_n(j)\}=\Nat{n}$, then the code $\code{C}$ is an
  MDS-PIR capacity-achieving code.
\end{theorem}
\begin{IEEEproof}
  Since any $[n,k]$ code $\code{C}$ contains at least one information set $\set{I}$, the automorphisms
  $\{\pi_i\}_{i\in\Nat{n}}$ guarantee that
  \begin{equation*}
    \set{I}_i\eqdef\{\pi_i(j)\colon j\in\set{I}\},\quad i\in\Nat{n},
  \end{equation*}    
  are all information sets of $\code{C}$. By assumption, for a given $j\in\set{I}$, we have
  $\{\pi_1(j),\ldots,\pi_n(j)\}=\Nat{n}$. Since there are in total $k$ coordinates in $\set{I}$,  every coordinate
  appears exactly $k$ times in $\{\set{I}_i\}_{i\in\Nat{n}}$, and hence an MDS-PIR capacity-achieving matrix
  $\mat{\Lambda}_{k,n}(\code{C})$ satisfying Definition~\ref{def:MDS-PIRcapacity-achieving-codes} exists.
\end{IEEEproof}
}

Using their known code automorphisms and Theorem~\ref{thm:sufficient_MDS-PIRcapacity-achieving-codes}, it is easy to
prove that the families of cyclic codes and RM codes achieve the MDS-PIR capacity.

\subsection{Cyclic Codes}
\label{sec:cyclic-codes_MDS-PIRcapacity-achieving}

\begin{corollary}
  \label{cor:MDS-PIRcapacity-achiving_cyclic-codes}
  Cyclic codes are MDS-PIR capacity-achieving codes.
\end{corollary}
\begin{IEEEproof}
Let $\pi_i$ denote the automorphism of an $[n,k]$ cyclic code $\code{C}$ that cyclically shifts each coordinate to the right by $i$ positions. Clearly, for every code coordinate $j\in\Nat{n}$, $\{\pi_1(j),\ldots,\pi_n(j)\}=\Nat{n}$, and the result follows from Theorem~\ref{thm:sufficient_MDS-PIRcapacity-achieving-codes}.
\end{IEEEproof}

\subsection{Reed-Muller Codes}
\label{sec:reed-muller_PIRcapacit-achieving}

\begin{corollary}
  \label{cor:MDS-PIRcapacity-achieving_RMcodes}
  RM codes are MDS-PIR capacity-achieving codes.
\end{corollary}

\begin{IEEEproof}
  Consider an arbitrary RM code $\code{R}(v,m)$ with $v\in\{0\}\cup\Nat{m}$ for some $m\in\Nat{}$. 
  Consider the $n$ distinct automorphisms $g_i(\vect{\mu})\eqdef\vect{\mu}+\vect{\sigma}_i$, where
  $\vect{\sigma}_i$ is the $i$-th $m$-tuple in $\GF(2)^{m\times 1}$, $i\in\Nat{n}$, $n=2^m$
  (see Section~\ref{sec:reed-muller-codes}).
  For any $\vect{\mu}\in\GF(2)^{m\times 1}$,
  \begin{IEEEeqnarray*}{rCl}
    \{g_1(\vect{\mu}),\ldots,g_n(\vect{\mu})\} = \{\vect{\mu}+\vect{\sigma}_1,\ldots,\vect{\mu}+\vect{\sigma}_n\}
  \end{IEEEeqnarray*}
  forms the vector space $\GF(2)^{m\times 1}$,  
  and the result follows from 
  Theorem~\ref{thm:sufficient_MDS-PIRcapacity-achieving-codes}. 
\end{IEEEproof}

We remark here that because of the property of invertible and affine automorphisms for binary RM codes, it is not too
hard to see that Corollary~\ref{cor:MDS-PIRcapacity-achieving_RMcodes} can be extended to nonbinary generalized RM codes
\cite{DelsarteGoethalsMacWilliams70_1}. The detailed discussion is omitted. {\mg Furthermore, note that in the
  independent work \cite{FreijHollantiGnilkeHollantiHorlemannKarpukKubjas19_1app} it was also shown that RM codes can
  achieve the asymptotic MDS-PIR capacity, albeit with a protocol that requires a much larger $\beta$ and $d$.}

{\mg Besides cyclic codes and RM codes,} there exist other families of codes satisfying
Theorem~\ref{thm:sufficient_MDS-PIRcapacity-achieving-codes}, for instance, the class of low-density parity-check (LDPC) codes
constructed from array codes \cite{fan00,YangHelleseth03_1}. We further emphasize that the proof of
Theorem~\ref{thm:sufficient_MDS-PIRcapacity-achieving-codes} indicates that the automorphisms of an $[n,k]$ code are very
important to design an MDS-PIR capacity-achieving matrix.

\subsection{Local Reconstruction Codes}
\label{sec: Ematrix_LRC}

In this subsection, we prove that a certain family of LRCs achieves the MDS-PIR capacity by directly showing the
existence of an $(n-k)$-regular $n\times n$ matrix $\mat{E}$. 

Consider an $[n,k]$ distance-optimal $(r,\delta)$ information locality code (see Definition~\ref{def: OptLRCs}) for which
the $(n'-k) \times n'$ matrix
  \begin{align}
  \label{Eq: H_MDS}
  \left(\begin{array}{cccc|c}
          \bm P_1 & \bm P_2 & \cdots & \bm P_{L_{\mathsf c}} & \multirow{2}*{$\bm I_{n'-k}$} \\
          \bm M_1 & \bm M_2 & \cdots & \bm M_{L_{\mathsf c}} & \\
        \end{array}\right) \triangleq \bm H^{\mathsf{MDS}}
\end{align}
is the parity-check matrix of an $[n',k]$ MDS code over $\GF(q)$, where
$n'=n-(L_\mathsf{c}-1)(\delta-1)$.\footnote{Examples of codes that satisfy \eqref{Eq: H_MDS} are Pyramid codes, the LRCs
  in \cite{Hua12}, and codes from the parity-splitting construction of \cite{Kam14}.} For such a class of codes, we give
an explicit construction of the matrix $\bm E$ in order to design the PIR protocol.

Recall that $L=\left\lfloor\frac{n}{n_\mathsf{c}}\right\rfloor$, $n_\mathsf{c}=r+\delta-1$, and let
$\bar r\eqdef n\bmod n_\mathsf{c}$. We consider
\begin{IEEEeqnarray*}{rCl}
  \mat{E}=
  \begin{pmatrix}
    \mat{E}_{1,1}& \mat{E}_{1,2}& \ldots& \mat{E}_{1,L+1}
    \\
    \vdots& \vdots& \vdots& \vdots
    \\
    \mat{E}_{L+1,1}& \mat{E}_{L+1,2}& \ldots& \mat{E}_{L+1,L+1}
  \end{pmatrix}
\end{IEEEeqnarray*}
having $(L+1)^2$ submatrices $\mat{E}_{l,h}$, $l,h\in\Nat{L+1}$. For any $l,h\in\Nat{L}$,
the submatrices $\mat{E}_{l,h}$ have dimensions $n_\mathsf{c}\times n_\mathsf{c}$, $\mat{E}_{l,L+1}$ has dimensions 
  $n_\mathsf{c}\times\bar{r}$, $\mat{E}_{L+1,h}$ has dimensions $\bar{r}\times n_\mathsf{c}$, and $\mat{E}_{L+1,L+1}$ has dimensions 
  $\bar{r}\times \bar{r}$. We denote by $\bm e_i^{(l)}$, $l\in\Nat{L+1}$, the $i$-th row of $\bigl(\mat{E}_{l,1}|\ldots|\mat{E}_{l,L+1}\bigr)$. The coordinates of $\bm e_i^{(l)}$ represent the coordinates of the code
$\mathcal C$ defined by its parity-check matrix in \eqref{Eq: H_opt_LRC}. Furthermore, each row vector is subdivided into
$L+1$ subvectors $\bm e^{(l)}_{i,j}$, $j\in\Nat{L+1}$, as
\begin{align*}
  \bm e_i^{(l)}=(e_{i,1}^{(l)},\ldots,e_{i,n}^{(l)})
  =(\bm e_{i,1}^{(l)},\ldots, \bm e_{i,L}^{(l)},\bm e_{i,L+1}^{(l)}).
\end{align*}
The subvectors $\bm e^{(l)}_{i,1},\ldots, \bm e^{(l)}_{i,L}$ are of length $n_\mathsf{c}$, while $\bm e^{(l)}_{i,L+1}$ is of
length $\bar r$. Correspondingly, we can think about $\mat{E}$ as partitioned into $L+1$ column partitions, where the first $L_\mathsf{c}$ partitions correspond to the $L_\mathsf{c}$ local codes and the remaining $L+1-L_\mathsf{c}$ partitions correspond to global parities (see also \eqref{eq:parities_LRC}). We can write $\mat{E}$ as
\begin{align*}
\Scale[0.95]{
  \bm E\eqdef\left(\begin{matrix}
      \bm e_1^{(1)}\\
      \vdots\\
      \bm e_{n_\mathsf{c}}^{(1)}\\
      \vdots\\
      \bm e_{n_\mathsf{c}}^{(L)}\\[1mm]
      \bm e_{1}^{(L+1)}\\
      \vdots\\
      \bm e_{\bar r}^{(L+1)}
    \end{matrix}\right)=\left(\begin{matrix}
      \bm e^{(1)}_{1,1} & \bm e^{(1)}_{1,2} & \cdots & \bm e^{(1)}_{1,L} & \bm e^{(1)}_{1,L+1}\\
      \vdots  & \vdots  & \cdots & \vdots & \vdots\\
      \bm e^{(1)}_{n_\mathsf{c},1} & \bm e^{(1)}_{n_\mathsf{c},2} & \cdots & \bm e^{(1)}_{n_\mathsf{c},L} & \bm e^{(1)}_{n_\mathsf{c},L+1}\\
      \vdots  & \vdots  & \cdots & \vdots & \vdots\\
      \bm e^{(L)}_{n_\mathsf{c},1} & \bm e^{(L)}_{n_\mathsf{c},2} & \cdots & \bm e^{(L)}_{n_\mathsf{c},L} & \bm e^{(L)}_{n_\mathsf{c},L+1}\\[1mm]
      \bm e^{(L+1)}_{1,1} & \bm e^{(L+1)}_{1,2} & \cdots & \bm e^{(L+1)}_{1,L} & \bm e^{(L+1)}_{1,L+1}\\
      \vdots  & \vdots  & \cdots & \vdots & \vdots\\
      \bm e^{(L+1)}_{\bar r,1} & \bm e^{(L+1)}_{\bar r,2} & \cdots & \bm e^{(L+1)}_{\bar r,L} & \bm e^{(L+1)}_{\bar r,L+1}
    \end{matrix}\right)}.
\end{align*}
We refer to the set of rows $\bm e_1^{(l)},\ldots, \bm e_{n_\mathsf{c}}^{(l)}$ as the $l$-th row partition of $\bm E$.

For convenience, we divide $\bm E$ into four submatrices $\tilde{\bm E}$, $\bm W$, $\bm Z$, and $\bm O$ defined as
\begin{IEEEeqnarray}{rCl}
\Scale[0.95]{
  \tilde{\mat{E}}}& \eqdef &
  \Scale[0.95]{\begin{pmatrix}
    \bm e^{(1)}_{1,1} & \bm e^{(1)}_{1,2} & \cdots & \bm e^{(1)}_{1,L}\\
    \bm e^{(1)}_{2,1} & \bm e^{(1)}_{2,2} & \cdots & \bm e^{(1)}_{2,L}\\
    \vdots  & \vdots  & \cdots & \vdots\\
    \bm e^{(L)}_{n_\mathsf{c},1} & \bm e^{(L)}_{n_\mathsf{c},2} & \cdots & \bm e^{(L)}_{n_\mathsf{c},L}
  \end{pmatrix}},
  \Scale[0.95]{\bm Z\eqdef
  \begin{pmatrix}
    \bm e^{(1)}_{1,L+1}\\
    \bm e^{(1)}_{2,L+1}\\
    \vdots\\
    \bm e^{(L)}_{n_\mathsf{c},L+1}
  \end{pmatrix}},
  \nonumber\\[1mm]
  \Scale[0.95]{\bm W}& \Scale[0.95]{\eqdef} &
  \Scale[0.95]{\begin{pmatrix}
    \bm e^{(L+1)}_{1,1} & \bm e^{(L+1)}_{1,2} & \cdots & \bm e^{(L+1)}_{1,L}\\
    \vdots  & \vdots  & \cdots & \vdots\\
    \bm e^{(L+1)}_{\bar r,1} & \bm e^{(L+1)}_{\bar r,2} & \cdots & \bm e^{(L+1)}_{\bar r,L}
  \end{pmatrix},
  \bm O\eqdef
  \begin{pmatrix}
    \bm e^{(L+1)}_{1,L+1}\\
    \vdots\\
    \bm e^{(L+1)}_{\bar r,L+1}
  \end{pmatrix}},\IEEEeqnarraynumspace \label{Eq: SubMats_E}
\end{IEEEeqnarray}
where $\tilde{\bm E}$ is an $n_\mathsf{c}L\times n_\mathsf{c}L$ matrix, having $L^2$ submatrices $\mat{E}_{l,h}$,
$l,h\in\Nat{L}$.

In the following, we give a systematic construction of $\bm E$ such that it is $(n-k)$-regular.
The construction involves two steps.
\begin{enumerate}
\item[a)] \textbf{Initialize matrices $\tilde{\bm E}$, $\bm W$, $\bm Z$, and $\bm O$.} Matrix $\bm Z$ is initialized to
  the all-zero matrix of dimensions $n_\mathsf{c}L\times \bar r$.  Matrices $\mat{W}$ and $\mat{O}$ are initialized by
  setting $e^{(L+1)}_{i,j}=1$, $i\in\Nat{\bar r}$, $j\in\set{P}=\bigcup_{j'=1}^{L+1}\set{P}_{j'}$, where $\set{P}$
  corresponds to the parity coordinates of $\code{C}$ and the sets $\set{P}_{j'}$ are defined in
  Section~\ref{sec:local-reconstr-codes} (see \eqref{eq:parities_LRC}). Let $m=\bigl\lfloor\frac{n-k}{L}\bigr\rfloor$,
  $m_1= m+1$, $\rho_1=\cdots=\rho_t=m_1$, and $\rho_{t+1}=\cdots=\rho_{L}=m$, where $t=(n-k)\bmod L$. Matrix
  $\tilde{\bm E}$ is initialized with the structure
\begin{align}
  \label{Eq: Ehat_structure}
  \tilde{\bm E}=
\left(\begin{matrix}
      \bm \pi_1  & \bm \pi_2 & \cdots & \bm \pi_L\\
      \bm \pi_L  & \bm \pi_1 & \cdots & \bm \pi_{L-1}\\
      \vdots     & \vdots    & \cdots & \vdots\\
      \bm \pi_2  & \bm \pi_3 & \cdots & \bm \pi_1
    \end{matrix}\right),
\end{align}
where each matrix entry $\bm\pi_{l}$, $l\in\Nat{L}$, is a $\rho_l$-regular square matrix of dimensions $n_\mathsf{c} \times n_\mathsf{c}$. Notice
that due to the structure in \eqref{Eq: Ehat_structure}, $\tilde{\bm E}$  has row and column weight equal to $n-k$,
and subsequently each row of $\bm E$ has weight $n-k$. Note also that the columns of $\bm E$ with coordinates in $\mathcal P_j$, $j\in\Nat{L}$, have column weight $n-k+\bar{r}$, while the columns with coordinates in $\mathcal P_{L+1}$ have weight $\bar r$.

\item[b)]\mg \textbf{Swapping elements between $\tilde{\bm E}$ and $\bm Z$.} 
The swapping of elements is performed iteratively with $\bar r$ iterations. For each iteration, in the $i$-th row partition and $j$-th column partition, we consider a set of row coordinates $\mathcal R^{(i)}_j$ of size $|\mathcal{P}_j|$ from which $s_{j}^{(i)}\in\{0,1\}$ ones from columns with coordinates in $\mathcal P_j$, $j \in \Nat{L}$,  are swapped with zeroes in the corresponding rows of $\bm Z$. For convenience, we define $\bm s^{(i)}=(s_{1}^{(i)},\ldots,s_{L}^{(i)})$ and require that  $\sum_{j=1}^{L}s^{(i)}_j = 1$.  Note that  $\mathcal R^{(i)}_j$ and $\bm s^{(i)}$ depend on the iteration number. We describe the procedure for iteration $j' \in \Nat{\bar{r}}$. For the first row partition, select $\bm s^{(1)}$ with $s^{(1)}_j=1$ and $s^{(1)}_z=0$, $\forall\, z\in\Nat{L}\backslash\{j\}$, for some $j\in\Nat{L}$, such that if $j \in \Nat{L_{\mathsf c}}$ there exist $\delta-1$ rows in the first row partition and $j$-th column partition such that their individual weight is strictly larger than $\delta-1$, and otherwise if $j \in \Nat{L_{\mathsf c}+1:L}$, all rows in the first row partition and $j$-th column partition must have weight  larger than or equal to $\max(1,m-(\delta-1))$. This will ensure that the resulting erasure patterns after the swap (as described next) are correctable by $\mathcal{C}$ (see Appendix~\ref{sec:proof_step_b}). Such an $\bm s^{(1)}$ will also always exist   for all $\bar{r}$ iterations as shown in Appendix~\ref{sec:proof_step_b}. Next, for all $i'\in\mathcal R^{(1)}_j$ and $p\in\mathcal P_j$ (where different $p$'s are chosen for different $i'$'s, and index $j$ is such that $s^{(1)}_j=1$) the one at coordinate  $(i',p)$ of $\tilde{\bm E}$ is swapped with a zero at coordinate $(i',j')$ of $\bm Z$  (this corresponds to coordinate $(i',n_{\mathsf c}L+j')$ of $\bm E$). Then, for the remaining row partitions $i=2,\ldots,L$, consider $\bm s^{(i)}$ to be  the $(i-1)$-th right cyclic shift of $\bm s^{(1)}$ and repeat the swapping procedure for the first row partition. 
Due to the specific selection of $\bm s^{(1)}$, the corresponding erasure patterns for all row partitions after the swaps are correctable by $\mathcal{C}$ (see  Appendix~\ref{sec:proof_step_b}).
Note that we have performed  $\sum_{j=1}^{L} \mathcal |\mathcal{P}_j|=n-k-\bar r$ swaps from the columns of $\tilde{\bm E}$ with coordinates in the set  $\cup_{j=1}^{L} \mathcal P_j$ to the $j'$-th column of $\bm Z$. Thus, each column in $\cup_{j=1}^{L} \mathcal P_j$ has column weight $n-k+\bar r-1$ and the $(n_{\mathsf c}L+j')$-th column has column weight $n-k-\bar r+\bar r=n-k$. Letting $j'=j'+1$ and repeating the above procedure $\bar r$ times ensures $\bm E$ to be $(n-k)$-regular.
\end{enumerate}
This completes the construction of $\bm E$, which has row and column weight $n-k$. In the following theorem, we show
that each row of $\bm E$ (considered as an erasure pattern) can be corrected by any code from the class of 
  distance-optimal $(r,\delta)$ information locality codes whose parity-check matrices are as in \eqref{Eq: H_opt_LRC}
and are compliant with \eqref{Eq: H_MDS}. Thus, this class of codes is MDS-PIR capacity-achieving.
\begin{theorem}
  \label{th: LRCcap_proof}
  An $[n,k]$ distance-optimal $(r,\delta)$ information locality code $\code{C}$ with parity-check matrix as in
  \eqref{Eq: H_opt_LRC} and satisfying \eqref{Eq: H_MDS} is an MDS-PIR capacity-achieving code.
\end{theorem}
\begin{IEEEproof}
  See Appendix~\ref{Appendix: Proof2}.
\end{IEEEproof}

In the following, we present an example to illustrate the construction of {the} matrix $\bm E$. The existence of such
a matrix ensures that the PIR protocols presented in Sections~\ref{sec:file-dep-PIR} and \ref{sec:file-indep-PIR} achieve the finite
MDS-PIR capacity $\const{C}_f$ in \eqref{eq:MDS-PIRcapacity} and the asymptotic MDS-PIR capacity $\const{C}_\infty$ in
\eqref{eq:PIRasympt-capacity}, respectively.

\begin{example}
  Consider an $[n=7,k=4]$ Pyramid code $\mathcal C$ that is constructed from an $[n'=6,4]$ Reed-Solomon (RS) code over
  $\GF(2^3)$ with parity-check matrices
  \begin{align*}
    \begin{split}
      \bm H^{\mathcal C}&=\left(\begin{matrix}
          z^3 & 1 & 1 & 0   & 0   & 0 & 0\\
          0   & 0 & 0 & z^3 & z   & 1 & 0\\
          z^4 & 1 & 0 & z^5 & z^5 & 0 & 1
        \end{matrix}\right)
              \end{split}
        \end{align*}
          and
        \begin{align*}
    \begin{split}
              \bm H^{\mathsf{MDS}}&=\left(\begin{matrix}
          z^3 & 1 & z^3 & z & 1 & 0\\
          z^4 & 1 & z^5 & z^5 & 0 & 1
        \end{matrix}\right),
    \end{split}
  \end{align*}
  respectively, where $z$ denotes a primitive element of $\GF(2^3)$. It is easy to see that $\code{C}$ is a
  distance-optimal $(r=2,\delta=2)$ information locality code. We have $n_\mathsf{c}=3$, $L=L_\mathsf{c}=2$, and
  $\bar r\eqdef n\bmod n_\mathsf{c}=1$. Since $\rho_1=2$ and $\rho_2=1$, we get
  \begin{align*}
    \tilde{\bm E}=\left(\begin{matrix}
        \bm \pi_1 & \bm \pi_2\\
        \bm \pi_2 & \bm \pi_1
      \end{matrix}\right)=\left(\begin{array}{ccc|ccc}
                                  1 & 1 & 0 & 1 & 0 & 0\\
                                  0 & 1 & 1 & 0 & 1 & 0\\
                                  1 & 0 & 1 & 0 & 0 & 1\\
                                  \hline
                                  1 & 0 & 0 & 1 & 1 & 0\\
                                  0 & 1 & 0 & 0 & 1 & 1\\
                                  0 & 0 & 1 & 1 & 0 & 1\\
                                \end{array}\right), \;\bm Z=\left(\begin{matrix}
                                  0\\
                                  0\\
                                  0\\
                                  0\\
                                  0\\
                                  0
                                \end{matrix}\right),
  \end{align*}
  {\mg where $\bm \pi_1$ is a  $2$-regular $3 \times 3$ matrix and $\bm \pi_2$ is picked as the identity matrix. 
  The set of parity coordinates is $\mathcal P=\{3,6,7\}$, and we set $e_{1,3}^{(3)}=e_{1,6}^{(3)}=e_{1,7}^{(3)}=1$. As such,
  we get
  \begin{align*}
    \bm W=\left(\begin{matrix}
        0 & 0 & 1 & 0 & 0 & 1
      \end{matrix}\right)\text{ and }
                            \bm O=\left(\begin{matrix}
                                1
                              \end{matrix}\right).
  \end{align*}
  This completes Step a) of the construction above. Note that each row of $\bm E$ has now weight $3$. The second step of the procedure (Step b)) is as follows. Consider the first iteration, $j'=1$. In the first row partition we choose $\bm s^{(1)}=(s^{(1)}_1=1,s^{(1)}_2=0)$. 
  Taking $\mathcal R^{(1)}=\{2\}$, we do the swap between the coordinates $(i'=2,p=3\in\mathcal P_1)$ and $(i',6+j')$. For the second row partition we have $\bm s^{(2)}=(0,1)$ which is a right cyclic shift of $\bm s^{(1)}$. Taking $\mathcal R^{(2)}=\{6\}$, we do the swap between the coordinates $(i'=6,p=6\in\mathcal P_2)$ and $(i',6+j')$. Thus, we have 
\begin{align*}
\begin{split}
	e_{2,3}^{(1)}=0, \;e_{2,7}^{(1)}=1,\\
	e_{3,6}^{(2)}=0, \;e_{3,7}^{(2)}=1.
\end{split}
\end{align*}
Since $\bar r=1$, this completes Step b), which results in}
\begin{align*}
  \bm E=\left(\begin{array}{ccc|ccc|c}
                1 & 1 &   0 & 1 & 0 & 0 &   0\\
                0 & 1 &\r 0 & 0 & 1 & 0 &\r 1\\
                1 & 0 &   1 & 0 & 0 & 1 &   0\\
                \hline
                1 & 0 & 0 & 1 & 1 &   0&   0\\
                0 & 1 & 0 & 0 & 1 &   1&   0\\
                0 & 0 & 1 & 1 & 0 &\r 0&\r 1\\
                \hline
                0 & 0 & 1 & 0 & 0 & 1 & 1
              \end{array}\right).
\end{align*}
The entries in red indicate the swapped values within each row. It can easily be verified that each row of $\bm E$ is an
erasure pattern that is correctable by code $\mathcal C$.
\end{example}

\section{Optimizing the PIR Rate for the Noncolluding Case}
\label{sec:SingleColl_Analysis}


\setlength{\textfloatsep}{14pt}
\begin{algorithm}[t]
\SetKwFunction{CompEraPat}{ComputeErasurePatternList}
\SetKwFunction{CompEraMat}{ComputeMatrix}
\SetKwFunction{CompInfSet}{ComputeInformationSetList}
\SetKwInOut{Input}{Input}
\SetKwInOut{Output}{Output}

\Input{Distributed storage code $\mathcal{C}$ of length $n$}
\Output{Optimized matrix $\mat{E}_{\mathsf{opt}}$ and largest possible $\Gamma$}
$\Gamma\leftarrow\min\left(k,d_\mathsf{min}^\code{C}-1\right)$\label{algcPoP:Init}\\
$\mat{E}_\mathsf{opt}\leftarrow \emptyset$, $\Gamma_\mathsf{opt} \leftarrow \Gamma$\\
$\set{L}_{n-k}\leftarrow$ \CompEraPat{$\code{C},n-k$}\label{algcPoP:subprocedure0.5}\\
\While{$\Gamma\leq n-k$}{\label{algcPoP:outerWhile}
  $\mathcal{L}_\Gamma \leftarrow$ \CompEraPat{$\mathcal{C},\Gamma$} \label{algcPoP:subprocedure1}\\
  \If{$\mathcal L_\Gamma \not=\emptyset$}{
    $\bm{E} \leftarrow$ \CompEraMat{$\mathcal{L}_\Gamma,\mathcal L_{n-k}$} \label{algcPoP:subprocedure2}\\
    \eIf{$\mat{E} \neq \emptyset$}{
      $\bm E_{\mathsf{opt}} \leftarrow \mat{E}$, $\Gamma_{\mathsf{opt}} \leftarrow \Gamma$\\
    }{
      \KwRet{$(\bm E_{\mathsf{opt}}, \Gamma_{\mathsf{opt}})$}
    }
  }
  $\Gamma\leftarrow\Gamma+1$\\
}
\KwRet{$(\bm E_{\mathsf{opt}},  \Gamma_{\mathsf{opt}})$}
\caption{Optimizing the PIR rate}\label{alg:cPoP}
\end{algorithm}

For codes for which we are not able to prove that they achieve the MDS-PIR capacity, in this section we provide an
algorithm to optimize Protocols~1 and 2 in order to achieve the highest possible PIR rate $\const{R}(\code{C})$ for a
given code $\code{C}$ by taking the structure of the underlying code into consideration. The algorithm is given in
Algorithm~\ref{alg:cPoP} and is based on Theorem~\ref{th:GenPIR}. In particular, we need to find a $(d+\beta) \times n$
matrix $\mat{E}$ in \eqref{eq:PIRerasure-matrix} for which $\hat{\mat{E}}$ consists of erasure patterns of weight $\Gamma$
that are all correctable by $\mathcal C$, and for which $\bar{\mat{E}}$ corresponds to information sets of $\code{C}$
(the support of each row is the complement of an information set). In addition, it is required that each column weight
of $\hat{\mat{E}}$ is equal to the corresponding column weight of $\bm 1 - \bar{\mat{E}}$ (see
Section~\ref{sec:file-indep-PIR}). Note that from the resulting matrix $\bm E$ we can find a PIR achievable rate matrix
by taking its binary complement as in \eqref{eq:equivalent_R-E} (see Section~\ref{sec:file-indep-PIR}), thus optimizing
Protocols~1 and 2 in Sections~\ref{sec:file-dep-PIR} and \ref{sec:file-indep-PIR}, respectively.

The main issues that need to be addressed are the efficient enumeration of the set of erasure patterns of a given weight
$\Gamma$ (corresponding to the rows of $\hat{\mat{E}}$) and also of weight $n-k$ ($\bar{\mat{E}}$, corresponding to information sets) that can be
corrected by $\mathcal{C}$, and the efficient computation of the matrix $\bm E$. 
These issues are addressed by the subprocedures
\texttt{ComputeErasurePatternList}($\mathcal{C},\cdot$) and \texttt{ComputeMatrix}($\set{L}_\Gamma,\set{L}_{n-k}$), in Lines~\ref{algcPoP:subprocedure0.5}, \ref{algcPoP:subprocedure1}, and \ref{algcPoP:subprocedure2} of Algorithm~\ref{alg:cPoP}, and discussed below in
Sections~\ref{sec:ComputeList} and \ref{sec:computeE}, respectively. Here, $\set{L}_\Gamma$ and $\set{L}_{n-k}$ correspond to erasure
patterns for $\hat{\mat{E}}$ and $\bar{\mat{E}}$, respectively. We remark that the algorithm will always return a valid
$\mat{E}\neq\emptyset$, since initially
$\Gamma=\min(k,d^{\code{C}}_\mathsf{min}-1)$. 
This follows directly from the fact that we can construct an arbitrary $\hat{\bm E}$ with row weights $\Gamma$ such that its column weights match the corresponding weights in  $\bm{1}-\bar{\bm E}$. Each row in $\hat{\bm E}$ is an erasure pattern that is correctable by $\mathcal C$.
  
Let $d=k$ and $\beta=\Gamma$.  In the particular case of $\mathcal{C}$ being a rate $R^{\mathcal C}>1/2$ systematic MDS
code, $d_{\mathsf{min}}^{\mathcal C}=n-k+1$, and the algorithm will do exactly one iteration of the main loop. This
follows directly from the construction of $\bm E$: the matrix $\bm E$ can be constructed by taking the support of an
arbitrary information set of $\code{C} $ and cyclically shifting it $n$ times to construct an $n \times n$ PIR
achievable rate matrix, after which the resulting matrix is complemented as \eqref{eq:equivalent_R-E} to get $\bm E$. In
this case, the overall PIR scheme reduces to the scheme described in \cite[Sec.~IV]{TajeddineGnilkeElRouayheb18_1} for
systematic MDS codes of rate $R^{\mathcal C}>1/2$. Clearly, for general MDS codes (including nonsystematic codes) of
rate $R^{\mathcal C}>1/2$, the same construction of $\bm E$ works, and the algorithm will perform exactly one iteration
of the main loop also for nonsystematic MDS codes. In the case of $\code{C}$ being a rate $R^{\mathcal C}\leq 1/2$
general MDS code, the initial value of $\Gamma$ becomes $k$ (since
$\Gamma=\min(k,d^\code{C}_\mathsf{min}-1)=\min(k,n-k)=k$), but the algorithm will also find a valid matrix
$\bm E$ for $\Gamma=n-k\geq k$. Again, the existence of $\bm E$ follows from the same argument of cyclically shifting an
existing information set $n$ times. In the general case of $d \neq k$ and $\beta \neq \Gamma$, a similar argument to the
one above can be made.

\subsection{\texttt{ComputeErasurePatternList}($\mathcal{C},\cdot$)}
\label{sec:ComputeList}

Computing a list of erasure patterns that are correctable for a given short code can be done using any
maximum likelihood (ML) decoding algorithm. For small codes, all length-$n$ binary vectors (or erasure patterns) of weight $\Gamma$ (or $n-k$) that are
correctable can be found by exhaustive search, while for longer codes a random search can be performed, in the sense
of picking length-$n$ binary vectors (or erasure patterns) of weight $\Gamma$ (or $n-k$) at random, and then verifying whether they are
correctable or not. Alternatively, one can apply a random permutation $\pi$ to the columns of $\bm H^{\mathcal C}$,
apply the Gauss-Jordan algorithm to the resulting matrix to transform it into \emph{row echelon form}, collect a subset
of size $\Gamma$ of the column indices of \emph{leading-one-columns},\footnote{The leading-one-columns are the columns
  containing a \emph{leading one}, where the first nonzero entry in each matrix row of a matrix in row echelon form is
  called a leading one.} and finally apply the inverse permutation $\inv{\pi}$ to this subset of column indices. The
resulting set corresponds to erased coordinates in $\mathcal C$ that can be recovered by the code. Finally, one can
check whether all cyclic shifts of the added erasure pattern are correctable or not and add the correctable cyclic
shifts to $\mathcal{L}_\Gamma$ (or $\mathcal{L}_{n-k}$).

\subsection{\texttt{ComputeMatrix}($\mathcal{L}_\Gamma,\mathcal L_{n-k}$)}
\label{sec:computeE}

Given the lists $\mathcal{L}_\Gamma$ and $\mathcal L_{n-k}$ of erasure patterns of weight $\Gamma$ and $n-k$,
respectively, that are correctable for $\code{C}$, we construct a
$\bigl(|\set{L}_\Gamma|+|\set{L}_{n-k}|\bigr)\times n$ matrix, denoted by $\mat{\Psi}=(\psi_{i,j})$, in which each row
$i\in\Nat{|\set{L}_\Gamma|}$ is one of the erasure patterns from $\set{L}_\Gamma$ and each row
$i\in\Nat{|\set{L}_\Gamma|+1:|\set{L}_\Gamma|+|\set{L}_{n-k}|}$ is one of the erasure patterns from
$\set{L}_{n-k}$. The problem is now to find a $d \times n$ submatrix $\hat{\mat{\Psi}}$ of the upper part of $\mat{\Psi}$ (rows $1$ to $|\mathcal{L}_\Gamma|$) and a $\beta\times n$ submatrix $\bar{\mat{\Psi}}$ of the
  lower part  of $\mat{\Psi}$ (rows $|\mathcal{L}_\Gamma|+1$ to $|\mathcal{L}_\Gamma|+|\mathcal L_{n-k}|$) such that
  the column weight of each of the $n$ columns is the same for $\hat{\mat{\Psi}}$ and the binary complement of
  $\bar{\mat{\Psi}}$, where $\beta$ and $d$ are chosen such that $\beta k = \Gamma d$.
  
   This can be formulated as an integer program (in the integer variables
$\eta_1,\ldots,\eta_{|\mathcal{L}_\Gamma| + |\mathcal L_{n-k}|}$) in the following way,
\begin{IEEEeqnarray}{rCl}
  \text{maximize} &&\quad\sum_{i=1}^{|\mathcal{L}_\Gamma| + |\mathcal L_{n-k}|} \eta_i
  \nonumber\\
  \text{s.\,t.} &&\quad\sum_{i=1}^{|\mathcal{L}_\Gamma|} \eta_i
  \psi_{i,j} =
  \sum_{i=|\mathcal{L}_\Gamma|+1}^{|\mathcal{L}_\Gamma|
    + |\mathcal L_{n-k}|} \eta_i (1-\psi_{i,j}),
  \,\forall\,j\in\Nat{n},
  \nonumber\\
  &&\quad\eta_i\in\{0,1\},\,\forall\,
  i\in\Nat{|\mathcal{L}_\Gamma| +  |\mathcal
    L_{n-k}|},
  \IEEEeqnarraynumspace\label{eq:LIP}\\
  &&\quad\sum_{i=1}^{|\mathcal{L}_\Gamma|} \eta_i = d,
  \text{ and }
  \sum_{i=|\mathcal{L}_\Gamma|+1}^{|\mathcal{L}_\Gamma|
    + |\mathcal L_{n-k}|} \eta_i = \beta.\nonumber
\end{IEEEeqnarray}
A valid $(d+\beta)\times n$ matrix $\bm E$ is constructed from the rows of $\bm \Psi$ with $\eta_i$-values
equal to one in any feasible solution of (\ref{eq:LIP}). When $|\mathcal{L}_\Gamma| + |\mathcal L_{n-k}|$ is large, solving
(\ref{eq:LIP}) may become impractical (solving a general integer program is known to be NP-hard), in which case one can
take several random subsets (of some size) of the lists $\mathcal{L}_\Gamma$ and $\mathcal L_{n-k}$, construct the
corresponding matrices $\bm \Psi$, and try to solve the program in (\ref{eq:LIP}).

\section{Multiple Colluding Nodes}
\label{sec:MultipleCollNodePIR}

In this section, we consider the scenario where $T>1$ nodes act as spies and have the ability to collude. In particular,
{\mg we propose a protocol for this scenario that improves upon the PIR protocol in
  \cite{FreijHollantiGnilkeHollantiKarpuk17_1}}. We refer to the protocol in \cite{FreijHollantiGnilkeHollantiKarpuk17_1} as the \emph{$(\code{C},\bar{\code{C}})$-retrieval protocol (or scheme)},
since it is based on two linear codes: an $[n,k]$ code $\code{C}$ and an $[n,\bar{k}]$ code $\bar{\code{C}}$, where
$\mathcal C$ is the underlying storage code of the DSS and {$\bar{\code{C}}$ defines the queries. 
 Furthermore, the retrieval process is defined by an $[n,\tilde{k}]$ code $\tilde{\mathcal C}$ that is the Hadamard product of $\mathcal C$ and $\bar{\code{C}}$, $\tilde{\mathcal C}=\mathcal C\circ\bar{\mathcal C}$. The protocol yields privacy against at most $T=d_{\mathsf{min}}^{\bar{\mathcal C}^\perp}-1$ colluding nodes 
under the assumption that  the code  $\bar{\mathcal C}$ with $d_{\mathsf{min}}^{\bar{\mathcal C}^\perp}=T+1$   exists for the given $T$} (existing in the sense that the Hadamard product of $\code{C}$ and $\bar{\code{C}}$ has rate strictly smaller than $1$).

Originally, the protocol was designed to work with GRS codes, a class of MDS codes, i.e., both {codes $\code{C}$ and  $\bar{\code{C}}$ are GRS codes. In this case $\bar{\code{C}}$ has parameters  $[n,\bar{k}=T]$, the retrieval code $\tilde{\code{C}}$ has parameters  $[n,\tilde{k}=k+T-1]$, and the PIR rate is
\begin{align*}
	\const{R}_{\mathsf{GRS}}=\frac{n-(k+T-1)}{n}.
\end{align*}
}%
{For non-MDS codes, the protocol achieves a PIR rate 
\begin{align*}
	\const{R}(\C,\Cbar)=\frac{d^{\tilde{\mathcal C}}_\mathsf{min}-1}{n},	
\end{align*}
 which is lower than $\const{R}_{\mathsf{GRS}}$. 
In general, when the underlying codes are arbitrary codes, it can be shown that the PIR rate of the $(\code{C},\bar{\code{C}})$-retrieval protocol is upperbounded by}
{\begin{align}
    \label{eq:RuB}
    \const{R}_{\mathsf{UB}}\eqdef\frac{n-\tilde{k}}{n}.
\end{align}
In particular, the  $(\code{C},\bar{\code{C}})$-retrieval protocol in \cite{FreijHollantiGnilkeHollantiKarpuk17_1} achieves a PIR rate $\const{R}(\C,\Cbar)<\const{R}_{\mathsf{UB}}$ for non-MDS codes.}
%
%
%
%
Furthermore, it was shown in \cite{Hol17} that if $\mathcal C$ is either a GRS code or an RM code, then
$\bar{\mathcal C}$ always exists for any $T \leq n-k$.  In this section, we look at this protocol from the perspective
of arbitrary linear codes $\mathcal C$ and propose an improved protocol, referred to as Protocol~3, that achieves a
higher PIR rate $\const{R}_{\mathsf{P3}}(\C,\Cbar)$, where
$\const{R}(\C,\Cbar)\leq \const{R}_{\mathsf{P3}}(\C,\Cbar) \leq\const{R}_{\mathsf{UB}}\leq \const{R}_{\mathsf{GRS}}$. In particular,
we show that the upper bound $\const{R}_{\mathsf{UB}}$ can be achieved for some non-MDS codes. Also, for a given $T$ we
present a code family for $\mathcal C$ for which $\bar{\mathcal C}$ exists.

\subsection{Protocol~3: The Multiple Colluding Nodes Case}
\label{sec:CCbarprotocol}


The protocol presented here, referred to as Protocol~3, can be seen as an extension of Protocol~2 in
Section~\ref{sec:file-indep-PIR}. We assume that each file $\bm X^{(m)}=\bigl(x_{i,j}^{(m)}\bigr)$, $m\in\Nat{f}$, of size
$\beta\times k$,  is stored using an $[n,k]$ code $\mathcal C$ over $\GF(q)$, where $x^{(m)}_{i,j}\in\GF(q^\ell)$ for some
$\ell\in\Nat{}$.  Let $\bar{\mathcal C}$ be an $[n,\bar{k}]$ code over $\GF(q)$. {\mg The code $\bar{\mathcal C}$ is
  used to design the query matrix $\bm Q^{(l)}$, of dimensions $d\times \beta f$, where $\bm q_i^{(l)}$ is the $i$-th
  subquery of $\bm Q^{(l)}$ (see Section~\ref{sec:privacy}).} Furthermore, $\bar{\mathcal C}$ characterizes $T$, i.e., the
maximum number of colluding nodes the PIR protocol can handle whilst maintaining information-theoretic privacy. {\mg As
  for Protocol~2, $\beta$ and $d$ are taken as small as possible according to \eqref{Eq: Beta_and_d_DEF}.} The response
vector corresponding to the $i$-th subquery $\bm q_i^{(l)}$, denoted by $\bm \rho^{(i)}=(r_{1,i}, \ldots,r_{n,i})^{\textup{\textsf{\tiny T}}}$,
is a collection of the $n$ response symbols $r_{l,i}$ from the $n$ storage nodes and is related to the codewords of an
$[n,\tilde{k}]$ code $\tilde{\mathcal C}=\mathcal C\circ\bar{\mathcal C}$. Furthermore, $\tilde{\mathcal C}$
characterizes the PIR rate of the protocol.

\subsubsection{Query Construction}

The protocol requires that the user constructs queries by choosing $\beta f$ codewords
$\bar{\bm c}^{(m)}_i=(\bar c^{(m)}_{i,1}, \ldots,\bar c^{(m)}_{i,n})$, $i\in\Nat{\beta}$ and $m \in \Nat{f}$,
drawn independently and uniformly at random from the code $\bar{\mathcal C}$. It then constructs the
vector
\begin{align*}
  \mathring{\bm c}_{l}=(\mathring{\bm c}^{(1)}_l,\ldots,\mathring{\bm c}^{(f)}_l),\quad l\in\Nat{n},
\end{align*}
where $\mathring{\bm c}^{(m)}_l=(\bar c^{(m)}_{1,l},\ldots,\bar c^{(m)}_{\beta,l})$. Thus, the vector
$\mathring{\bm c}_{l}$ is of length $\beta f$. The vector $\mathring{\bm c}^{(m)}_l$ is a collection of the entries of
the $l$-th coordinates of the codewords $\bar{\bm c}^{(m)}_1,\ldots,\bar{\bm c}^{(m)}_\beta$ that pertain to the $m$-th
file. We denote by $\mathcal J_i\subseteq\Nat{n}$, $i\in\Nat{d}$, $|\mathcal J_i|=\Gamma$, the set of nodes from which
the protocol obtains code symbols pertaining to the $m$-th file {\mg from the $i$-th subresponses.} 

Similar to Protocol~2 presented in Section~\ref{sec:file-indep-PIR} for the case of noncolluding nodes, we need to construct a
matrix $\hat{\bm E}$ and $\beta$ information sets $\{\set{I}_i\}_{i\in\Nat{\beta}}$. The matrix $\hat{\bm E}$ is a
$d\times n$ binary matrix where each row represents an erasure pattern of weight $\Gamma$ correctable by
$\tilde{\mathcal C}=\mathcal C\circ\bar{\mathcal C}$. The column weight profile of $\hat{\bm E}$ is determined from
$\{\set{I}_i\}_{i\in\Nat{\beta}}$ as in Section~\ref{sec:file-indep-PIR}. Note that $\mathcal J_i$ is the support of the $i$-th
row vector of $\hat{\bm E}$. Let $m$ denote the index of the requested file. {\mg Then, the $i$-th subquery to node $l$
  is constructed as}
 \begin{align}
  \label{Eq: CollQuery_designa}
 \bm q_i^{(l)}=\mathring{\vect{c}}_{l}+\bm \delta_i^{(l)}, 
 \end{align}
 where
   \begin{align}
    \label{Eq: CollQuery_design}
    \bm \delta_i^{(l)}=\begin{cases}
      \bm \omega_{\beta(m-1)+s_i^{(l)}} & \text{if }
      l\in\mathcal J_{i},
      \\
     \bm \omega_{0} & \text{otherwise},
    \end{cases}
  \end{align}
for $l\in\Nat{n}$, where $\bm \omega_t$, $t\in\Nat{\beta f}$, is the $t$-th $(\beta f)$-dimensional unit vector and
$\bm \omega_0=\bm 0_{1\times\beta f}$. The index $s_i^{(l)}$ is defined as
\begin{align}
  \label{Eq: User_file_index}
  s_i^{(l)}\in\mathcal F_l=\{t\in\Nat{\beta}\colon l\in\mathcal I_t\}
\end{align}
and $s_i^{(l)}\neq s_{i'}^{(l)}$ for $i\neq i'$, $i,i'\in\Nat{d}$. The index $s_i^{(l)}$ denotes the symbol downloaded
from the $s_i^{(l)}$-th row of the chunk pertaining to $\bm X^{(m)}$ of the $l$-th node {\mg in response to the $i$-th
  subquery}. Clearly, we see that the symbols downloaded from all nodes form $\beta$ information sets as
$\sum_{i=1}^{d} |\mathcal J_i|=\sum_{i=1}^{\beta} |\mathcal I_i|=\beta k$.

{\mg Note that in \eqref{Eq: CollQuery_designa}, the vector $\mathring{\bm c}_{l}$ introduces randomness such that
  privacy is ensured, while the vector $\bm\omega$ is deterministic and is properly designed such that the requested file can be
  recovered by the user.}

\subsubsection{Response Construction}

For the $i$-th {\mg subquery}, the response symbol from the $l$-th node is constructed as
\begin{align}
\label{Eq: Holl_response}
  r_{l,i}=\langle \bm q_i^{(l)}, (c_{1,l}^{(1)},\ldots,c_{\beta,l}^{(f)})\rangle.
\end{align}
The response symbol in \eqref{Eq: Holl_response} is the dot product between the subquery vector to the $l$-th node and its 
content. The user obtains a response vector $\bm \rho^{(i)}$, consisting of response symbols from $n$ nodes
as
\begin{align}
\label{Eq: PIR_coll_response}
  \bm \rho^{(i)}=\left(\begin{matrix}
      r_{1,i}\\
      r_{2,i}\\
      \vdots\\
      r_{n,i}
    \end{matrix}\right)=\underbrace{
  \sum_{i'=1}^\beta\sum_{m'=1}^f\left(\begin{matrix}
      \bar c_{i',1}^{(m')}c_{i',1}^{(m')}\\
      \bar c_{i',2}^{(m')}c_{i',2}^{(m')}\\
      \vdots\\
      \bar c_{i',n}^{(m')}c_{i',n}^{(m')}\\
    \end{matrix}\right)}_{\in\bigl\{\vect{x}\in(\GF(q^\ell))^n\colon\mat{H}^{\tilde{\code{C}}}\vect{x}=\bm 0\bigr\}}
  +\left(\begin{matrix}
      o_1^{(i)}\\
      o_2^{(i)}\\
      \vdots\\
      o_{n}^{(i)}
  \end{matrix}\right),
\end{align}
where $\bm H^{\tilde{\mathcal C}}$ is a parity-check matrix of the code $\tilde{\mathcal C}$,\footnote{Note that the
  upload cost of the PIR scheme in \cite{FreijHollantiGnilkeHollantiKarpuk17_1,Hol17} grows linearly with $f$. 
  However, when the file size is large (i.e., when $\ell$ is large; $q^{\ell}$ is the field size of the message symbols) the upload cost can be ignored.} the symbol
$o^{(i)}_l$ denotes the symbol obtained from the $l$-th node {\mg corresponding to the $i$-th subquery}, and
\begin{align*}
  o^{(i)}_l=\begin{cases}
    c^{(m)}_{i',l} & \text{if}\,\, l\in\mathcal J_i,\\
    0 & \text{otherwise},\\
  \end{cases}
\end{align*}
where $i'=s_i^{(l)}$. These symbols are obtained by post-processing \eqref{Eq: PIR_coll_response} as follows,
\begin{align}
  \label{Eq: User_PostProcessing}
  \bm H^{\tilde{\mathcal C}}\bm \rho^{(i)}=\bm H^{\tilde{\mathcal C}}\left(\begin{matrix}
      o_1^{(i)}
      \\
      o_2^{(i)}
      \\
      \vdots
      \\
      o_{n}^{(i)}
    \end{matrix}\right).
\end{align}

This completes the construction of the PIR protocol. In the following, we prove that this protocol satisfies the PIR
conditions \eqref{eq:cond1} and \eqref{eq:cond2} in Definition~\ref{def:cond}.
\begin{lemma}
  \label{lem: CollPrivacy}
  Consider a DSS that uses an $[n,k]$ code with subpacketization $\alpha$ to store $f$ files, each divided into $\beta$
  stripes, and assume the privacy model of Section~\ref{sec:privacy} with a set
  $\set{T}=\{t_1,\ldots,t_{\card{\set{T}}}\}\subset\Nat{n}$ of
  $\card{\set{T}}\leq T \leq d^{\bar{\code{C}}^\perp}_\mathsf{min}-1$ colluding nodes. Then, the subqueries
  $\bm q^{(l)}_i$, $l\in\Nat{n}$, $i\in\Nat{d}$, designed as in \eqref{Eq: CollQuery_designa} and \eqref{Eq:
    CollQuery_design} satisfy $\mathsf H\bigl(m|\bm Q^{(t_1)},\ldots,\bm Q^{(t_{\card{\set{T}}})}\bigr)=\mathsf{H}(m)$.
\end{lemma}
\begin{IEEEproof}
  The addition of a deterministic vector in \eqref{Eq: CollQuery_designa} does not change the probability distribution
  of the vectors $\bm q^{(t_1)}_i,\ldots,\bm q^{(t_{\card{\set{T}}})}_i$. The same can be said about their joint
  distribution. Furthermore, in each query matrix $\bm Q^{(l)}$, $l\in\{t_1,\ldots,t_{\card{\set{T}}}\}$, the subqueries
  $\vect{q}^{(l)}_i$ are independent of each other. Thus, the proof follows the same lines as the proof of
  \cite[Th.~8]{FreijHollantiGnilkeHollantiKarpuk17_1}.
\end{IEEEproof}

{\mg
\begin{theorem}
  \label{th:CollPIR}
  Consider a DSS that uses an $[n,k]$ code $\code{C}$ with subpacketization $\alpha$ to store $f$ files, each divided
  into $\beta$ stripes. Let $\bar{\mathcal C}$ be an $[n,\bar{k}]$ code such that there exists an $[n,\tilde{k}]$ code
  $\tilde{\mathcal C}=\mathcal C\circ\bar{\mathcal C}$ of rate $R^{\tilde{\mathcal C}} < 1$. 
  %
  If there exists a $\Gamma$-row regular $d \times n$ binary matrix $\bm{\hat{E}}$ in which each row is a correctable
  erasure pattern for $\tilde{\mathcal C}$ and satisfying condition $\mathsf{C3}$, then
  $\mathsf H(\bm X^{(m)}|\bm \rho^{(1)},\ldots, \bm \rho^{(d)})=0$ and the PIR rate
  \begin{align}
    \label{eq:UB_PIR_rate}
    \const{R}_{\mathsf{P3}}(\C,\Cbar)=\frac{\Gamma}{n}\leq \const{R}_{\mathsf{UB}}
  \end{align}
  is achievable.
\end{theorem}}
\begin{IEEEproof}
  {By assumption there exists a matrix $\hat{\bm E}$ of size $d\times n$ having row weight $\Gamma$.} Furthermore,
  again by assumption, each row of $\hat{\bm E}$ is an erasure pattern that is correctable by $\tilde{\mathcal C}$. From
  \eqref{Eq: CollQuery_design}, \eqref{Eq: User_PostProcessing} results in
  \begin{IEEEeqnarray*}{rCl}
    \bm H^{\tilde{\mathcal C}}\bm \rho^{(i)} = \bm H^{\tilde{\mathcal C}}|_{\chi(\hat{\bm e}_i)} 
    \left(\begin{matrix}
      o_{l_1}^{(i)}
      \\
      o_{l_2}^{(i)}
      \\
      \vdots
      \\
      o_{l_{|\mathcal{J}_i|}}^{(i)}
    \end{matrix}\right),
  \end{IEEEeqnarray*}%
  where $\hat{\bm e}_i$ is the $i$-th row of $\hat{\bm E}$ and $l_j$, $j \in \Nat{|\mathcal{J}_i|}$, denotes the elements of $\mathcal{J}_i$. The above linear system of equations is full rank as
  $\bm H^{\tilde{\mathcal C}}|_{\chi(\hat{\bm e}_i)}$ is full rank. This is because $\hat{\bm e}_i$ is a correctable
  erasure pattern for $\tilde{\mathcal C}$. As such,
  the $\Gamma$ symbols $\{ o^{(i)}_l \}_{l \in \mathcal{J}_i}$ are obtained.  From all responses, the user obtains
  $\Gamma d=\beta k$ code symbols of the code $\mathcal C$. Furthermore, from \eqref{Eq: User_file_index}, these
  $\Gamma d$ symbols are part of the $\beta$ information sets $\{\set{I}_i\}_{i\in\Nat{\beta}}$ of $\code{C}$. Thus,
  $\mathsf H\bigl(\bm X^{(m)}|\bm \rho^{(1)},\ldots, \bm \rho^{(d)}\bigr)=0$.
\end{IEEEproof}

Unlike \cite{FreijHollantiGnilkeHollantiKarpuk17_1}, where the authors consider sets $\mathcal J_i$ with a fixed
structure, we generalize the sets to match arbitrary codes $\mathcal C$, $\bar{\mathcal C}$, and $\tilde{\mathcal
  C}$. In particular, the sets in \cite{FreijHollantiGnilkeHollantiKarpuk17_1} were constructed targeting MDS codes,
{\mg in which case the PIR rate of the $(\code{C},\bar{\code{C}})$-retrieval protocol is upperbounded by
  $\const{R}_{\mathsf{UB}}$ in \eqref{eq:RuB}, as mentioned earlier}. However, the use of these sets for arbitrary codes
$\mathcal C$ and $\bar{\mathcal C}$ does not allow to obtain the requested file $\bm X^{(m)}$. Thus,
Theorem~\ref{th:CollPIR} can be seen as a generalization of \cite[Th.~7]{FreijHollantiGnilkeHollantiKarpuk17_1}, where
the PIR rate {\mg for non-MDS codes} was shown to be
$\const{R}(\C,\Cbar)=(d^{\tilde{\mathcal C}}_\mathsf{min}-1)/n < \const{R}_{\mathsf{UB}}$. Our proposed protocol can
achieve higher rates as illustrated in the following corollary. In particular, we will show that the upper bound
$\const{R}_{\mathsf{UB}}$ is achievable for some classes of non-MDS codes.

\begin{corollary}
  If for an $[n,k]$ code $\code{C}$ and an $[n,\bar{k}]$ code $\bar{\mathcal C}$ there exists an $[n,\tilde{k}]$ code
  $\tilde{\mathcal C}=\mathcal C\circ\bar{\mathcal C}$ of rate $R^{\tilde{\mathcal C}} < 1$ and an $(n-\tilde{k})$-row
  regular $d \times n$ binary matrix $\bm{\hat{E}}$ in which each row is a correctable erasure pattern by
  $\tilde{\mathcal C}$ and satisfying condition $\mathsf{C3}$, then Protocol~3 achieves the upper bound
  $\const{R}_{\mathsf{UB}}$.
\end{corollary}

As for Protocol~2, the parameters $\Gamma$,
$\beta$, and $d$ mentioned in Theorem~\ref{th:CollPIR} 
have to be carefully selected such $\beta k = \Gamma d$ and such that a
$\Gamma$-row regular matrix $\hat{\bm E}$ (satisfying condition $\mathsf{C}3$) actually exists with a valid collection
of information sets $\{\set{I}_i\}_{i\in\Nat{\beta}}$ for $\code{C}$.

\subsection{Example}
\label{sec:ColludingCase}

Lemma~\ref{lem: CollPrivacy} proves that the proposed protocol provides privacy up to
$T=d^{\bar{\code{C}}^\perp}_{\mathsf{min}}-1$ colluding nodes. This is illustrated in the example below.

Consider a DSS with $n=12$ nodes that stores a single file $\bm X^{(1)}$ of size $1\times4$. $\bm X^{(1)}$ is encoded
using the $[12,4,6]$ binary code $\mathcal C$ with parity-check matrix
\begin{align*}
  \bm H^{\mathcal C}=
  \left(\begin{array}{cccccccccccc}
          0 & 1 & 1 & 0 & 1 & 0 & 0 & 0 & 0 & 0 & 0 & 0\\
          1 & 0 & 1 & 0 & 0 & 1 & 0 & 0 & 0 & 0 & 0 & 0\\
          1 & 1 & 1 & 0 & 0 & 0 & 1 & 0 & 0 & 0 & 0 & 0\\ 
          1 & 1 & 0 & 1 & 0 & 0 & 0 & 1 & 0 & 0 & 0 & 0\\ 
          1 & 0 & 1 & 1 & 0 & 0 & 0 & 0 & 1 & 0 & 0 & 0\\ 
          0 & 1 & 1 & 1 & 0 & 0 & 0 & 0 & 0 & 1 & 0 & 0\\ 
          1 & 1 & 1 & 1 & 0 & 0 & 0 & 0 & 0 & 0 & 1 & 0\\ 
          0 & 0 & 1 & 1 & 0 & 0 & 0 & 0 & 0 & 0 & 0 & 1
        \end{array}\right).
\end{align*}
Let $\bar{\mathcal C}=\mathcal C$, and the code $\tilde{\mathcal C}=\mathcal C\circ\bar{\mathcal C}$ has parity-check
matrix
\begin{align*}
  \bm H^{\tilde{\mathcal C}}=
  \left(\begin{array}{cccccccccccc}
          1 & 1 & 1 & 1 & 0 & 0 & 1 & 1 & 1 & 1 & 0 & 0\\ 
          1 & 1 & 0 & 1 & 1 & 1 & 0 & 1 & 0 & 0 & 1 & 1 
        \end{array}\right).
\end{align*}

Note that the dual code $\bar{\mathcal C}^\perp$ has minimum Hamming distance $d_{\mathsf{min}}^{\bar{\mathcal C}^\perp}=3$,
thus Protocol~3 protects against $T=d^{\bar{\mathcal C}^\perp}_{\mathsf{min}}-1=2$ colluding nodes. Choosing
$\Gamma=d^{\tilde{\mathcal C}}_\mathsf{min}-1=1$, one can use the PIR protocol as presented in
\cite{FreijHollantiGnilkeHollantiKarpuk17_1} to get a PIR rate of $\const{R}(\C,\Cbar)=\frac{1}{12}$. However, we can set
$\Gamma=2$ and use Protocol~3 to achieve a higher PIR rate. Note that the
value of $\Gamma$ cannot be greater than $2$ as the number of redundant symbols in $\tilde{\mathcal C}$ is $2$. We choose
\begin{align*}
  \hat{\bm E}&=\left(\begin{matrix}
      0 & 0 & 0 & 0 & 0 & 0 & 0 & 0 & 1 & 0 & 0 & 1\\
      0 & 1 & 1 & 0 & 0 & 0 & 0 & 0 & 0 & 0 & 0 & 0
    \end{matrix}\right)
\end{align*}
and $\set{I}_1=\{2,3,9,12\}$. Thus, $\beta=1$ and $d=2$.  Note that each row of $\hat{\bm E}$ is an erasure pattern
that is correctable by the $[12,10,2]$ code $\tilde{\mathcal C}$, and that $\set{I}_1$ is an information set of
$\mathcal C$. In order to form all the queries (each query consists of $d=2$ subqueries), we need to choose $s_1^{(9)}$, $s_1^{(12)}$,
$s_2^{(2)}$, and $s_2^{(3)}$. From \eqref{Eq: User_file_index}, we have
\begin{align*}
  s_1^{(9)}=1, s_1^{(12)}=1, s_2^{(2)}=1, \text{ and } s_2^{(3)}=1.
\end{align*}
Now, consider {\mg the first subqueries}. The query vectors $\bm q_{1}^{(9)}$ and $\bm q_{1}^{(12)}$ are
\begin{align*}
  \begin{split}
    \bm q_{1}^{(9)} &=\mathring{\bm c}_{9}+ \bm \omega_{1}=\mathring{\bm c}_{9}+1,
    \\
    \bm q_{1}^{(12)}&=\mathring{\bm c}_{12}+\bm \omega_{1}=\mathring{\bm c}_{12}+1,
  \end{split}
\end{align*}
and $\bm q^{(l)}_{1}=\mathring{\vect{c}}_l$, $\forall\,l\in\Nat{12}\backslash\{9,12\}$. {\mg The corresponding response
  vector is}
\begin{align*}
  \bm \rho^{(1)}=\underbrace{
  \sum_{i'=1}^1\sum_{m'=1}^1\left(\begin{matrix}
      \bar c_{i',1}^{(m')}c_{i',1}^{(m')}\\
      \bar c_{i',2}^{(m')}c_{i',2}^{(m')}\\
      \vdots\\
      \bar c_{i',12}^{(m')}c_{i',12}^{(m')}\\
    \end{matrix}\right)}_{\in\bigl\{\vect{x}\in(\GF(q^\ell))^{12}\colon\mat{H}^{\tilde{\code{C}}}\vect{x}=\bm 0\bigr\}}
  +\left(\begin{matrix}
      0\\
      \vdots\\
      0\\
      c_{1,9}^{(1)}\\
      0\\
      0\\
      c_{1,12}^{(1)}
    \end{matrix}\right).
\end{align*}
Finally, the user computes
\begin{align*}
  \bm H^{\tilde{\mathcal C}}\bm \rho^{(1)}=\bm H^{\tilde{\mathcal C}}\left(\begin{matrix}
      0\\
      \vdots\\
      0\\
      c_{1,9}^{(1)}\\
      0\\
      0\\
      c_{1,12}^{(1)}
    \end{matrix}\right)=\left(\begin{matrix}
      c_{1,9}^{(1)}\\
      c_{1,12}^{(1)}
    \end{matrix}\right).
\end{align*}
In a similar manner, {\mg from $\bm \rho^{(2)}$} the user obtains $c_{1,2}^{(1)}$ and $c_{1,3}^{(1)}$.
Clearly, the indices of the symbols downloaded by the user form the information set $\mathcal I_1$, from which we can
obtain the requested file $\bm X^{(1)}$. The PIR rate of the scheme is
$\const{R}_{\mathsf{P3}}(\C,\Cbar)=\frac{4}{24}=\frac{1}{6}$, i.e., double of the PIR rate of the protocol in
\cite{FreijHollantiGnilkeHollantiKarpuk17_1}. Furthermore, it achieves the upper bound in \eqref{eq:UB_PIR_rate}.

A limiting factor for {\mg Protocol~3} is that the upper bound on the PIR rate $\const{R}_{\mathsf{UB}}$ in
\eqref{eq:UB_PIR_rate} depends on the dimension of $\tilde{\mathcal C}$. Furthermore, in order to achieve the PIR
property with large $T$, one requires $\bar{\mathcal C}$ to be of large dimension such that $\bar{\mathcal C}^\perp$ has
large minimum Hamming distance. Therefore, for arbitrary $\mathcal C$ and $\bar{\mathcal C}$ the chances that
$\tilde{\mathcal C}=\mathcal C\circ\bar{\mathcal C}$ has rate $R^{\tilde{\mathcal C}}=1$ (the code does not even correct
a single erasure) are quite high for large values of $T$, since the Hadamard product is highly nonlinear. In other
words, the probability that $\const{R}_{\mathsf{UB}}=0$ is high. Below, we present code constructions for which
$\const{R}_{\mathsf{UB}}>0$.

\subsection{Codes for Protocol~3}

As seen in the preceding subsections, the codes $\mathcal C$ and $\bar{\mathcal C}$ must be chosen such that the
code $\tilde{\mathcal C}=\mathcal C \circ \bar{\mathcal C}$ has rate $R^{\tilde{\mathcal C}}<1$ for the PIR protocol to
work. In this  subsection, we provide a family of codes $\mathcal C$ and $\bar{\mathcal C}$ that satisfy $R^{\tilde{\mathcal C}}<1$ for a given $T$.

In particular, we show that a special class of UUV codes can be used for the codes $\mathcal C$ and
$\bar{\mathcal C}$ to obtain a valid $\tilde{\mathcal C}$. Let $\mathcal U$ be an $[n_1,k_1]$ code and $\mathcal V$ an $[n_1,1]$ repetition code, both over $\GF(2)$. We construct the
$[n=2n_1,k=k_1+1]$ code $\mathcal C=(\mathcal U\mid\mathcal U+\mathcal V)$ with generator matrix
\begin{align}
  \label{Eq: UUV const}
  \bm G^{\mathcal C}=\left(\begin{matrix}
      \bm G^{\mathcal U}    & \bm G^{\mathcal U}\\
      \vect{0}_{1\times n_1}& \vect{1}_{1\times n_1}
    \end{matrix}\right).
\end{align}

\begin{theorem}
  \label{th: CstarDproof}
  Let $\mathcal U$ be an $[n_1,k_1]$  binary code where $n_1\geq k_1+2$ and $\mathcal V$ an $[n_1,1]$ binary 
  repetition code. Then, the $[n=2n_1,k=k_1+1]$ codes $\mathcal C$ and $\bar{\mathcal C}$ constructed using \eqref{Eq:
    UUV const} ensure that the vector space $\tilde{\mathcal C}=\mathcal C\circ\bar{\mathcal C}$ of length $n$ is a
  linear code of dimension strictly less than $n$, i.e., $R^{\tilde{\mathcal C}}<1$.
\end{theorem}
\begin{IEEEproof}
  See Appendix~\ref{Appendix: Proof3}.
\end{IEEEproof}

Theorem~\ref{th: CstarDproof} proves that for an arbitrary linear code $\mathcal U$, the UUV code ensures that
$\tilde{k}<n$ and thus $\tilde{\mathcal C}$ is a valid code. The fact that any code $\mathcal U$ can
be used in the protocol makes the UUV construction attractive. Also, the UUV construction may produce
$d_\mathsf{min}$-optimal binary linear codes. For instance, the codes $\mathcal C$ and $\bar{\code{C}}$ in
Section~\ref{sec:ColludingCase} are $d_\mathsf{min}$-optimal binary linear codes that can be constructed through the UUV construction. One drawback of the UUV construction, however, is that
the constructed codes are in general low rate codes.

In \cite{Hol17}, the authors showed that choosing $\mathcal C$ and $\bar{\mathcal C}$ to be RM codes with carefully
selected parameters ensures that $\tilde{\mathcal C}$ is also an RM code of dimension $\tilde{k}<n$. However, the PIR
rate is very low \cite[Th.~15]{Hol17}. In the following subsection, we show that RM codes can indeed achieve a higher
PIR rate of $\const{R}_{\mathsf{P3}}(\C,\Cbar)=(n-\tilde{k})/n=\const{R}_{\mathsf{UB}}$.

\subsection{Codes Achieving the Maximum PIR Rate of {\mg Protocol~3}}
\label{sec:codes_CstarD-PIRmaxRate}

In order to consider the codes achieving the maximum possible PIR rate for {\mg Protocol~3}, we give a definition
similar to Definition~\ref{def:PIRachievable-rate-matrix} in Section~\ref{sec:file-dep-PIR}.

\begin{definition}
  \label{def:codes_CstartD-PIRmaxRate}
  Let $\code{C}$ be an $[n,k]$ code and $\bar{\code{C}}$ an $[n,\bar{k}]$ code. Denote by
  $\tilde{\code{C}}=\code{C}\circ\bar{\code{C}}$ the $\tilde{k}$-dimensional code generated by the Hadamard product of
  $\code{C}$ and $\bar{\code{C}}$. A $(k+n-\tilde{k})\times n$ binary matrix $\tilde{\mat{\Lambda}}_{k,k+n-\tilde{k}}$
  is called a PIR maximum rate matrix for {\mg Protocol~3} if the following conditions are satisfied.
  \begin{enumerate}
  \item \label{item:3} $\tilde{\mat{\Lambda}}_{k,k+n-\tilde{k}}$ {\mg is a $k$-column regular matrix}, and
  \item \label{item:4} there are exactly $k$ rows $\{\vect{\lambda}_i\}_{i\in\Nat{k}}$ and $n-\tilde{k}$ rows
    $\{\vect{\lambda}_{i+k}\}_{i\in\Nat{n-\tilde{k}}}$ of $\tilde{\mat{\Lambda}}_{k,k+n-\tilde{k}}$ such that
    $\forall\,i\in\Nat{k}$, $\chi(\vect{\lambda}_i)$ {\mg is an information set} for $\tilde{\code{C}}$ and
    $\forall\,i\in\Nat{n-\tilde{k}}$, $\chi(\vect{\lambda}_{i+k})$ {\mg is an information set} for $\code{C}$.
  \end{enumerate}
\end{definition}

Similar to the case of noncolluding nodes in Section~\ref{sec:file-indep-PIR}, it is not difficult to show that the
existence of a $k\times n$ matrix $\hat{\mat{E}}$ for the code $\tilde{\code{C}}=\code{C}\circ\bar{\code{C}}$ and an
$(n-\tilde{k})\times n$ matrix $\bar{\mat{E}}$ for the code $\code{C}$ is equivalent to the existence of
$\tilde{\mat{\Lambda}}_{k,k+n-\tilde{k}}$.

The following corollary follows immediately from a similar reasoning as  for
Theorem~\ref{thm:general-d_MDS-PIRcapacity-achieving-codes}.
\begin{corollary}
  \label{cor:general-d_CstarD-PIRmaxRate-codes}
  If a PIR maximum rate matrix $\tilde{\mat{\Lambda}}_{k,k+n-\tilde{k}}$ exists for {\mg Protocol~3}, then
  \begin{IEEEeqnarray*}{rCll}
    d_s^{\code{C}}& \geq &\frac{n-\tilde{k}}{k}s, &\quad\forall\,s\in\Nat{k},
    \\
    d_s^{\tilde{\code{C}}}& \geq &s, &\quad\forall\,s\in\Nat{\tilde{k}}.
  \end{IEEEeqnarray*}
\end{corollary}
\begin{IEEEproof}
  Using an argumentation similar to the proof of Theorem~\ref{thm:general-d_MDS-PIRcapacity-achieving-codes}, the existence of
  a PIR maximum rate matrix for {\mg Protocol~3} implies that there exist $k$ information sets
  $\{\tilde{\set{I}}_i\}_{i\in\Nat{k}}$ of $\tilde{\code{C}}$ and $n-\tilde{k}$ information sets
  $\{\set{I}_{i'}\}_{i'\in\Nat{n-\tilde{k}}}$ of $\code{C}$ such that each coordinate $j$ of $\code{C}$ appears exactly
  $k$ times in {\mg $\{\tilde{\set{I}}_i\}_{i\in\Nat{k}}\cup\{\set{I}_{i'}\}_{i'\in\Nat{n-\tilde{k}}}$},
  $j\in\Nat{n}$. Hence, we obtain
  \begin{IEEEeqnarray*}{rCl}
    k\card{\chi(\set{D})}& =
    &\underbrace{\sum_{i=1}^k\bigcard{\tilde{\set{I}}_i\cap\chi(\code{D})}}_{\geq 0}+
    \sum_{i'=1}^{n-\tilde{k}}\bigcard{\set{I}_{i'}\cap\chi(\code{D})}
    \\
    & \geq
    &\sum_{i'=1}^{n-\tilde{k}}\bigcard{\set{I}_{i'}\cap\chi(\code{D})}\geq (n-\tilde{k})s;
    \\
    k\card{\chi(\tilde{\set{D}})}& = &\sum_{i=1}^k\bigcard{\tilde{\set{I}}_i\cap\chi(\tilde{\code{D}})}+
    \underbrace{\sum_{i'=1}^{n-\tilde{k}}\bigcard{\set{I}_{i'}\cap\chi(\tilde{\code{D}})}}_{\geq 0}
    \\
    & \geq
    &\sum_{i=1}^k\bigcard{\tilde{\set{I}}_i\cap\chi(\tilde{\code{D}})}\geq k s,
  \end{IEEEeqnarray*}
  where $\code{D}$ is an $[n,s]$ subcode of $\code{C}$, $s\in\Nat{k}$, and $\tilde{\code{D}}$ is an $[n,s]$ subcode of
  $\tilde{\code{C}}$, $s\in\Nat{\tilde{k}}$.
\end{IEEEproof}

It can be seen from the proof above that we can only have $\ecard{\tilde{\set{I}}\cap\chi(\code{D})}\geq 0$ for an
information set $\tilde{\set{I}}$ of $\tilde{\code{C}}$ and a subcode $\code{D}\subseteq\code{C}$ (or
$\ecard{\set{I}\cap\chi(\tilde{\code{D}})}\geq 0$ for an information set $\set{I}$ of $\code{C}$ and a subcode
$\tilde{\code{D}}\subseteq\tilde{\code{C}}$). Hence, unlike in Conjecture~\ref{conj:general-d_MDS-PIRcapacity-achiving-codes}, we
do not conjecture this necessary condition to be sufficient.

Similar to Theorem~\ref{thm:sufficient_MDS-PIRcapacity-achieving-codes} for the noncolluding case, we provide a
sufficient condition for codes to achieve the maximum possible PIR rate of Protocol~3 by using code automorphisms of
$\code{C}$ and $\tilde{\code{C}}$.
\begin{theorem}
  \label{thm:sufficient_PIRmaxRte}
  Let $\code{C}$ be an $[n,k]$ code, $\bar{\code{C}}$ an $[n,\bar{k}]$ code, and
  $\tilde{\code{C}}=\code{C}\circ\bar{\code{C}}$. If there exist $k$ information sets
  $\tilde{\set{I}}_1,\ldots,\tilde{\set{I}}_k$ of $\tilde{\code{C}}$, an information set $\set{I}$ of $\code{C}$, and
  $n-\tilde{k}$ distinct automorphisms of $\code{C}$ such that for every code coordinate $j_i\in\set{I}$, $i\in\Nat{k}$, 
  \begin{IEEEeqnarray*}{rCl}
    \tilde{\set{I}}_i\cup\{\pi_1(j_i),\ldots,\pi_{n-\tilde{k}}(j_i)\bigr\}=\{1,2,\ldots,n\},
  \end{IEEEeqnarray*}
  then the codes $\code{C}$ and $\Cbar$ achieve the maximum possible PIR rate of Protocol~3, i.e.,
  $\const{R}_{\mathsf{UB}}$.
\end{theorem}
\begin{IEEEproof}
  Since there exist $n-\tilde{k}$ distinct automorphisms of $\code{C}$ such that
  $\set{I}_j\eqdef\{\pi_j(j_i)\colon j_i\in\set{I}\}$, $j\in\Nat{n-\tilde{k}}$,  are information sets of $\code{C}$, and
  for every code coordinate $j_i\in\set{I}$, $i\in\Nat{k}$,
  \begin{IEEEeqnarray*}{rCl}
    \tilde{\set{I}}_i\cup\{\pi_1(j_i),\ldots,\pi_{n-\tilde{k}}(j_i)\bigr\}=\{1,2,\ldots,n\}, 
  \end{IEEEeqnarray*}
  each code coordinate $h\in\Nat{n}$ appears exactly $k$ times in
  $\{\tilde{\set{I}}_i\}_{i\in\Nat{k}}\cup\{\set{I}_{j}\}_{j\in\Nat{n-\tilde{k}}}$, which shows the existence of a PIR
  maximum rate matrix $\tilde{\mat{\Lambda}}_{k,k+n-\tilde{k}}$ for Protocol~3.
\end{IEEEproof}

We now show that  RM codes achieve the maximum PIR rate of {\mg Protocol~3}.
\begin{corollary}
  \label{cor:PIRmaxRate_RMcodes}
  Let $\code{C}$ be an $[n,k]$ RM code $\code{R}(v,m)$, $\bar{\code{C}}$ an $[n,\bar{k}]$ RM code $\code{R}(\bar{v},m)$,
  and $\tilde{k}=k+\bar{k}$, where $n=2^m$, $k=\sum_{i=0}^v\binom{m}{i}$, and
  $\bar{k}=\sum_{i=0}^{\bar{v}}\binom{m}{i}$. Then, a PIR maximum rate matrix $\tilde{\mat{\Lambda}}_{k,k+n-\tilde{k}}$
  exists for {\mg Protocol~3}, and its PIR rate is
  \begin{IEEEeqnarray*}{rCl}
  \const{R}_{\mathsf{P3}}(\C,\Cbar) =\frac{n-\tilde{k}}{n}=\Rub.
  \end{IEEEeqnarray*}
\end{corollary}

\begin{IEEEproof}
  It can be easily shown that $\tilde{\code{C}}=\code{C}\circ\bar{\code{C}}$ is an RM code $\code{R}(\tilde{v},m)$ with
  $\tilde{v}=v+\bar{v}$. Consider two information sets 
  $\mathcal I$ and $\tilde{\mathcal I}$ 
  of $\code{C}$ and $\tilde{\code{C}}$, respectively. (Lemma~\ref{lem:det-infoSets_RMcode} gives one way to construct
  these two information sets.) We construct the $k+n-\tilde{k}$ information sets
  \begin{IEEEeqnarray*}{rCll}
    \tilde{\set{I}}_i& \eqdef &\{\vect{\sigma}+\vect{\mu}_i\colon\vect{\sigma}\in\tilde{\set{I}}\},\quad&i\in\Nat{k},
    \\
    \set{I}_j& \eqdef &\{\vect{\mu}+\bar{\vect{\sigma}}_j\colon\vect{\mu}\in\set{I}\}, \quad&j\in\Nat{n-\tilde{k}},
  \end{IEEEeqnarray*}
  for $\tilde{\code{C}}$ and $\code{C}$, respectively, 
  where $\{\vect{\mu}_i\}_{i\in\Nat{k}}$ and $\{\bar{\vect{\sigma}}_j\}_{j\in\Nat{n-\tilde{k}}}$ are the numbered binary
  $m$-tuples in $\set{I}$ and $\GF(2)^{m\times 1}\setminus\tilde{\set{I}}$, respectively.

  Without loss of generality, the $i$-th information set $\tilde{\set{I}}_i$, $i\in\Nat{k}$, can be written as
  \begin{IEEEeqnarray*}{rCl}
    \tilde{\set{I}}_i=\{\vect{\mu}_i+\vect{\sigma}_1,\ldots,\vect{\mu}_i+\vect{\sigma}_{\tilde{k}}\},
  \end{IEEEeqnarray*}
  where $\vect{\sigma}_{j'}\in\tilde{\set{I}}$, $j'\in\Nat{\tilde{k}}$. Furthermore, consider the $i$-th elements across
  all sets $\set{I}_j$, $j\in\Nat{n-\tilde{k}}$. They have the form 
    $\vect{\mu}_i+\bar{\vect{\sigma}}_j$,	
  where $\vect{\mu}_i\in\set{I}$. Since $\bar{\vect{\sigma}}_j\in\GF(2)^{m\times 1}\setminus\tilde{\set{I}}$ and
  $\vect{\sigma}_i\in\tilde{\set{I}}$, the set 
  \begin{IEEEeqnarray*}{rCl}
    \{\vect{\mu}_i+\vect{\sigma}_1,\ldots,\vect{\mu}_i+\vect{\sigma}_{\tilde{k}}\}
    \cup\{\vect{\mu}_i+\bar{\vect{\sigma}}_1,\ldots,\vect{\mu}_i+\bar{\vect{\sigma}}_{n-\tilde k}\}
  \end{IEEEeqnarray*}
  with cardinality $n=2^m$ is equal to $\GF(2)^{m\times 1}$, i.e., the set containing the elements of the $i$-th
  information set $\tilde{\mathcal I}_i$ and the $i$-th elements $\vect{\mu}_i+\bar{\vect{\sigma}}_j$ in all sets
  $\set{I}_j$ is equal to the set of all binary $n=2^m$ tuples. Therefore, we are able to find $k$ information sets
  $\{\tilde{\set{I}}_i\}_{i\in\Nat{k}}$ of $\tilde{\code{C}}$, an information set $\set{I}$ of $\code{C}$, and
  $n-\tilde{k}$ distinct automorphisms $\pi_{j}(\vect{\mu})=\vect{\mu}+\bar{\vect{\sigma}}_j$ of $\code{C}$,
  $j\in\Nat{n-\tilde{k}}$, satisfying Theorem~\ref{thm:sufficient_PIRmaxRte}. This completes the proof.
\end{IEEEproof}

We remark again that Corollary~\ref{cor:PIRmaxRate_RMcodes} can be extended to nonbinary generalized RM codes. Finally,
note that in the independent work \cite{FreijHollantiGnilkeHollantiHorlemannKarpukKubjas19_1app} it was also shown that
RM codes can achieve the maximum possible PIR rate of the $(\code{C},\Cbar)$-retrieval protocol, i.e.,
$\const{R}_{\mathsf{UB}}$, for transitive codes. However, it is important to highlight that our Protocol~3 requires a
much smaller $\beta$ (number of stripes) and a significant smaller $d$ (number of subqueries). Indeed, the protocol in
\cite{FreijHollantiGnilkeHollantiHorlemannKarpukKubjas19_1app} requires very large $\beta$ and $d$ (in the order of
$10000$ for the example provided), and thus our protocol is more practical.

\subsection{Optimizing the PIR rate}
\label{sec:optC}

{\mg For those codes for which we do not have a proof that $\const{R}_{\mathsf{UB}}$ is achieved}, we now provide an
algorithm to optimize the PIR rate $\const{R}_{\mathsf{P3}}(\C,\Cbar)$ for a given storage code $\C$ and query code $\Cbar$ such that it comes closer to the upper bound
$\const{R}_{\mathsf{UB}}$. The algorithm is identical to Algorithm~\ref{alg:cPoP} for the case of noncolluding nodes with some
key differences which we highlight below.
\begin{itemize}
\item In Line~\ref{algcPoP:Init}, $\Gamma$ is initialized to $1$.
\item The while loop in Line~\ref{algcPoP:outerWhile} runs up to $n-\tilde{k}$.
\item The first argument to the subprocedure \texttt{ComputeErasurePatternList}($\cdot,\Gamma$) is changed from
  $\code{C}$ to $\mathcal C\circ\bar{\mathcal C}$ in Line~\ref{algcPoP:subprocedure1}.
\end{itemize}

With these minor modifications, Algorithm~\ref{alg:cPoP} can be used to optimize the PIR rate in the case of $T$
colluding nodes. Numerical results are presented below in Section~\ref{sec:results}. Note that $\Gamma$ is initialized
to $1$ as opposed to $\min(k,d_\mathsf{min}^{\code{C}}-1)$ in the case of noncolluding nodes. This is because
$\hat{\bm E}$ and $\bar{\bm E}$ of $\bm E$ are based on different codes. This also guarantees that the algorithm always
returns $\bm E_{\mathsf{opt}} \neq \emptyset$ (assuming $d_{\mathsf{min}}^{\tilde{\code{C}}} \geq 2$), since in this
case all weight-$1$ erasure patterns are correctable by $\tilde{\code{C}}$, and a valid matrix $\bm E$ can be trivially
constructed.

\section{Numerical Results}
\label{sec:results}

In this section, we present maximized PIR rates for the PIR protocols described in
Sections~\ref{sec:file-dep-PIR}, \ref{sec:file-indep-PIR}, and \ref{sec:MultipleCollNodePIR}. Unless specified otherwise, these protocols are
optimized using Algorithm~\ref{alg:cPoP} with minimum possible values for the parameters $\beta$ and $d$ as given in \eqref{Eq:
  Beta_and_d_DEF}.
In contrast to Sections~\ref{sec:MDS-PIRcapacity-achiving-codes} to \ref{sec: Ematrix_LRC}, where different classes of
codes were proved to be MDS-PIR capacity-achieving, we consider here other codes (with two exceptions as detailed below)
and their highest possible PIR rates.  The results are tabulated in
Tables~\ref{table_of_codes_t1_above12} and \ref{table_of_codes_t1_below12} for the case of noncolluding nodes, and in
Table~\ref{table_of_codes_colluding_case} for the colluding case.  Results in Table~\ref{table_of_codes_t1_above12} are for code
rates strictly larger than $1/2$, while codes of rate at most $1/2$ are tabulated in Table~\ref{table_of_codes_t1_below12}.

In Tables~\ref{table_of_codes_t1_above12} and \ref{table_of_codes_t1_below12}, $\const{C}_{\infty}$ (see \eqref{eq:PIRasympt-capacity})
is the asymptotic MDS-PIR capacity
and $\const{R}_\mathsf{opt}$ is the optimized PIR rate computed from Algorithm~\ref{alg:cPoP}. In
Table~\ref{table_of_codes_t1_above12}, $\const{R}_\mathsf{non-opt}=(d^{\mathcal{C}'}_\mathsf{min}-1)/n$, while in
Table~\ref{table_of_codes_t1_below12}, $\const{R}_\mathsf{non-opt}=(d^{\mathcal{C}}_\mathsf{min}-1)/n$. In
Table~\ref{table_of_codes_colluding_case}, $\const{C}_{\mathsf{LB},\infty}\eqdef (n-(k+T-1))/n$ is a lower bound (taken
from \cite{FreijHollantiGnilkeHollantiKarpuk17_1}) on the asymptotic MDS-PIR capacity in the case of at most $T$
colluding nodes, while $\const{R}_\mathsf{opt}$ is the optimized PIR rate computed from Algorithm~\ref{alg:cPoP} and
$\const{R}_\mathsf{non-opt}=(d^{\tilde{\mathcal{C}}}_\mathsf{min}-1)/n$.
  
  \setlength{\textfloatsep}{18.60004pt plus 2.39996pt minus 4.79993pt}
\begin{table}[!t]
     \caption{Optimized values for the PIR rate for different codes having code rates strictly larger than $1/2$ for the case of noncolluding nodes.}
    \label{table_of_codes_t1_above12}
    \centering
    \def\Hline{\noalign{\hrule height 2\arrayrulewidth}}
    \vskip -2.0ex 
    \begin{tabular}{@{}lccccc@{}}
        \Hline \\ [-2.0ex]
        Code & $d_\mathsf{min}^{\code{C}}$ & $d_\mathsf{min}^{\code{C}'}$ & $\const{R}_\mathsf{non-opt}$ & $\const{R}_\mathsf{opt}$ & $\const{C}_{\infty}$ \\[0.25ex]
        \hline
        \\ [-2.0ex] \hline  \\ [-2.0ex]
        $\mathcal C_1:[5,3]$ (Example~\ref{ex:5_3_code}) & $2$ & $3$ &  $0.4$ & $0.4$ & $0.4$ \\
        $\mathcal C_2:[11,6]$  & $4$ & $4$ & $0.2727$ & $0.4545$ & $0.4545$ \\
        $\mathcal C_3:[12,8]$ Pyramid & $4$ & $4$ & $0.25$ & $0.3333$ & $0.3333$ \\
        $\mathcal C_4:[18,12]$ Pyramid & $5$ & $5$ & $0.2222$ & $0.3333$ & $0.3333$ \\
        $\mathcal C_5:[16,10]$ LRC & $5$ & $5$ & $0.25$ & $ 0.3750$ & $0.3750$ \\
        
        $\mathcal C_6:[154,121]$ LRC & $4$ & $6$ & $0.0325$ & $0.2013$ & $0.2143$ \\
      $\mathcal C_7:[187,121]$ LRC & $7$ & $16$ & $0.0802$ & $0.3262$ & $0.3529$ \\
      \hline
    \end{tabular}
\end{table}

\begin{table}[!t]
     \caption{Optimized values for the PIR rate for different codes having code rates at most $1/2$ for the case of noncolluding nodes.}
    \label{table_of_codes_t1_below12}
    \centering
    \def\Hline{\noalign{\hrule height 2\arrayrulewidth}}
    \vskip -2.0ex 
    \begin{tabular}{@{}lcccc@{}}
        \Hline \\ [-2.0ex]
        Code & $d_\mathsf{min}^{\code{C}}$ & $\const{R}_{\mathsf{non-opt}}$ & $\const{R}_{\mathsf{opt}}$ & $\const{C}_{\infty}$ \\[0.25ex]
        \hline
        \\ [-2.0ex] \hline  \\ [-2.0ex]
        $\mathcal C_8: [7,3]$ (Example~\ref{ex: 7_3_code}) & 4 & 0.4286 & 0.5714 & 0.5714\\
        $\mathcal C_{9}: [9,4]$ LRC (\hspace{-0.05ex}\cite[Ex.~1]{tam14})   & 5 & 0.4444 & 0.5555 & 0.5555\\
        $\mathcal C_{10}: [12,6]$ LRC (\hspace{-0.05ex}\cite[Ex.~2]{tam14}) & 6  & 0.4167 & 0.5 & 0.5\\
        \hline
    \end{tabular}
\end{table}

The code $\mathcal C_1$ in Table~\ref{table_of_codes_t1_above12} is from Example~\ref{ex:5_3_code}, $\mathcal C_2$ is an $[11,6]$
binary linear code with optimum minimum Hamming distance, while codes $\mathcal C_3$ and $\mathcal C_4$ are Pyramid codes, taken
from \cite{Hua07}, of locality $4$ and $6$, respectively, 
$\mathcal C_5$ is an LRC of locality $5$ borrowed from \cite{Sat13}. In \cite{hao16}, a construction of optimal (in
terms of minimum Hamming distance) binary LRCs with multiple repair groups was given. In particular, in \cite[Constr.~3]{hao16},
a construction based on array LDPC codes was provided. 
The minimum Hamming distance of array LDPC codes is known for certain sets of parameters (see, e.g., \cite{ros14} and
references therein). Codes $\mathcal{C}_6$ and $\mathcal{C}_7$ in Table~\ref{table_of_codes_t1_above12} are \emph{optimal}
LRCs based on array LDPC codes constructed using \cite[Constr.~3]{hao16} and having information locality $11$. The protocols for these two underlying codes have $\beta=\Gamma$ and $d=k$.

\begin{table*}[!t]
     \caption{Optimized values for the PIR rate for different codes for the colluding case with $T=2$ and $T=3$.}
    \label{table_of_codes_colluding_case}
    \centering
    \def\Hline{\noalign{\hrule height 2\arrayrulewidth}}
    \vskip -2.0ex 
    \begin{tabular}{@{}lllcccccc@{}}
      \Hline \\ [-2.0ex]
      Code $\code{C}$ & $\bar{\code{C}}$ & $\tilde{\code{C}}$ & $d_\mathsf{min}^{\code{C}}$ &  $T$ & $\const{R}_{\mathsf{non-opt}}$ & $\const{R}_{\mathsf{opt}}$ & $\const{R}_{\mathsf{UB}}$ & $\const{C}_{\mathsf{LB},\infty}$ \\[0.25ex]
      \hline
      \\ [-2.0ex] \hline  \\ [-2.0ex]
      $\mathcal C_{9}: [9,4]$ LRC (\hspace{-0.05ex}\cite[Ex.~1]{tam14})  & ${\rm {RS}}[9,2]$  & ${\rm {RS}}[9,6]$ & 5  & 2 & 0.3333  & 0.3333 & 0.3333 & 0.4444\\
      $\mathcal C_{10}: [12,6]$ LRC (\hspace{-0.05ex}\cite[Ex.~2]{tam14})  & ${\rm {RS}}[12,2]$ & ${\rm {RS}}[12,8]$ & 6  &2 & 0.3333  & 0.3333 & 0.3333 & 0.4167\\
      $\mathcal C_{11}: [12,4]$ (Section~\ref{sec:ColludingCase})  & $\code{C}$  &$[12,10,2] $ & 6 &2  &0.0833  &0.1667 & 0.1667 & 0.5833\\
      $\mathcal C_{12}: [12,4]$ LRC (\hspace{-0.05ex}\cite[Ex.~5]{tam14})  & ${\rm {RS}}[12,2]$  &$[12,7,5] $ & 6  &2 & 0.3333 &0.4167 & 0.4167 & 0.5833\\
      $\mathcal C_{13}: [26,9]$  $(\mathcal{U} \mid \mathcal{U}+\mathcal{V})$ & $\code{C}$  &$[26,22,1] $ & 8  &3 & 0 & 0.1538 & 0.1538 & 0.5769 \\
      $\mathcal C_{14}: [32,6]$  $(\mathcal{U} \mid \mathcal{U}+\mathcal{V})$ & $\code{C}$  &$[32,16,8] $ & 16  &3 & 0.2188 &0.5 & 0.5 & 0.75 \\
      \hline
    \end{tabular}
\end{table*}

Code $\code{C}_8$ in Table~\ref{table_of_codes_t1_below12} is the dual code of the $[7,4,3]$ Hamming code and is taken from Example~\ref{ex: 7_3_code}, while the codes $\code{C}_{9}$ and $\code{C}_{10}$ are $d_\mathsf{min}$-optimal LRCs over $\GF(13)$ of all-symbol locality $2$ and $3$, respectively, taken from \cite{tam14} (see Examples~1 and 2, respectively,  in \cite{tam14}). 
%
%
%
These two codes are also tabulated in Table~\ref{table_of_codes_colluding_case}. The corresponding $\bar{\code{C}}$
codes are RS codes and their parameters are given in Table~\ref{table_of_codes_colluding_case} (an RS code of length $n$
and dimension $k$ is denoted by ${\rm {RS}}[n,k]$). Code $\code{C}_{11}$ (from
Table~\ref{table_of_codes_colluding_case}) is taken from Section~\ref{sec:ColludingCase}, while code $\code{C}_{12}$
(also from Table~\ref{table_of_codes_colluding_case}) is taken from \cite{tam14} (see Example~5 in \cite{tam14}). Note
that $\code{C}_{12}$ is an LRC of length $12$ over $\GF(13)$ with two disjoint recovering sets of sizes $2$ and $3$,
respectively, for every symbol of the code (all-symbol locality). Code $\code{C}_{13}$ (from
Table~\ref{table_of_codes_colluding_case}) is a $[26,9,8]$ binary UUV code that is close to an optimal binary linear
code (the best known code for these parameters has a minimum Hamming distance of $9$), while the code $\code{C}_{14}$ is
a UUV code where $\mathcal{U}$ is a $[16,5,8]$ RM code (the code $\mathcal{R}(1,4)$). Note that $\code{C}_{14}$ becomes
the RM code $\mathcal{R}(1,5)$.  Due to the high computational complexity of Algorithm~\ref{alg:cPoP} for
$\code{C}_{14}$, we are unable to compute the maximum rate of Protocol~3 for the minimum values of $\beta$ and
$d$. Instead, we take $\beta=\Gamma$ and $d=k$ and use Corollary~\ref{cor:PIRmaxRate_RMcodes} to obtain the maximum rate
of the protocol.

It is observed in Tables~\ref{table_of_codes_t1_above12} and \ref{table_of_codes_t1_below12} that in the case of
noncolluding nodes, {\mg the optimized PIR rate} $\const{R}_\mathsf{opt}$ is equal to the asymptotic capacity
$\const{C}_{\infty}$ for all tabulated codes except $\mathcal C_6$ and $\mathcal C_7$. {\mg Note that the codes $\C_1$,
  $\C_2$, and $\C_5$--$\C_{10}$ do not fall within the code families that we proved are MDS-PIR capacity-achieving (see
  Section~\ref{sec:MDS-PIRcapacity-achiving-codes}). Thus, the results in Tables~\ref{table_of_codes_t1_above12} and
  \ref{table_of_codes_t1_below12} show that, interestingly, other codes can achieve the asymptotic MDS-PIR capacity as
  well. On the other hand, $\C_3$ and $\C_4$ satisfy the conditions of Theorem~\ref{th: LRCcap_proof}. Thus, they are
  MDS-PIR capacity-achieving {\mg with $\beta=\Gamma$ and $d=k$}. The results in the table show that they also achieve
  $\const{C}_{\infty}$ for $\beta$ and $d$ as in \eqref{Eq: Beta_and_d_DEF}.}  Also, note that by the nature of the
optimization procedure (see Remark~\ref{remark:1}), MDS-PIR capacity-achieving matrices $\mat{\Lambda}_{\kappa,\nu}$ with
$\frac{\kappa}{\nu}=\frac{k}{n}$ of all tabulated codes except $\mathcal C_6$ and $\mathcal C_7$ are found. This implies
that they are also MDS-PIR capacity-achieving codes for any finite number of files and must satisfy the necessary
condition based on generalized Hamming weights in Theorem~\ref{thm:general-d_MDS-PIRcapacity-achieving-codes}. Due to
the high computational complexity of Algorithm~\ref{alg:cPoP}, it is difficult to maximize the PIR rates of
$\mathcal C_6$ and $\mathcal C_7$. Therefore, it is an open problem whether or not they are MDS-PIR capacity-achieving
codes. For the colluding case (see Table~\ref{table_of_codes_colluding_case}) the lower bound
$\const{C}_{\mathsf{LB},\infty}$ on the asymptotic MDS-PIR capacity is not achieved, even after optimization. To the
best of our knowledge, GRS codes are the only known class of codes where this bound is actually achieved
\cite{FreijHollantiGnilkeHollantiKarpuk17_1}. On the other hand, the upper bound $\const{R}_{\mathsf{UB}}$ (from
\eqref{eq:UB_PIR_rate}) is attained in all cases.

\section{Conclusion}
\label{sec:conclusion}

We presented three different PIR protocols, namely Protocol~1, Protocol~2, and Protocol~3, for DSSs where data is stored
using an arbitrary linear code. We first considered the case where no nodes in the DSS collude. Under this scenario,
Protocols~1 and 2 achieve the PIR property. We proved that, for certain non-MDS codes, Protocol~1 achieves the finite
MDS-PIR capacity (and also the asymptotic MDS-PIR capacity) and Protocol~2, which is a much simpler protocol compared to
Protocol~1, achieves the asymptotic MDS-PIR capacity. Thus, the MDS property is not necessary in order to achieve the
MDS-PIR capacity (both finite and asymptotic). We also provided a necessary and a sufficient condition for codes to be
MDS-PIR capacity-achieving with Protocols~1 and 2. The necessary condition is based on generalized Hamming weights while the sufficient
condition is obtained from code automorphisms of the linear storage code. We proved that cyclic codes, RM codes, and a class of
distance-optimal information locality codes are MDS-PIR capacity-achieving codes. For other codes, we provided an
optimization algorithm that optimizes Protocols~1 and 2 in order to maximize their PIR rates. We also considered the
scenario where a subset of nodes in the DSS collude. For such a scenario, we proposed Protocol~3, which is an
improvement of the PIR protocol by Freij-Hollanti \emph{et al.}. The improvement allows the protocol to achieve 
higher PIR rates, and the PIR rates for non-MDS codes are no longer limited by the minimum Hamming distance of the retrieval
code. Subsequently, we presented an optimization algorithm to optimize the PIR rate of the protocol, and a family of
codes based on the classical $(\mathcal{U} \mid \mathcal{U} + \mathcal{V})$ construction that can be used with this
protocol. Furthermore, as for the noncolluding case, we provided a necessary and a sufficient condition to achieve
  the maximum possible PIR rate of Protocol~3. Moreover, we proved that RM codes satisfy the sufficient condition and
  can achieve much higher PIR rates than previously reported by Freij-Hollanti \emph{et al.}. Finally, we presented some
numerical results on the PIR rates for several linear codes, including distance-optimal all-symbol locality LRCs
constructed by Tamo and Barg.

\appendices

\section{Proof of Lemma~\ref{lem:det-infoSets_RMcode}}
\label{sec:proof_det-infoSets_RMcode}

We need to ensure that given a $k\times n$ generator matrix $\mat{G}$ of $\code{R}(v,m)$ with
$k=\sum_{i=0}^v\binom{m}{i}$ and $n=2^m$, the $k\times k$ matrix $\mat{G}|_{\set{I}}$ that comprises the columns of the
generator matrix indexed by the coordinates of $\set{I}$ is invertible. We are going to elaborate on this by considering
all the monomials $z_1^{\mu_1}\cdots z_m^{\mu_m}$, $\mu_i\in\GF(2)$, in a so-called \emph{graded lexicographic order},
where each vector $\trans{\vect{\mu}=(\mu_1,\ldots,\mu_m)} \in\GF(2)^{m\times 1}$ defines a column of the generator
matrix $\mat{G}$ according to \eqref{eq:transformation}. Formally speaking, denote $z_1^{\mu_1}\cdots z_m^{\mu_m}$ by
$\vect{z}^{\vect{\mu}}$. We say $\vect{z}^{\vect{\mu}}\prec\vect{z}^{\vect{\mu}'}$ either if
$\Hwt{\vect{\mu}}<\Hwt{\vect{\mu}'}$ or if $\Hwt{\vect{\mu}}=\Hwt{\vect{\mu}'}$ and the topmost nonzero entry of
$\vect{\mu}-\vect{\mu}'$ (subtraction is over the reals) is positive. For instance, in graded lexicographical ordering we have
$z_1\prec z_2\prec z_3\prec z_1z_2\prec z_1z_3\prec z_2z_3\prec z_1z_2z_3$ for $m=3$.

Now we are ready for the proof. It is noted that a basis of $\code{R}(v,m)$ can be viewed as
$\set{B}\eqdef\{1,z_1,z_2,z_3,\ldots\}=\{\vect{z}^{\vect{\mu}'}\colon \Hwt{\vect{\mu}'}\leq v\}$. Let us list the
monomials in $\set{B}$ in graded lexicographical order, and let the $\ell$-th monomial $f_\ell(\vect{z})$ of the ordered
list represent the $\ell$-th row of $\mat{G}$, $\ell\in\Nat{k}$. According to the generator matrix construction of
$\code{R}(v,m)$, it is known that the $(\ell,\vect{\mu})$ entry of $\mat{G}$ is equal to the value of the $\ell$-th
monomial $f_\ell(\vect{z})$ at $\vect{z}=\vect{\mu}$ \cite[Ch.~13]{MacWilliamsSloane77_1}. Furthermore, given a column
coordinate $\vect{\mu}\in\set{I}$, for $\ell\in\Nat{k}$, we have
\begin{IEEEeqnarray*}{rCl}
  f_{\ell}(\vect{\mu})=
  \begin{cases}
    1 & \text{if } \vect{z}^{\vect{\mu}}=f_{\ell}(\vect{z}),
    \\
    0 & \text{if } \vect{z}^{\vect{\mu}}\prec f_{\ell}(\vect{z}).
  \end{cases}
\end{IEEEeqnarray*}
Thus, the $(\vect{z}^{\vect{\mu}},\vect{\mu})$ entry can be seen as a pivot of $\mat{G}|_{\set{I}}$ and
$\mat{G}|_{\set{I}}$ is obviously invertible.

\section{Proof of Theorem~\ref{thm:PIRachievable-rate_code}}
\label{sec:achievablity-proof}

The proof is completed by showing that the following statements are true.
\begin{description}[leftmargin=0cm]
\item[File symmetry within each storage node.] For all repetitions, we investigate file symmetry for every possible
  combination of files in each round within each storage node.  In the first round ($\ell=1$) of all $\kappa$
  repetitions, it follows from \eqref{eq:undesired-files_l} that, for each $m'\in\Nat{2:f}$, the downloaded number of
  undesired symbols $y^{(m')}_{s,j}$ is equal to $\kappa\const{U}(1)=\kappa^{f-1}$, while for the desired symbols, from
  \eqref{eq:desired-files_1}, it follows that the user requests $\kappa^{f-1}$ code symbols $y^{(1)}_{s,j}$. In the
  $(\ell+1)\text{-th}$ round of all $\kappa$ repetitions, $\ell\in\Nat{f-1}$, arbitrarily choose a combination of files indexed by
  $\set{M}\subseteq\Nat{2:f}$, where $\card{\set{M}}=\ell$. It follows from \eqref{eq:desired-files_l} that the total
  number of requested desired symbols for files pertaining to $\{1\}\cup\set{M}$ is equal to
  \begin{IEEEeqnarray*}{rCl}
    \IEEEeqnarraymulticol{3}{l}{%
      (\nu-\kappa)\bigl[\bigl(\const{U}(\ell)-1\bigr)-\const{U}(\ell-1)+1\bigr]}\nonumber\\*\quad%
    & = &
    (\nu-\kappa)\kappa^{f-(\ell+1)}(\nu-\kappa)^{\ell-1}=
    \kappa^{f-(\ell+1)}(\nu-\kappa)^{\ell}.
  \end{IEEEeqnarray*}
  On the other hand, for the undesired symbols, it follows from \eqref{eq:undesired-files_l} that in the
  $(\ell+1)\text{-th}$ round the user requests
  \begin{IEEEeqnarray*}{rCl}
    \IEEEeqnarraymulticol{3}{l}{%
      \kappa\bigl[\bigl(\const{U}(\ell+1)-1\bigr)-\const{U}(\ell)+1\bigr]}\nonumber\\*\quad%
    & = &
    \kappa\kappa^{f-(\ell+2)}(\nu-\kappa)^{\ell}
    =\kappa^{f-(\ell+1)}(\nu-\kappa)^{\ell}
  \end{IEEEeqnarray*}
  linear sums for a combination of files indexed by $\set{M}\subseteq\Nat{2:f}$, $\card{\set{M}}=\ell+1$. Thus, in rounds
  $\Nat{f-1}$, an equal number of linear sums for all combinations of files indexed by $\mathcal M\subseteq\Nat{f}$ are downloaded. By
  construction, these are linear sums of unique code symbols pertaining to $f$ files. Thus, symmetry in all
  $f-1$ rounds is ensured. In the $f$-th round, only desired symbols are downloaded. Since each desired symbol is a linear
  combination of code symbols from all $f$ files, an equal number of linear sums is again downloaded for the combination of files indexed by $\Nat{f}$. Therefore, symmetry within each node and in each round is ensured.
  
\item[The $\beta\times k$ file $\mat{X}^{(1)}$ can be reliably decoded.] In the first round ($\ell=1$) of all $\kappa$
  repetitions, $\forall\,s\in\Nat{\kappa^{f-1}}$, the user has downloaded the matrix
  \begin{IEEEeqnarray*}{rCl}
    \begin{pmatrix}
      y^{(1)}_{\kappa^{f-1}({\r a_{1,1}}-1)+s,1}&\cdots &
      y^{(1)}_{\kappa^{f-1}({\r a_{1,n}}-1)+s,n}
      \\
      \vdots&\cdots &\vdots
      \\
      y^{(1)}_{\kappa^{f-1}({\r a_{\kappa,1}}-1)+s,1}&\cdots& y^{(1)}_{\kappa^{f-1}({\r a_{\kappa,n}}-1)+s,n}
    \end{pmatrix}
  \end{IEEEeqnarray*}
  of code symbols. Given an $a\in\Nat{\nu}$, recalling Definitions~\ref{def:PIRinterference-matrices} and \ref{def:aSet_A}, it
  follows that for each $s\in\Nat{\kappa^{f-1}}$, the coordinate set
  $\set{S}\bigl(\kappa^{f-1}(a-1)+s\big |\kappa^{f-1}(\mat{A}_{\kappa\times n}-\mat{1}_{\kappa\times
    n})+s\mat{1}_{\kappa\times n}\bigr)$ contains an information set. Hence, the
  $(\kappa^{f-1}(a-1)+1)\text{-th},\ldots,(\kappa^{f-1}(a-1)+\kappa^{f-1})\text{-th}$ stripes are recovered. Since
  $a_{i,j} \in \Nat{\nu}$, we know until now that the user has obtained the
  $1\text{-st},2\text{-nd},\ldots,(\kappa^{f-1}(\nu-1)+\kappa^{f-1})\text{-th}$ stripes. Note that
  $\kappa^{f-1}(\nu-1)+\kappa^{f-1}=\const{D}(0)\nu$. Moreover, owing to \eqref{eq:desired-interference}, in the
  $(\ell'=\ell+1)\text{-th}$ round of all $\kappa$ repetitions with $\ell\in\Nat{f-1}$,
  $\forall\,s\in\Nat{\const{D}(\ell-1):(\const{D}(\ell)-1)}$ the matrices
  \begin{IEEEeqnarray*}{rCl}
    \begin{pmatrix}
      y^{(1)}_{s\cdot\nu+{\r a_{1,1}},1}&\cdots &y^{(1)}_{s\cdot\nu+{\r a_{1,n}},n}
      \\
      \vdots &\cdots &\vdots
      \\
      y^{(1)}_{s\cdot\nu+{\r a_{\kappa,1}},1}&\cdots &y^{(1)}_{s\cdot\nu+{\r a_{\kappa,n}},n}\\
    \end{pmatrix}
  \end{IEEEeqnarray*}
  of code symbols are downloaded. Similarly, fix an $s\in\Nat{\const{D}(\ell-1):(\const{D}(\ell)-1)}$. Then,
  $\forall\,a\in\Nat{\nu}$, the coordinate set
  $\set{S}\bigl(s\nu+a\big |s\nu\mat{1}_{\kappa\times n}+\mat{A}_{\kappa\times n}\bigr)$ must contain an information
  set, and the user can recover the $(s\nu+1)\text{-th},\ldots,(s\nu+\nu)\text{-th}$ stripes. Observe that in the last
  $(\ell'=(f-1)+1)\text{-th}$ round, the row index of the last recovered stripe is equal to
  $(\const{D}(f-1)-1)\nu+\nu$. Hence, the total number of stripes the user has recovered is
  \begin{IEEEeqnarray*}{rCl}
    \IEEEeqnarraymulticol{3}{l}{%
      \bigl(\const{D}(f-1)-1\bigr)\nu+\nu}\nonumber\\*\quad%
    & = &\left[\sum_{\ell=0}^{f-1}\binom{f-1}{\ell}
    \kappa^{f-(\ell+1)}(\nu-\kappa)^{\ell}-1\right]\nu+\nu
    \\
    & = &(\nu^{f-1}-1)\nu+\nu=\nu^f.
  \end{IEEEeqnarray*}
  This indicates that the user has recovered all $\nu^f$ stripes for $\mat{X}^{(1)}$, and $\mat{X}^{(1)}$ is in fact
  reliably reconstructed.
  
\item[The PIR achievable rate is expressed as \eqref{eq:PIRachievable-rate_code}.] According to
  \eqref{eq:undesired-files_l}, since there are $\binom{f-1}{\ell}$ combinations of files other than the first file with
  index $m=1$, the user has downloaded
  \begin{IEEEeqnarray*}{rCl}
    \IEEEeqnarraymulticol{3}{l}{%
      \kappa\binom{f-1}{\ell}\bigl[\const{U}(\ell)-1-\const{U}(\ell-1)+1\bigr] }\nonumber\\*\quad%
    & = &
    \kappa\binom{f-1}{\ell}\kappa^{f-(\ell+1)}(\nu-\kappa)^{\ell-1}\nonumber\\
    & = & \binom{f-1}{\ell}\kappa^{f-\ell}(\nu-\kappa)^{\ell-1}
  \end{IEEEeqnarray*}
  undesired symbols from each storage node in the $\ell\text{-th}$ round, $\ell\in\Nat{f-1}$, of each repetition. Moreover, from
  \eqref{eq:desired-files_1} and \eqref{eq:desired-files_l}, the user has downloaded $\kappa^{f-1}$ desired symbols from
  each storage node in round $\ell=1$ of each repetition, and
  \begin{IEEEeqnarray*}{c}
    \const{D}(\ell)-1-\const{D}(\ell-1)+1 =\binom{f-1}{\ell}\kappa^{f-(\ell+1)}(\nu-\kappa)^{\ell}
  \end{IEEEeqnarray*}
  extra desired symbols from each storage node in the $(\ell+1)\text{-th}$ round, $\ell\in\Nat{f-1}$, of each repetition. In summary, the
  total download cost for Protocol~1 using $\mat{\Lambda}_{\kappa,\nu}(\code{C})$ is equal to
  \begin{IEEEeqnarray*}{rCl}
    nd& = &\text{total number of undesired symbols}\nonumber\\
    && \>+\text{ total number of desired symbols}
    \\
    & = &\kappa n\sum_{\ell=1}^{f-1}\binom{f-1}{\ell}
    \kappa^{f-\ell}(\nu-\kappa)^{\ell-1}\nonumber\\
    && \>+\kappa n\sum_{\ell=0}^{f-1}\binom{f-1}{\ell}
    \kappa^{f-(\ell+1)}(\nu-\kappa)^{\ell}
    \\
    & = &\kappa n\Biggl[\frac{\kappa}{\nu-\kappa}\sum_{\ell=1}^{f-1}\binom{f-1}{\ell}
    \kappa^{f-(\ell+1)}(\nu-\kappa)^{\ell}\nonumber\\
    &&\qquad \>+\sum_{\ell=0}^{f-1}\binom{f-1}{\ell}
    \kappa^{f-(\ell+1)}(\nu-\kappa)^{\ell}\Biggr]
    \\
    & = &\kappa n\left[\frac{\kappa}{\nu-\kappa}(\nu^{f-1}-\kappa^{f-1})+\nu^{f-1}\right]
    \\
    & = &\frac{\kappa n}{\nu-\kappa}\Bigl[\kappa\nu^{f-1}-\kappa^f+\nu^f-\kappa\nu^{f-1}\Bigr]
    \\
    & = &\frac{\kappa n}{\nu-\kappa}\Bigl[\nu^f-\kappa^f\Bigr].
  \end{IEEEeqnarray*}
  Therefore, the PIR achievable rate $\const{R}(\code{C})$ is given by
  \begin{IEEEeqnarray*}{rCl}
    \const{R}(\code{C})& = &\frac{\beta k}{n d}
    =\frac{\nu^f k}{\frac{\kappa n}{\nu-\kappa}\Bigl[\nu^f-\kappa^f\Bigr]}
    =\frac{\frac{(\nu-\kappa)k}{\kappa n}}{\Bigl[1-\bigl(\frac{\kappa}{\nu}\bigr)^f\Bigr]}.
  \end{IEEEeqnarray*}
\end{description}

\section{Proof of Lemma~\ref{lem:lower-nu_k}}
\label{sec:proof_lower-nu_k}

By setting $\kappa=k$ and using Definition~\ref{def:PIRinterference-matrices}, we will prove the existence of $\mat{\Lambda}_{k,\nu}$ with $\nu=k+\min(k,d^{\code{C}}_\mathsf{min}-1)$. In fact, given an $[n,k,d^{\code{C}}_\mathsf{min}]$ code $\code{C}$,
observe that for an interference matrix $\mat{A}_{k\times n}$ derived from a valid $\mat{\Lambda}_{k,\nu}$,
$\set{S}(a|\mat{A}_{k\times n})$ must contain an information set $\forall\,a\in\Nat{\nu}$. We first choose
$\Gamma=\min(k,d^{\code{C}}_\mathsf{min}-1)$ information sets of $\code{C}$. Note that since every code contains
at least one information set, one can always arbitrarily choose $\Gamma$ information sets even if some of them are
repeatedly chosen. Let us denote the selected information sets by $\set{I}_i$, $i\in\Nat{\Gamma}$, and start to
construct the corresponding matrix $\mat{A}_{k\times n}$ with
\begin{IEEEeqnarray}{rCl}
  a_{i,j}& = &k+i, \text{ if } j\in\set{I}_i,\,i\in\Nat{\Gamma}.
  \label{eq:entries_A1}
\end{IEEEeqnarray}
In this way, $k \Gamma$ entries of $\mat{A}_{k\times n}$ are constructed. Next, denote the remaining nonconstructed
entries in each column of $\mat{A}_{k\times n}$ by
\begin{IEEEeqnarray*}{rCl}
  \set{A}_j\eqdef\Bigl\{a_{i^{(j)}_1,j},\ldots,a_{i^{(j)}_{s(j)},j}\Bigr\},\quad j\in\Nat{n},
\end{IEEEeqnarray*}
where $s(j)\leq k$ is the total number of nonconstructed entries in each column. Hence, there are in total
$kn-k\Gamma=k(n-\Gamma)$ nonconstructed entries as follows,
\begin{IEEEeqnarray}{rCl}
  \Bigl\{a_{i^{(1)}_1,1},\ldots,a_{i^{(1)}_{s(1)},1},\ldots,a_{i^{(n)}_1,n},\ldots,a_{i^{(n)}_{s(n)},n}\Bigr\}.
  \label{eq:remaining-entries_A}
\end{IEEEeqnarray}
If we consecutively assign $1,\ldots,k$ to the entries of $\mat{A}_{k\times n}$ in \eqref{eq:remaining-entries_A}
and repeat this process $n-\Gamma$ times, the remaining $k(n-\Gamma)$ entries of $\mat{A}_{k\times n}$ will certainly be
constructed. Note that since we consecutively assign values of $\Nat{k}$ and the largest number of empty entries of each column of 
$\mat{A}_{k\times n}$ is $k$, it is impossible to have repeated values of $\Nat{k}$ in each column of the constructed
$\mat{A}_{k\times n}$. From \eqref{eq:entries_A1} and \eqref{eq:remaining-entries_A}, it can be seen that each
$a\in\Nat{k}$ occurs in $n-\Gamma$ columns of $\bm A_{k\times n}$. From Proposition~\ref{prop:infoS_nkd}, we can then
say that the set $\set{S}(a|\bm A_{k\times n})$ of cardinality $n-\Gamma\geq n-(d^{\code{C}}_{\mathsf{min}}-1)$ contains an information
set. For the remaining $a\in\Nat{k+1:k+\Gamma}$, \eqref{eq:entries_A1} ensures that $\set{S}(a|\bm A_{k\times n})$ contains an
information set. Thus, this procedure will result in a valid PIR interference matrix $\mat{A}_{k\times n}$. The proof is
then completed, since we can construct a PIR achievable rate matrix $\mat{\Lambda}_{k,k+\Gamma}$ from
$\mat{A}_{k\times n}$.

\section{Proof of Theorem~\ref{th:GenPIR}}
\label{Appendix: Proof1}

Consider the $i$-th subresponse of each response $\bm r_l$. Out of the $n$ subresponses generated from the $n$ storage
nodes, there are $\Gamma$ subresponses originating from a subset of nodes $\mathcal J\subset\Nat{n}$,
$|\mathcal J|=\Gamma$, of the form
\begin{align}
  \label{Eq: P1E1}
  r_{l,i}=Y_{l}+c^{(m)}_{s,l},\quad \forall\,l\in\mathcal J,\,s\in\Nat{\beta}.
\end{align}
$Y_{l}$ is referred to as \emph{code interference symbol}. Considering $\bm G^{\mathcal C}=(g_{i',l})$, for
$l\in\Nat{n}$, each code symbol and code interference symbol have the form
\begin{align}
  \label{Eq: P1E1.5}
  c^{(m)}_{s,l}&=\sum_{i'=1}^{k}g_{i',l}x_{s,i'}^{(m)},\\
  Y_{l}&=\sum_{i'=1}^{k}g_{i',l}I_{(i-1)k+i'},\label{Eq: P1E1.6}
\end{align}	
where $x_{s,i'}^{(m)}$ is an information symbol of $\mathcal C$, and
$I_{(i-1)k+i'}=\sum_{m=1}^{f}\sum_{i''=(m-1)\beta+1}^{m\beta} u_{i,i''}x_{i''-(m-1)\beta,i'}^{(m)}$ is an interference
symbol. To obtain $\Gamma$ code symbols from \eqref{Eq: P1E1}, the user requires the knowledge of the code interference symbols $Y_l$. This is obtained
from the remaining $n-\Gamma$ subresponses of the nodes in $\widebar{\mathcal J}\eqdef\Nat{n}\setminus\mathcal J$, which  are
\begin{align}
  \label{Eq: P1E2}
  \begin{split}
    r_{l,i}=Y_{l},\quad\forall\,l\in\widebar{\mathcal J}%
  \end{split}.%
\end{align}%
From \eqref{Eq: P1E1.5} and \eqref{Eq: P1E1.6} we can observe that the interference symbols $Y_l$ have the same form as the code symbols of $\mathcal
C$. Since there are $\Gamma$ unknowns, solving \eqref{Eq: P1E2} resembles ML decoding of the code $\mathcal C$. \eqref{Eq: P1E2}
is a full rank system in the unknowns $I_{(i-1)k+1},\ldots, I_{ik}$ (from the third requirement $\mathsf{C3}$ of $\hat{\bm E}$ in
Section~\ref{sec:file-indep-PIR}) in $\GF(q^\ell)$. Hence, knowing the interference symbols allows the recovery of $\Gamma$
\emph{unique} (from the first requirement $\mathsf{C1}$ for $\hat{\bm E}$ in Section~\ref{sec:file-indep-PIR}) code symbols from the
$i$-th subquery as the user has the knowledge of $Y_l$, $l\in\mathcal J$. In a similar way, from all subqueries, the
user obtains $d\Gamma=\beta k$ \emph{unique} code symbols pertaining to file $\bm X^{(m)}$. %
%
%
%
%
%
These $\beta k$ code symbols are part of $\beta$ information sets (from the second requirement $\mathsf{C2}$ of $\hat{\bm E}$ in
Section~\ref{sec:file-indep-PIR} and \eqref{Eq: IndexDownload}). {\mg Furthermore, since each information set is implicitly
  linked to a unique stripe of the requested file and $s \in \Nat{\beta}$ (see \eqref{Eq: P1E1}) is selected (without
  repetition) from $\mathcal{F}_l$ (see \eqref{Eq: IndexDownload}), $k$ code symbols from each stripe are obtained, and}
the user can recover the whole file $\bm X^{(m)}$, from which it follows that
$\mathsf H\bigl(\bm X^{(m)}|\bm r_1,\ldots, \bm r_n\bigr)=0$.

\section{Proof of Lemma~\ref{lem:infoS-suppD}}
\label{sec:proof_infoS-suppD}

We prove the inequality by using the well known Sylvester's rank inequality:\footnote{The proof of this inequality is
  available in the literature on linear algebra, so here we omit the proof.} If $\mat{U}$ is an $s\times k$ matrix and
$\mat{G}$ is a matrix of size $k\times n$, then
\begin{IEEEeqnarray*}{rCl}
  \rank{\mat{U}\mat{G}}\geq \rank{\mat{U}}+\rank{\mat{G}}-k.
\end{IEEEeqnarray*}

Let $\code{C}$ be an $[n,k]$ code with generator matrix
$\mat{G}$. Given an arbitrary information set $\set{I}$, $\mat{G}|_{\set{I}}$ is by definition
invertible (see Definition~\ref{def:infoS}). We next choose an arbitrary subcode
$\code{D}\subseteq\code{C}$ of dimension $s$ that can be generated by $\mat{U}\mat{G}$
for some $s\times k$ matrix $\mat{U}$ of rank $s$.

Applying Sylvester's rank inequality, we have
\begin{IEEEeqnarray*}{rCl}
  \rank{\mat{U}(\mat{G}|_\set{I})}\geq s+k-k=s.
\end{IEEEeqnarray*}
Because each basis vector of the space $\mat{U}(\mat{G}|_\set{I})$ must at least contain
one nonzero component, this leads to
\begin{IEEEeqnarray*}{rCl}
  \card{\set{I}\cap\chi(\code{D})}=\card{\chi(\code{D}|_{\set{I}})} =\card{\chi(\mat{U}(\mat{G}|_\set{I}))}\geq s,
\end{IEEEeqnarray*}
where $\chi(\code{D})$ is the support of $\code{D}$ (see Definition~\ref{def:suppD}).

\section{Proof of Theorem~\ref{th: LRCcap_proof}}
\label{Appendix: Proof2}
{\mg The proof is a two-step procedure. First, we prove that all rows in $\bm E$ after Step a)
  are correctable by $\mathcal C$. Secondly, we prove that the swaps in certain rows in Step b) ensure that the
  resulting rows are correctable erasure patterns. We start by proving two key lemmas}
(Lemmas~\ref{lem: Capacity_LRC} and \ref{lem: Correctable_e} below), which will form the basis of the overall proof of the theorem.
\begin{lemma}
  \label{lem: Capacity_LRC}
  Let $\mathcal C$ be an $[n,k]$ distance-optimal $(r,\delta)$ information locality code consisting of
  ${L_{\mathsf c}}$ local codes and with parity-check matrix as in \eqref{Eq: H_opt_LRC}. Additionally, it adheres to the
  condition in \eqref{Eq: H_MDS}. Then, $\mathcal C$ can simultaneously correct $\delta-1+\nu_j$ erasures, $\nu_j\geq0$,
  in each local code $\code{C}|_{\set{S}_j}$ provided that the number of global parities available is at least
  $\nu_1+\cdots+\nu_{L_{\mathsf{c}}}$.
\end{lemma}
\begin{IEEEproof}
  We begin by defining $\bm H^{\mathcal C}|_{\mathcal J}^{\mathcal I}$ as the submatrix of $\bm H^{\mathcal C}$
  restricted in columns by the set $\mathcal J$ and in rows by the set $\mathcal I$. For $j\in\Nat{L_{\mathsf c}}$,
  consider the $j$-th local code. Let $\mathcal E_j$ denote the set of coordinates that are erased in the $j$-th local
  code, where $|\mathcal E_j|=\delta-1+\nu_j$. Let
    \begin{align*}
    \mathcal R_j=\{(\delta-1)(j-1)+1,\ldots,(\delta-1)j\}\cup\set{A}_j
  \end{align*}
   be a set of rows of $\bm H^{\mathcal C}$ of cardinality
  $|\mathcal R_j|=|\mathcal E_j|$, where $\set{A}_j\subset\Nat{L_{\mathsf c}(\delta-1)+1:(n-k)}$, $\card{\set{A}_j}=\nu_j$, is a set of rows of $\bm H^{\mathcal C}$ (which correspond to  parity-check equations of the available global parities). In order to
  prove the lemma one needs to prove that
  \begin{align}
    \label{Eq: L2_condition}
    \Bigrank{\bm H^{\mathcal C}\big |_{\cup_j\mathcal E_j}^{\cup_j\mathcal R_j}}
    =\sum_{j=1}^{L_{\mathsf c}}(\delta-1+\nu_j).
  \end{align}
  
  For each $j\in\Nat{L_{\mathsf c}}$ and $j'\neq j$, assume that there exists a set
  $\set{A}_{j'}\subset\Nat{L_{\mathsf c}(\delta-1)+1:(n-k)}$ such that ${\mathcal A}_j\cap{\mathcal
    A}_{j'}=\emptyset$. Then, it follows that $\mathcal R_j\cap\mathcal R_{j'}=\emptyset$, and since
  $\mathcal E_j\cap\mathcal E_{j'}=\emptyset$,
  \begin{align*}
    \Bigrank{\bm H^{\mathcal C}\big |_{\cup_j\mathcal E_j}^{\cup_i\mathcal R_j}}=\sum_{j=1}^{L_{\mathsf c}}
    \bigrank{\bm H^{\mathcal C}|_{\mathcal E_j}^{\mathcal R_j}}.
  \end{align*}
  Thus, to show \eqref{Eq: L2_condition} it is sufficient to show that
  \begin{align}
    \label{Eq: L2ranklocal}
    \bigrank{\bm H^{\mathcal C}|_{\mathcal E_j}^{\mathcal R_j}}=\delta-1+\nu_j
  \end{align}
  for all $j\in\Nat{L_{\mathsf c}}$.
  
  To show this, consider now the $[n',k]$ MDS code $\code{C}'$ whose parity-check matrix is given by
  $\bm H^{\mathsf{MDS}}$ in \eqref{Eq: H_MDS}. Let $\set{S}'_j\subset\Nat{n'}$ denote a set of coordinates of
  $\mathcal C'$ of cardinality $\card{\set{S}'_j}=k+\delta-1+\nu_j$. More
  specifically, 
  \begin{align*}
    \set{S}'_j=\{1,\ldots,k\}\cup\{k+1,\ldots,k+\delta-1\}\cup\set{B}_j,
  \end{align*}
  where $\set{B}_j=\{a-L_\mathsf{c}(\delta-1)+(\delta-1)+k\colon a\in\set{A}_j\} \subset\Nat{\delta+k:n'}$. In other
  words, the set $\set{S}'_j$ consists of $k$ systematic coordinates and $\delta-1+\nu_j$ parity coordinates of
  $\mathcal C'$. The punctured code $\code{C}'_j=\code{C}'|_{\set{S}'_j}$ is defined by a parity-check
  matrix $\mat{H}^{\code{C}_j'}$ of dimensions $(\delta-1+\nu_j)\times(k+\delta-1+\nu_j)$ that is
  a submatrix of $\mat{H}^\mathsf{MDS}$. Since the punctured code of an MDS code is also an MDS code~\cite{Feyling93_1},
  $\code{C}'_j$ has minimum Hamming distance $d_{\mathsf{min}}^{\code{C}'_j}=\delta+\nu_j=\delta+\card{\set{A}_j}$.  Note that
  for some column index set $\mathcal J\subset\Nat{k+\delta-1+\nu_j}$, $|\mathcal J|=|\mathcal E_j|$, one can build
  $\mat{H}^\code{C}|_{\set{E}_j}^{\set{R}_j}=\mat{H}^{\code{C}'_j}|_\set{J}$. From the MDS property, it follows that
\begin{align*}
\bigrank{\mat{H}^{\code{C}}|_{\set{E}_j}^{\set{R}_j}} =\bigrank{\mat{H}^{\code{C}'_j}|_\set{J}}=\delta-1+\nu_j.
\end{align*}

Finally, if the total number of global parities is at least  $\sum_{j=1}^{L_{\mathsf c}} \nu_j$, we can assign to the set $\set{A}_j$, $j\in\Nat{L_{\mathsf c}}$, a set of $\nu_j$ rows of $\mat{H}^{\code{C}}$ corresponding to global parity-checks such that the sets $\set{A}_j$ are all disjoint, hence  \eqref{Eq: L2ranklocal} holds  for all $j\in\Nat{L_{\mathsf c}}$, and 
 \eqref{Eq: L2_condition} follows, which completes the proof.
\end{IEEEproof}

\begin{lemma}
  \label{lem: Correctable_e}
  Consider an erasure pattern $\bm e$ of length $n$ of the form
  \begin{align*}
    \bm e=(e_1,\ldots,e_n)=(\bm e_1,\ldots, \bm e_{L},\bm e_{L+1}),	
  \end{align*}
  where the subvectors $\bm e_1,\ldots, \bm e_{L}$ are all of length $n_\mathsf{c}=r+\delta-1$ and $\bm e_{L+1}$ is of
  length $\bar r=n\bmod n_\mathsf{c}$. Let $\chi{({\bm e}_j)}$, $j\in\Nat{L+1}$, be the support of $\bm e_j$ and
  $t=(n-k)\bmod L$. If $|\chi{({\bm e}_1)}|=\cdots=|\chi{({\bm e}_t)}|=m_1$,
  $|\chi{({\bm e}_{t+1})}|=\cdots=|\chi{({\bm e}_L)}|=m$, and $|\chi{({\bm e}_{L+1})}|=0$, where
  $m=\bigl\lfloor\frac{n-k}{L}\bigr\rfloor$ and $m_1=m+1$, then $\bm e$ is correctable by $\mathcal C$.
\end{lemma}
\begin{IEEEproof}
  The erasure pattern $\bm e$ is divided into $L+1$ partitions represented by
  $\bm e_j=(e_{n_\mathsf{c}(j-1)+1},\ldots, e_{n_\mathsf{c} j})$, $j\in\Nat{L}$, and $\bm e_{L+1}=(e_{n_\mathsf{c}L+1},\ldots, e_{n})$, 
  where 
  $\bm e_{j}$, $j\in\Nat{L_\mathsf{c}}$, corresponds to the coordinates of the $j$-th local code, and
  $\bm e_{L_{\mathsf c}+1},\ldots, \bm e_{L+1}$ correspond to the coordinates of the global parities of $\mathcal C$. 
  
 The set $\chi{({\bm e}_j)}$, $j\in\Nat{L+1}$, is the set of coordinates erased from the $j$-th partition, and we construct the erasure patterns ${\bm e}_j$, $j \in \Nat{L}$,  such that
  $|\chi{({\bm e}_j)}|=\delta-1+\nu_j$ with
  \begin{align*}
    \nu_j=\begin{cases}
      m_1-(\delta-1) & \text{if } j\in\Nat{t},
      \\
      m-(\delta-1) & \text{if } j\in\Nat{t+1:L},
    \end{cases}
  \end{align*}
  where $t=(n-k)\bmod L$, and let $\chi{({\bm e}_{L+1})}=\emptyset$. In other words, we construct the erasure patterns such
  that the erasures are distributed as equally as possible across the first $L$ partitions.
  
  {\mg From Definition~\ref{def: OptLRCs}}, 
  it follows that $n-k\geq(\delta-1)L_\mathsf{c}+(L-L_\mathsf{c})(r+\delta-1) \geq L(\delta-1)$ (where the last
  inequality follows from $L \geq L_\mathsf{c}$), hence $\delta-1$ is an integer satisfying the inequality
  $L(\delta-1)\leq n-k$, and subsequently $\delta-1\leq \frac{n-k}{L}$. The integer $m$ is the largest integer such that
  $m\leq\frac{n-k}{L}$. Therefore, $\delta-1\leq m$.  To show that $\bm e$ is correctable it is enough to show that the
  erasures in the $L_{\mathsf c}$ local codes can be corrected, since in this case we have a nonerased information set
  for $\code{C}$, which allows to correct the remaining erasures in $\bm e$.
 
  From Lemma~\ref{lem: Capacity_LRC}, to correct $\delta-1+\nu_j$ erasures in the $j$-th local code for all
  $j\in\Nat{L_{\mathsf c}}$, the number of global parities available, $ \gamma_{\mathsf{tot}}+\bar r$, must be
  \begin{align}
    \label{Th: Th3_Cond}
    \begin{split}
      \gamma_{\mathsf{tot}}+\bar r&\geq\\
      \sum_{j=1}^{L_\mathsf{c}}\nu_j&=\begin{cases}
        m_1 t+m(L_{\mathsf c}-t)-L_\mathsf{c}(\delta-1) & \text{if }t\leq L_{\mathsf c},
        \\
        m_1 L_{\mathsf c}-L_{\mathsf c}(\delta-1)& \text{if }t>L_{\mathsf c},
      \end{cases}
    \end{split}
  \end{align}
where $\gamma_{\mathsf{tot}}$ is the number of global parities available in the
  $(L_\mathsf{c}+1)\text{-th},\ldots, L\text{-th}$ partitions and $\bar r=n-n_\mathsf{c} L$ is the number of global parities in the $(L+1)\text{-th}$ partition. By counting the
  number of global parities not erased in $L-L_{\mathsf c}$ partitions, we get
  \begin{align}
    \label{Th: Th3_Gtot}
    \gamma_{\mathsf{tot}}=
    \begin{cases}
      (n_\mathsf{c}-m)(L-L_{\mathsf c}) & \text{if }t\leq L_{\mathsf c},
      \\[5pt]
      (n_\mathsf{c}-m_1)(t-L_{\mathsf c})+(n_\mathsf{c}-m)(L-t) & \text{if }t> L_{\mathsf c}.
    \end{cases}
  \end{align}
  By substituting \eqref{Th: Th3_Gtot} into \eqref{Th: Th3_Cond}, we get (after performing some simple arithmetic) the
  condition
  \begin{align*}
    n-k-mL\geq t,
  \end{align*}
  which is valid for both cases of $t$ ($t \leq L_{\mathsf c}$ and $t > L_{\mathsf c}$).  By definition of $t$ and $m$,
  the above inequality is met with equality, and it follows that $\bm e$ is a
  correctable erasure pattern.
\end{IEEEproof}

\subsection{Proof of Step a)}

Let $\tilde{\bm E}$, $\bm W$, $\bm Z$, and $\bm O$ be submatrices of $\bm E$ as shown in \eqref{Eq: SubMats_E}. We begin
the proof by proving that each of the $n_\mathsf{c} L$ rows of the matrix $\bigl(\tilde{\bm E}\mid\bm Z\bigr)$ is a 
correctable erasure pattern, where $\tilde{\bm E}$ is defined in \eqref{Eq: Ehat_structure}. This is proved by induction
on the row partitions of $\bigl(\tilde{\bm E}\mid\bm Z\bigr)$.

\begin{description}[leftmargin=0cm]
\setlength\itemsep{1ex}
\item[Base Case.]
Consider the first row partition of $\bigl(\tilde{\bm E}\mid\bm Z\bigr)$, given by
\begin{align*}
  \left(\begin{matrix}
      \bm \pi_1 & \bm\pi_2 & \cdots & \bm \pi_L & \bm 0_{n_\mathsf{c}\times\bar r}
    \end{matrix}\right).
\end{align*}
For each row vector $\bm e_i^{(1)}$, $i\in\Nat{n_\mathsf{c}}$, in this row partition, where the subscript $i$ indicates
the row index and the superscript the row partition, consider the subvectors
$\bm e_{i,1}^{(1)},\ldots,\bm e_{i,L}^{(1)}$. From Step a) in Section~\ref{sec: Ematrix_LRC}, for all $i\in\Nat{n_\mathsf{c}}$,
the $j$-th subvectors $\bm e_{i,j}^{(1)}$ have support of cardinality $|\chi(\bm e_{i,j}^{(1)})|=m_1$ for all
$j\in\Nat{t}$, where $t=(n-k)\bmod L$, $|\chi(\bm e_{i,j}^{(1)})|=m$ for $j\in\Nat{t+1:L}$, and
$|\chi(\bm e_{i,L+1}^{(1)})|=0$. Thus, the vectors $\bm e_{i}^{(1)}$ in the first row partition of
$\bigl(\tilde{\bm E}\mid\bm Z\bigr)$ have the same structure as the erasure pattern $\bm e$ from Lemma~\ref{lem:
  Correctable_e} and are therefore erasure patterns that are correctable by $\mathcal C$. Note that the number of global
parities available in the $(L_\mathsf{c}+1)\text{-th},\ldots,L\text{-th}$ subvectors of vector $\vect{e}^{(1)}_i$,
$\gamma^{(1)}_\mathsf{tot}$, is $\gamma^{(1)}_\mathsf{tot}=\gamma_\mathsf{tot}$, hence
$\gamma^{(1)}_\mathsf{tot}+\bar{r}=\gamma_\mathsf{tot}+\bar{r}\ge \sum_{j=1}^{L_\mathsf{c}}\nu_j$ and from {\mg the
  proof of} Lemma~\ref{lem: Correctable_e} the error pattern $\vect{e}^{(1)}_i$ is correctable.

\item[Inductive Step.] Assume that the vectors $\bm e_{i}^{(l)}$, $i\in\Nat{n_\mathsf{c}}$, in the $l$-th row 
  partition of $\bigl(\tilde{\mat{E}}|\mat{Z}\bigr)$ are correctable by $\mathcal C$ and that each local code
  $\code{C}|_{\set{S}_j}$ can correct $\delta-1+\nu^{(l)}_j$ erasures, $j\in\Nat{L_\mathsf{c}}$. The row vectors are
  taken from the matrix
\begin{align*}
  \left(\begin{matrix}
      \bm\pi_{\sigma^{l-1}(1)}& \bm\pi_{\sigma^{l-1}(2)}& \cdots& \bm\pi_{\sigma^{l-1}(L)}& \bm 0_{n_\mathsf{c}\times\bar r}
    \end{matrix}\right),
\end{align*}
where $\sigma\eqdef(L\,(L-1)\cdots 1)$ denotes a \emph{cycle} whose mapping is
$L\mapsto (L-1)\mapsto \cdots\mapsto 1\mapsto L$. The $(L+1)$-th subvectors satisfy $|\chi(\bm e_{i,L+1}^{(l)})|=0$. From Lemma~\ref{lem: Capacity_LRC}, the
underlying characteristic of the vectors $\vect{e}^{(l)}_{i}$ is that they are correctable erasure patterns if the number of global
parities not erased in $\vect{e}^{(l)}_i$, $\gamma^{(l)}_\mathsf{tot}+\bar r$, is larger than or equal to $\sum_{j=1}^{L_\mathsf{c}} \nu_{j}^{(l)}$.
%

In the $(l+1)$-th row partition of $\bigl(\tilde{\mat{E}}|\mat{Z}\bigr)$, the $n_\mathsf{c}$ rows have the form
\begin{IEEEeqnarray*}{rCl}
  \begin{pmatrix}
    \bm\pi_{\sigma^l(1)}& \bm\pi_{\sigma^l(2)}& \cdots& \bm\pi_{\sigma^l(L)} &\bm 0_{n_\mathsf{c}\times\bar r}
  \end{pmatrix}. 
\end{IEEEeqnarray*}
Due to the cyclic shifts, for $j\in\Nat{L}$, all the $j$-th subvectors of the vectors $\bm e_{i}^{(l+1)}$ in row partition $l+1$, $l\in\Nat{L-1}$, have support size
$|\chi(\bm e_{i,j}^{(l+1)})|=|\chi(\bm e_{i,\sigma^{-1}(j)}^{(l)})|$. Thus, there exist two indices
$j',j''\in\Nat{L}$, $j'\neq j''$, such that
\begin{align}
  \label{Eq: Th4_inductive_cond}
  \begin{split}
    |\chi(\bm e_{i,j'}^{(l+1)})|-|\chi(\bm e_{i,j'}^{(l)})|&= |\chi(\bm e_{i,j''}^{(l)})|-|\chi(\bm e_{i,j''}^{(l+1)})|,
    \\
    |\chi(\bm e_{i,j}^{(l+1)})|&=|\chi(\bm e_{i,j}^{(l)})|,\quad\forall\,j\in\Nat{L}\setminus\{j',j''\}.
  \end{split}
\end{align}
One can see that there are at most $4$ (depending on $t$ and $L_\mathsf{c}$) choices for the pair $(j',j'')$ as follows.  
\begin{description}[leftmargin=0cm]
\item[Case 1.] $j',j''\in\Nat{L_c}$: From \eqref{Eq: Th4_inductive_cond}, it follows that
  $\nu^{(l+1)}_{j'}-\nu^{(l)}_{j'}=\nu^{(l)}_{j''}-\nu^{(l+1)}_{j''}$, $\nu_{j}^{(l+1)}=\nu_{j}^{(l)}$, and
  $\gamma_{\mathsf{tot}}^{(l)}=\gamma_{\mathsf{tot}}^{(l+1)}$. Thus, we have
  $\sum_{j=1}^{L_{\mathsf c}}\nu_j^{(l+1)}=\sum_{j=1}^{L_{\mathsf c}}\nu_j^{(l)}=\gamma_{\mathsf{tot}}^{(l)}+\bar
  r=\gamma_{\mathsf{tot}}^{(l+1)}+\bar r$.
\item[Case 2.] $j',j''\in\Nat{L_{\mathsf c}+1:L}$: From \eqref{Eq: Th4_inductive_cond}, it follows that
  $\gamma_{\mathsf{tot}}^{(l+1)}=\gamma_{\mathsf{tot}}^{(l)}$ and
  $\sum_{j=1}^{L_{\mathsf c}}\nu_j^{(l+1)}=\sum_{j=1}^{L_{\mathsf c}}\nu_j^{(l)}$. Therefore,
  $\sum_{j=1}^{L_{\mathsf c}}\nu_j^{(l+1)}=\gamma_{\mathsf{tot}}^{(l+1)}+\bar r$.
\item[Case 3.] $j'\in\Nat{L_{\mathsf c}}$, $j''\in\Nat{L_{\mathsf c}+1:L}$: From \eqref{Eq: Th4_inductive_cond}, it follows
  that $\nu^{(l+1)}_{j'}-\nu^{(l)}_{j'}=\gamma_{\mathsf{tot}}^{(l+1)}-\gamma_{\mathsf{tot}}^{(l)}$. Moreover, it can be seen that
  $\sum_{j\neq j',j\in\Nat{L_{\mathsf c}}}\nu_j^{(l+1)}=\sum_{j\neq j',j\in\Nat{L_{\mathsf c}}}\nu_j^{(l)}$. Hence, we have
  \begin{IEEEeqnarray*}{rCl}
    \sum_{j=1}^{L_{\mathsf c}}\nu_j^{(l+1)}& = &\sum_{j\neq j',j\in\Nat{L_{\mathsf c}}}\nu_j^{(l+1)}+\nu_{j'}^{(l+1)}
    \\
    & = &\sum_{j\neq j',j\in\Nat{L_{\mathsf c}}}\nu_j^{(l+1)}+(\nu^{(l+1)}_{j'}-\nu^{(l)}_{j'})+\nu_{j'}^{(l)}
    \\
    & = &\sum_{j\neq j',j\in\Nat{L_{\mathsf c}}}\nu_j^{(l)}+(\gamma_{\mathsf{tot}}^{(l+1)}-\gamma_{\mathsf{tot}}^{(l)})
    +\nu_{j'}^{(l)}
    \\
    & = &\sum_{j=1}^{L_{\mathsf c}}\nu_j^{(l)}+(\gamma_{\mathsf{tot}}^{(l+1)}-\gamma_{\mathsf{tot}}^{(l)})
    \\
    & \overset{(b)}{=} &\gamma_{\mathsf{tot}}^{(l)}+\bar{r}+(\gamma_{\mathsf{tot}}^{(l+1)}-\gamma_{\mathsf{tot}}^{(l)})
    \\[1mm]
    & = &\gamma_{\mathsf{tot}}^{(l+1)}+\bar{r},\nonumber
  \end{IEEEeqnarray*}
  where $(b)$ holds since $\sum_{j=1}^{L_{\mathsf c}}\nu_j^{(l)}=\gamma_{\mathsf{tot}}^{(l)}+\bar{r}$.
\item[Case 4.] $j'\in\Nat{L_{\mathsf c}+1:L}$, $j''\in\Nat{L_{\mathsf c}}$: Following an argumentation similar to Case $3$,
  we have $\sum_{j=1}^{L_{\mathsf c}}\nu_j^{(l+1)}=\gamma_{\mathsf{tot}}^{(l+1)}+\bar r$.
\end{description}
In each of the above cases we see that the condition $\gamma_{\mathsf{tot}}+\bar r \ge \sum_{j=1}^{L_{\mathsf c}}\nu_j^{(l+1)}$
is satisfied (with equality). From the proof of Lemma~\ref{lem: Correctable_e}, the $n_\mathsf{c}$ rows in the $(l+1)\text{-th}$ row 
partition of $\bigl(\tilde{\mat{E}}|\mat{Z}\bigr)$ are correctable by $\mathcal C$, which completes the inductive step.
\end{description}

The rows of $(\bm W\mid\bm O)$ as shown in Step a) in Section~\ref{sec: Ematrix_LRC} have support corresponding to only the parity
symbols of $\mathcal C$. Thus, these rows are all correctable by $\mathcal C$, and it follows from the above arguments
that each row of $\bm E$ is an erasure pattern that is correctable by $\mathcal C$.

\subsection{Proof of Step b)} \label{sec:proof_step_b}

We now address the second part of the proof. Note that the columns with coordinates in $\mathcal P_j$, $j\in\Nat{L}$, have column weight $n-k+\bar{r}$ after Step a). Step b) involves the swapping of one entries from these coordinates with zero entries in the column coordinates of $\bm Z$. 
The swapping is done to ensure that the column weight of the columns indexed by $\mathcal P_j$, $j\in\Nat{L}$, is reduced to $n-k$, while those of the columns of $\bm Z$ are increased to $n-k-\bar{r}$. Since $\bm O$ is an all-one matrix, the columns of $\bm E$ with indices in $\mathcal P_{L+1}$ have also weight $n-k$. It is possible to show that such a swapping always exists. Overall, the resulting matrix $\bm E$ is $(n-k)$-column regular. To ensure that the erasure patterns are correctable, we use Lemma~\ref{lem: Capacity_LRC}. For each row, 
\begin{align}
\label{Eq: CountingArg}
	\sum_{j=1}^{L_{\mathsf c}}\nu_j\leq \gamma_{\mathsf{tot}}+   \gamma_{L+1},
\end{align} 
where $\gamma_{L+1}$ is number of nonerased parity coordinates in column partition $L+1$, must hold. 
Clearly, if for a certain row of $(\tilde{\bm E}\mid\bm Z)$ a one from a column from a column partition in $\Nat{L_{\mathsf c}+1:L}$ (corresponding to $\tilde{\bm E}$) is swapped with a zero in a column from partition $L+1$ (corresponding to $\bm Z$), then the resulting erasure pattern is still correctable by $\mathcal C$ as \eqref{Eq: CountingArg} is still valid. On the other hand, for $j\in\Nat{L_{\mathsf c}}$, if for a certain row of $(\tilde{\bm E}\mid\bm Z)$ a one from the $j$-th column partition is swapped with a zero  in the $(L+1)$-th column partition, then such a row is still a correctable erasure pattern provided that $\nu_j > 0$ before the swap. This is easy to see as the swapping procedure reduces $\nu_j$ and  $\gamma_{L+1}$ by one. Thus, \eqref{Eq: CountingArg} is still satisfied. From the aforementioned arguments and the fact that each row of any row partition of $(\tilde{\bm E}\mid\bm Z)$ has at most $\bar r$ swaps of ones occurring from the set of $\Nat{L}$ column partitions and zeroes from the $(L+1)$-th partition, it follows that the swaps according to Step b)  are valid over all $\bar{r}$ iterations  (valid in the sense that the resulting erasure patterns are correctable by $\mathcal{C}$) if 
\begin{align}
\label{Eq: SwapCond}
	\sum_{j=1}^{L_{\mathsf c}}\nu_j+\sum_{j=L_{\mathsf c}+1}^{L}(m-(\delta-1))\geq\bar r.
\end{align}
This is a counting argument, where according to Step b) {for each row} we restrict swapping $\nu_j$ coordinates in the $j$-th column partition, $j\in\Nat{L_{\mathsf c}}$, and $m-(\delta-1)$ coordinates in the column partitions $\Nat{L_{\mathsf c}+1:L}$ to make sure (following the arguments above) that the resulting erasure pattern after the swap is correctable by $\mathcal{C}$.  Using that $\nu_j=\rho_j-(\delta-1)$ and $t=n-k-mL$, it can be shown that the left hand side of  \eqref{Eq: SwapCond} can be lowerbounded by 
 $n-k-L(\delta-1)$ 
 when $t \leq L_{\mathsf c}$. Setting  $n=\bar r+L(r+\delta-1)$ and $k=L_{\mathsf c}r$, it follows that \eqref{Eq: SwapCond}   reduces to $L\geq L_{\mathsf c}$. 
By definition, this is always true. When $t > L_{\mathsf c}$, the left hand side of  \eqref{Eq: SwapCond}  is equal to $n-k-L(\delta-1) + L_{\mathsf c} - t$, and it can be shown that this is always larger than or equal to $\bar{r}$, since $t \leq L$ (details omitted for brevity). It follows that for all $\bar{r}$ iterations  and for all row partitions in the systematic procedure in Step b) there exists a valid swap such that  the resulting erasure patterns are still correctable by $\mathcal C$.

\section{Proof of Theorem~\ref{th: CstarDproof}}
\label{Appendix: Proof3}

To prove the theorem we need the following lemma.
\begin{lemma}
\label{th: BinUUVrank}
Let $\mathcal C$ be an  $[n=2n_1,k=k_1+1]$ binary code constructed from an $[n_1,k_1]$ code $\mathcal U$ through the
$(\mathcal{U} \mid \mathcal{U}+\mathcal{V})$ construction, where $\mathcal{V}$ is an $[n,1]$ binary repetition code. The
generator matrix $\bm G^{\mathcal C}$ of $\mathcal C$ is given in \eqref{Eq: UUV const}. Let
$\bar{\mathcal C}=\mathcal C$ and $\bm G^{\bar{\mathcal C}}=\bm G^{\mathcal C}$. Then, the code
$\tilde{\mathcal C}=\mathcal C\circ\bar{\mathcal C}$ is a vector space of dimension
\begin{align}
  \label{Eq: rank}
  \dim(\tilde{\mathcal C})\leq\begin{cases}
    k_1+n_1+1 & \text{if }n_1-k_1\leq\binom{k_1}{2},
    \\
    2k_1+\binom{k_1}{2}+1 & \text{otherwise}.
  \end{cases}
\end{align}
\end{lemma}

\begin{IEEEproof}
  From Definition~\ref{def: Hadamard Product}, we know that $\tilde{\bm c}\in\tilde{\mathcal C}$ has the form
  $\tilde{\bm c}=(c_1\bar{c}_1,\ldots,c_n\bar{c}_n)$, where $\bm c = (c_1,\ldots,c_n) \in \code{C}$ and
  $\bar{\bm c} = (\bar{c}_1,\ldots,\bar{c}_n) \in \bar{\code{C}}$. Considering $\bm G^{\mathcal C}=(g_{i,j}^{c})$ and
  $\bm G^{\bar{\mathcal C}}=(g_{i,j}^{\bar c})$, the vector space $\tilde{\mathcal C}$ is spanned by the row space of
\begin{align}
\label{Eq: BasisVecforHad}
\bm G^{\tilde{\mathcal C}}=\left(\begin{matrix}
	g_{1,1}^c\bm g^{\bar c}_1 & g_{1,2}^c\bm g^{\bar c}_2 & \cdots & g_{1,n}^c\bm g^{\bar c}_n\\
	g_{2,1}^c\bm g^{\bar c}_1 & g_{2,2}^c\bm g^{\bar c}_2 & \cdots & g_{2,n}^c\bm g^{\bar c}_n\\
	\vdots & \vdots & \cdots & \vdots\\
	g_{k,1}^c\bm g^{\bar c}_1 & g_{k,2}^c\bm g^{\bar c}_2 & \cdots & g_{k,n}^c\bm g^{\bar c}_n\\
\end{matrix}\right),
\end{align}
where the vector $\bm g^{\bar c}_{j}$, $j \in \Nat{n}$, denotes the $j$-th column vector of $\bm
G^{\bar{\code{C}}}$. The matrix $\bm G^{\tilde{\mathcal C}}$ is a matrix consisting of $k^2$ row vectors (corresponding to codewords of
$\tilde{\mathcal C}$) of length $n$. We divide $\mat{G}^{\tilde{\code{C}}}$ into $k$ submatrices
$\mat{G}^{\tilde{\code{C}}}_i$, where
$\mat{G}^{\tilde{\code{C}}}_i=
(g_{i,1}^c\vect{g}^{\bar{c}}_1|g_{i,2}^c\vect{g}^{\bar{c}}_2|\ldots|g_{i,n}^c\vect{g}^{\bar{c}}_n)$, $i\in\Nat{k}$ (see \eqref{Eq: BasisVecforHad}). From \eqref{Eq: UUV const} and since $\bm G^{\mathcal C}=\bm G^{\bar{\mathcal C}}$, we have
$g^c_{k,j}=g^{\bar c}_{k,j}=0$, $j \in \Nat{n_1}$, and $g^c_{k,j}=g^{\bar c}_{k,j}=1$, $j \in \Nat{n_1+1:n}$. Therefore,
\eqref{Eq: BasisVecforHad} can be expanded to
\begin{align}%
\label{Eq: HadProdUUV}
  \bm G^{\tilde{\mathcal C}}=\Scale[0.63]{\left(\begin{array}{@{}rrrrrr@{}}
		g_{1,1}^c\left(\begin{matrix} g_{1,1}^{\bar c}\\g_{2,1}^{\bar c}\\\vdots\\g_{k_1,1}^{\bar c}\\0\end{matrix}\right) & \cdots & g_{1,n_1}^c\left(\begin{matrix} g_{1,n_1}^{\bar c}\\g_{2,n_1}^{\bar c}\\\vdots\\g_{k_1,n_1}^{\bar c}\\0\end{matrix}\right) & g_{1,n_1+1}^c\left(\begin{matrix} g_{1,n_1+1}^{\bar c}\\g_{2,n_1+1}^{\bar c}\\\vdots\\g_{k_1,n_1+1}^{\bar c}\\1\end{matrix}\right) & \cdots & g_{1,n}^c\left(\begin{matrix} g_{1,n}^{\bar c}\\g_{2,n}^{\bar c}\\\vdots\\g_{k_1,n}^{\bar c}\\1\end{matrix}\right)\\[1cm]
		g_{2,1}^c\left(\begin{matrix} g_{1,1}^{\bar c}\\g_{2,1}^{\bar c}\\\vdots\\g_{k_1,1}^{\bar c}\\0\end{matrix}\right) & \cdots & g_{2,n_1}^c\left(\begin{matrix} g_{1,n_1}^{\bar c}\\g_{2,n_1}^{\bar c}\\\vdots\\g_{k_1,n_1}^{\bar c}\\0\end{matrix}\right) & g_{2,n_1+1}^c\left(\begin{matrix} g_{1,n_1+1}^{\bar c}\\g_{2,n_1+1}^{\bar c}\\\vdots\\g_{k_1,n_1+1}^{\bar c}\\1\end{matrix}\right) & \cdots & g_{2,n}^c\left(\begin{matrix} g_{1,n}^{\bar c}\\g_{2,n}^{\bar c}\\\vdots\\g_{k_1,n}^{\bar c}\\1\end{matrix}\right)\\
		\multicolumn{1}{c}{\vdots} & \cdots & \multicolumn{1}{c}{\vdots} & \multicolumn{1}{c}{\vdots} & \cdots & \multicolumn{1}{c}{\vdots}\\
		0\left(\begin{matrix} g_{1,1}^{\bar c}\\g_{2,1}^{\bar c}\\\vdots\\g_{k_1,1}^{\bar c}\\0\end{matrix}\right) & \cdots & 0\left(\begin{matrix} g_{1,n_1}^{\bar c}\\g_{2,n_1}^{\bar c}\\\vdots\\g_{k_1,n_1}^{\bar c}\\0\end{matrix}\right) & 1\left(\begin{matrix} g_{1,n_1+1}^{\bar c}\\g_{2,n_1+1}^{\bar c}\\\vdots\\g_{k_1,n_1+1}^{\bar c}\\1\end{matrix}\right) & \cdots & 1\left(\begin{matrix} g_{1,n}^{\bar c}\\g_{2,n}^{\bar c}\\\vdots\\g_{k_1,n}^{\bar c}\\1\end{matrix}\right)
	\end{array}\right)}.
\end{align}%
Furthermore, let $\bm G^{\mathcal U}=(g_{i,j}^u)$ be the generator matrix of $\mathcal U$. From \eqref{Eq: UUV const}, we
have $g^c_{i,j}=g_{i,j}^{\bar c}=g^u_{i,j}$ for $i\in\Nat{k_1}$ and $j\in\Nat{n_1}$. For $i,j\in\Nat{k}$, we denote the
$i$-th row of the $j$-th submatrix $\mat{G}^{\tilde{\mathcal C}}_j$ as $\bm w_i^{(j)}$. For $i\in\Nat{k-1}$, the $i$-th
row of the $i$-th submatrix $\mat{G}^{\tilde{\mathcal C}}_i$ is given as
\begin{align}
  \label{Eq: BasisVec_U}
  \bm w_{i}^{(i)}=(g^c_{i,1}g^{\bar c}_{i,1}, g^c_{i,2}g^{\bar c}_{i,2},\ldots,g^c_{i,n}g^{\bar c}_{i,n}).
\end{align}
Since $g_{i,j}^{\bar c}=g_{i,j}^c\in\GF(2)$, \eqref{Eq: BasisVec_U} reduces to
$\bm w_{i}^{(i)}=(g^c_{i,1}, g^c_{i,2},\ldots,g^c_{i,n})$. Furthermore, from \eqref{Eq: UUV const} we see that
$g^c_{i,j}=g^{c}_{i,n_1+j}=g^{u}_{i,j}$, $j\in\Nat{n_1}$, $i \in \Nat{k_1}$. Therefore, these $k_1=k-1$ rows form the $k_1$ basis vectors
of the code space $(\mathcal U,\mathcal U)$ and can be arranged in a matrix as
\begin{align}
  \label{Eq: proofP1}
  \left(\begin{matrix}
      \bm G^{\mathcal U} & \bm G^{\mathcal U}
    \end{matrix}\right).
\end{align}
The $k$-th row of $\mat{G}^{\tilde{\code{C}}}_i$ can be written as
\begin{align*}
\begin{split}
  \bm w_{k}^{(i)}&=(\overbrace{0,0,\ldots,0}^{n_1}, g^c_{i,n_1+1},g^c_{i,n_1+2},\ldots,g^c_{i,n})\\&\overset{(c)}{=}
  (0,0,\ldots,0,g_{i,1}^u,g_{i,2}^u,\ldots,g_{i,n_1}^u),
\end{split}
\end{align*}
where $(c)$ results from the structure of $\bm G^{\mathcal C}$ in \eqref{Eq: UUV const}. Stacking together the $k$-th row
of all $k_1$ submatrices $\mat{G}^{\tilde{\code{C}}}_i$, $i\in\Nat{k_1}$, results in the $k_1$ row vectors
\begin{align}
  \label{Eq: proofP2}
  \left(\begin{matrix}
      \bm 0_{k_1\times n_1}&\bm G^{\mathcal U}
    \end{matrix}\right).
\end{align}
In a similar way, the rows $\bm w_i^{(k)}$, $i\in\Nat{k}$, of the $k$-th submatrix $\mat{G}^{\tilde{\code{C}}}_k$ result
in the matrix
\begin{align}
  \label{Eq: proofP3}
  \left(\begin{matrix}
      \bm 0_{k_1\times n_1} & \bm G^{\mathcal U}\\
      \bm 0_{1\times n_1} & \bm 1_{1\times n_1}
    \end{matrix}\right).
\end{align}
Of the remaining $(k-1)(k-2)$ rows in \eqref{Eq: HadProdUUV}, since $\code{C}=\bar{\code{C}}$, there exist
$\binom{k_1}{2}$ distinct rows as follows,
\begin{align*}
  \bm{\Theta}=\left(\begin{matrix}
      g_{1,1}^cg_{2,1}^{\bar c} & g_{1,2}^cg_{2,2}^{\bar c} & \cdots & g_{1,n}^cg_{2,n}^{\bar c}\\
      g_{1,1}^cg_{3,1}^{\bar c} & g_{1,2}^cg_{3,2}^{\bar c} & \cdots & g_{1,n}^cg_{3,n}^{\bar c}\\
      \vdots & \vdots & \cdots & \vdots\\
      g_{1,1}^cg_{k_1,1}^{\bar c} & g_{1,2}^cg_{k_1,2}^{\bar c} & \cdots & g_{1,n}^cg_{k_1,n}^{\bar c}
      \\[5pt]      
      g_{2,1}^c g_{3,1}^{\bar c} & g_{2,2}^c g_{3,2}^{\bar c} & \cdots & g_{2,n}^c g_{3,n}^{\bar c}\\
      \vdots & \vdots & \cdots & \vdots\\
      g_{2,1}^cg_{k_1,1}^{\bar c} & g_{2,2}^cg_{k_1,2}^{\bar c} & \cdots & g_{2,n}^cg_{k_1,n}^{\bar c}\\[5pt]
      \vdots & \vdots & \cdots & \vdots\\[5pt]
      g_{k_1-1,1}^cg_{k_1,1}^{\bar c} & g_{k_1-1,2}^cg_{k_1,2}^{\bar c} & \cdots & g_{k_1-1,n}^cg_{k_1,n}^{\bar c}
    \end{matrix}\right).
\end{align*}
Furthermore, from the construction of $\bm G^{\mathcal C}$ in \eqref{Eq: UUV const}, we have
$(g^c_{i,1},\ldots,g^c_{i,n_1})=(g^c_{i,n_1+1},\ldots,g^c_{i,n})=(g^u_{i,1},\ldots,g^u_{i,n_1})$, $i\in\Nat{k_1}$ and
because $\bar{\code{C}}=\mathcal C$, we have
$(g^{\bar c}_{i,1},\ldots,g^{\bar c}_{i,n_1})=(g^{\bar c}_{i,n_1+1},\ldots,g^{\bar c}_{i,n})$. Therefore,
\begin{align}
  \label{Eq: proofP4}
  \bm \Theta=\left(\begin{matrix}
      \bm \theta_{\binom{k_1}{2}\times n_1} & \bm \theta_{\binom{k_1}{2}\times n_1}
    \end{matrix}\right),
\end{align}
where $\bm \theta_{\binom{k_1}{2}\times n_1}$ is a binary matrix of size $\binom{k_1}{2}\times n_1$. From \eqref{Eq:
  proofP1}--\eqref{Eq: proofP4}, $\bm G^{\tilde{\mathcal C}}$ can be written as
\begin{align*}
  \bm G^{\tilde{\mathcal C}}=\left(\begin{matrix}
      \bm G^{\mathcal U} & \bm G^{\mathcal U}\\
      \bm 0_{k_1\times n_1} & \bm G^{\mathcal U}\\
      \bm 0_{k_1\times n_1} & \bm G^{\mathcal U}\\
      \bm \theta_{\binom{k_1}{2}\times n_1} & \bm \theta_{\binom{k_1}{2}\times n_1}\\
      \bm 0_{1\times n_1} & \bm 1_{1\times n_1}
    \end{matrix}\right).
\end{align*}
Using Gaussian elimination, $\bm G^{\tilde{\mathcal C}}$ can be reduced to
\begin{align}
  \label{Eq: proofP5}
  \bm G^{\tilde{\mathcal C}}=\left(\begin{matrix}
      \bm G^{\mathcal U} & \bm 0_{k_1\times n_1}\\
      \bm 0_{k_1\times n_1} & \bm G^{\mathcal U}\\
      \bm 0_{k_1\times n_1} & \bm 0_{k_1\times n_1}\\
      \bm \theta_{\binom{k_1}{2}\times n_1} & \bm \theta_{\binom{k_1}{2}\times n_1}\\
      \bm 0_{1\times n_1} & \bm 1_{1\times n_1}
    \end{matrix}\right).
\end{align}
Let $\bm G^{\mathcal U}=\bigl(\bm I_{k_1}|\bm P_{k_1\times (n_1-k_1)}\bigr)$, where $\bm P_{k_1\times (n_1-k_1)}$ is the
parity matrix of size ${k_1\times (n_1-k_1)}$. We now count the number of independent rows in the matrix
\begin{align*}
  \left(\begin{matrix}
      \bm G^{\mathcal U}\\
      \bm \theta_{\binom{k_1}{2}\times n_1}
    \end{matrix}\right)=\left(\begin{matrix}
    \bm I_{k_1} & \bm P_{k_1\times (n_1-k_1)}
      \\[1mm]
      \multicolumn{2}{c}{\bm\theta_{\binom{k_1}{2}\times n_1}}
    \end{matrix}\right).
\end{align*}
Upon performing Gaussian elimination, we get
\begin{align*}
  \left(\begin{matrix}
      \bm I_{k_1} & \bm P_{k_1\times (n_1-k_1)} \\
      \bm 0_{(^{k_1}_2)\times k_1} & \bm \Delta_{\binom{k_1}{2}\times (n_1-k_1)}
    \end{matrix}\right),
\end{align*}
where $\bm \Delta_{\binom{k_1}{2}\times (n_1-k_1)}$ is a matrix of dimensions $\binom{k_1}{2}\times
(n_1-k_1)$ with elements in $\GF(2)$. Hence, we have $\rank{\mat{\Delta}}\leq\min\bigl(\binom{k_1}{2},(n_1-k_1)\bigr)$. From this and \eqref{Eq:
  proofP5}, we can easily see that
  \begin{align*}
    \tilde{k}&=\bigrank{\bm G^{\tilde{\mathcal C}}}  \\ 
  &=  k_1+k_1+\rank{\mat{\Delta}}+1 \\
  & \leq 
  \begin{cases}
    k_1+n_1+1 & \text{if } n_1-k_1 \leq\binom{k_1}{2},
    \\
    2k_1+\binom{k_1}{2}+1 & \text{otherwise}.
  \end{cases}
  \end{align*}
\end{IEEEproof}
Lemma~\ref{th: BinUUVrank} gives an upper bound on the dimension of $\tilde{\mathcal C}$. In order to prove
$\dim(\tilde{\mathcal C})<n$, we check when the upper bound in \eqref{Eq: rank} is at most $n-1$. For the first case in \eqref{Eq:
  rank}, we need to show
\begin{align*}
  \tilde{k}\leq k_1+n_1+1 \leq 2n_1-1.
\end{align*}
Clearly, this is true since $n_1\geq k_1+2$ by assumption. For the second case in \eqref{Eq: rank} we have to show
\begin{align*}
  \tilde{k}\leq 2k_1+\binom{k_1}{2}+1 \leq 2n_1-1.
\end{align*}
Since $n_1> \binom{k_1}{2}+k_1$, the above inequality reduces to
\begin{align*}
  \binom{k_1}{2} > 2.
\end{align*}
Clearly, this is true for $k_1\in\Nat{3:\infty}$. In the following, we argue for $k_1\in\Nat{2}$. Since $n_1\geq k_1+2$ by assumption,
we have
\begin{align*}
  2n_1-1\geq2(k_1+2)-1=2k_1+3>2k_1+\binom{k_1}{2}+1,
\end{align*}
for $k_1\in\Nat{2}$. Therefore, $\dim(\tilde{\mathcal C})<n$ for $n_1\geq k_1+2$.

\end{document}